\ifx\destination\undefined
  \def\destination{aanda}  
\fi
\def\arxiv{arxiv}
\def\aanda{aanda}
\def\publisher{publisher}
\RequirePackage{snapshot}
\PassOptionsToPackage{dvipsnames}{xcolor}
\PassOptionsToPackage{colorlinks=True,allcolors=RoyalBlue}{hyperlink}

\RequirePackage[2020-02-02]{latexrelease}

\ifx\destination\aanda
  \documentclass[ut8,T1,longauth]{aa}
  \usepackage{natbib}
  \bibpunct{(}{)}{;}{a}{}{,} 
\fi

\usepackage{supertabular}

\usepackage[utf8]{inputenc}
\usepackage[T1]{fontenc}

\usepackage{aas_macros} 
\usepackage{siunitx}
\DeclareSIPrefix\micro{\text{\textmu}}{-3}
\usepackage{graphicx}
\usepackage[title]{appendix}
\usepackage{booktabs}
\usepackage[skip=4pt]{caption}
\usepackage{here}
\graphicspath{{gfx/}}

\usepackage{lipsum}
\usepackage{pgf}
\usepackage{import}
\usepackage{tikz}

\usepackage{url}
\usepackage{xspace}
\usepackage[colorlinks=True,
            linkcolor=purple,         
            citecolor=RoyalBlue,    
            filecolor=purple,         
            urlcolor=purple           
           ]{hyperref}
\usepackage[nameinlink,capitalize]{cleveref}
\Crefname{section}{Sect.}{Sects.} 

\usepackage{textcomp}
\usepackage{ulem}

\usepackage[switch]{lineno}

\newcommand{\KM}[1]{\textcolor{magenta}{[#1]}}

\newcommand{\numb}[1]{\textcolor{orange}{#1}}
\renewcommand{\numb}[1]{\textcolor{black}{#1}}
\newcommand{\miriade}{\texttt{Miriade}\xspace}
\newcommand{\ssodnet}{\texttt{SsODNet}\xspace}
\newcommand{\rocks}{\texttt{rocks}\xspace}
\newcommand{\topcat}{\texttt{TOPCAT}\xspace}
\newcommand{\astropy}{\texttt{astropy}\xspace}

\newcommand{\Genoid}{\texttt{Genoid}\xspace}
\newcommand{\genoid}{\texttt{genoid}\xspace}

\renewcommand{\KM}[1]{\textcolor{black}{#1}}
\usepackage{placeins}

\begin{document}





\ifx\destination\aanda
  \title{
  A dynamical dichotomy in large binary asteroids}
  \subtitle{}
  \titlerunning{Dynamics of asteroid binaries}
  \authorrunning{Minker et al.}

  \author{%
    K.~Minker\inst{1}    \and 
    B.~Carry\inst{1}     \and 
    F.~Vachier\inst{2} \and 
    M.~Marsset\inst{3}   \and 
    J.~\v{D}urech\inst{4}   \and 
    J.~Hanu\v{s}\inst{4}   \and 
    L.~Liberato\inst{1}\and
    W.~J.~Merline\inst{5}  \and 
    J.~L.~Margot\inst{9}  \and 
    C.~Dumas\inst{13}  \and
    L.~M.~Close\inst{11} \and
    A.~Conrad\inst{12}\and
    W.~M.~Grundy\inst{6} \and     
    R.~Behrend\inst{7}   \and
    R.~Roy\inst{14}   \and 
    J.~Berthier\inst{2}   \and 
    I.~Sokova\inst{16} \and  
    E.~Sokov\inst{16}  \and       
    D.~Gorshanov\inst{16}  \and
    M.~Ferrais\inst{10}    \and     
    E.~Jehin\inst{8}      \and 
    A.~Martin\inst{15}
    \and
    K.~B.~Alton\inst{17}
    }

\institute{
    Universit{\'e} C{\^o}te d'Azur, Observatoire de la C{\^o}te d'Azur, CNRS, Laboratoire Lagrange, Bd de l'Observatoire, CS 34229, 06304 Nice cedex 4, France
    \email{kate.minker@oca.eu}
    \label{i:oca}
    \and 
     \KM{LTE, Observatoire de Paris, Université PSL, Sorbonne Université, Université de Lille, LNE, CNRS, 61 Avenue de l’Observatoire, 75014 Paris, France}
    \label{i:imcce}
    \and
    European Southern Observatory (ESO), Alonso de Cordova 3107, 1900, Casilla Vitacura, Santiago, Chile
    \label{i:ESO}
    \and
    Charles University, Faculty of Mathematics and Physics, Institute of Astronomy, V Hole\v sovi\v ck\'ach 2, 180 00 Prague, Czech Republic
    \label{i:prague}
    \and
    Southwest Research Institute, 1301 Walnut St. \#400, Boulder, CO 80302, USA
    \label{i:swri}
    \and
    Lowell Observatory, 1400 W. Mars Hill Rd., Flagstaff AZ 86001
    \label{i:lowell}
    \and
    Observatoire de Genève, 1290 Sauverny, Switzerland
    \label{i:geneva}
    \and
    Space sciences, Technologies \& Astrophysics Research (STAR) Institute, University of Liege, Liege, Belgium
    \label{i:trappist}
    \and
    UCLA Department of Earth, Planetary, and Space Sciences, Department of Physics and Astronomy
    \label{i:UClAastro}
    \and
    Florida Space Institute, University of Central Florida, 12354 Research Parkway, Partnership 1 building, Orlando, FL 32828, USA
    \label{i:florida}
    \and
    Steward Observatory, N420, Department of Astronomy, University of Arizona, 933 N. Cherry Ave. Tucson, AZ 85721
    \label{i:steward}
    \and  
    Large Binocular Telescope Observatory, University of Arizona, Tucson, AZ 85721, USA
    \label{i:lbto}
    \and
    Royal Observatory Edinburgh, Blackford Hill, Edinburgh, EH9 3HJ, United Kingdom
    \label{i:edinborough}
    \and
    Observatoire de Blauvac, 293 chemin de St Guillaume, F-84570 Blauvac, France
    \label{i:blauvac}
    \and
    Turtle Star Observatory, Friedhofstr. 15, 45478 Mülheim-Ruhr, Germany
    \label{i:turtle}
    \and
    The Central Astronomical Observatory of the Russian Academy of Sciences at Pulkovo, Pulkovo Observatory, 65 Pulkovskoye shosse, 196140 St. Petersburg
    \label{i:Pulcovo}
    \and
    UnderOak Observatory, Cedar Knolls, NJ USA
    \label{i:underoaks}
}

  \date{Received date / Accepted date }
   \keywords{Minor planets, asteroids: individual: (762) Pulcova, (283) Emma; Techniques: high angular resolution; Methods: observational }

  \abstract
   {
   No less than 15\% of large (diameter greater than 140\,km) asteroids have satellites. The commonly accepted mechanism for their
   formation is post-impact reaccumulation. However, the detailed physical and dynamical properties of these systems are not well understood, and many of them have not been studied in detail.
   }
   {
   \KM{We aim to study the population of large binary asteroid systems, in part through the characterization of (283) Emma and (762) Pulcova.}
   To do so, we compare the gravitational fields predicted from the shape of the primary
    body with the non-Keplerian gravitational components identified in orbital models of the satellites of each system.
    \KM{We also aim to contextualize these systems in the greater population of large binary systems, providing clues to asteroid satellite formation.}
    }
   {
   We reduce all historical high-angular-resolution adaptive-optics (AO) images from ground-based telescopes to conduct astrometric and photometric measurements of each system's components.
   We then determine orbital solutions for each system using the \genoid algorithm.
   We model the shapes of the system primaries using lightcurve-inversion techniques scaled with stellar occultations and AO images, and we develop internal structure models using SHTOOLS. \KM{Finally, we compare the distribution of the physical and orbital properties of the known binary asteroid systems.}
  }
   {
   We find a very low residual orbital solution for Emma with a
   gravitational quadrupole
   $J_2$ value significantly lower than what is expected from the shape model, implying that Emma has a significantly non-homogeneous internal
    structure, and an overall bulk density of $0.9\pm0.3$~g~cm-3$^{-3}$. The circular,
     co-planar orbit of Pulcova's satellite leaves substantial ambiguity in the orbital solution.
     We also
     find that the differences between these systems reflect an overall dichotomy within the population of large binary systems, with a strong correlation between \KM{primary elongation and satellite eccentricity observed in one group.}
     }
   {We determine that there may be two distinct \KM{formation pathways influencing the end-state dichotomy in these binary systems}, and that (762) Pulcova and (283) Emma belong to the two separate groups.
   }

  \maketitle

\fi


\section{Introduction}%
\label{sec:introduction}%

From the discovery of tiny Dactyl orbiting (243) Ida in 1993
\citep{1995Natur.374..783C} to the DART impact on Dimorphos in September 2022
\citep{2021PSJ.....2..173R}, asteroid satellites have proven time and time again
to be essential objects to understand the physical and dynamical
properties of the entire asteroid population
\citep{2015-AsteroidsIV-Margot}.
Notably, dynamical characterization of multiple asteroid systems provides
one of the only reliable ways in which to determine the mass and density
of the primary body from remote observations
\citep{2002-AsteroidsIII-4.2-Britt, 2012P&SS...73...98C, 2015-AsteroidsIV-Scheeres}.
In particular, direct observations of large binary systems are the most
efficient method of remotely probing asteroid internal structures
\citep[e.g.,][]{2021A&A...650A.129C, 2022A&A...662A..71F},
providing information about their formation and evolution that can then
be extrapolated to the entire asteroid population
\citep{2015-AsteroidsIV-Walsh, 2018NatAs...2..878N, 2021A&A...654A..56V}.

Increasing datasets and advancing instrumentation and image analysis techniques
have allowed us to develop progressively intricate models for asteroid shapes
\citep[e.g.,][]{2010Icar..205..460C, 2011-IPI-5-Kaasalainen, 2015A&A...576A...8V, 2015-AsteroidsIV-Durech}
and the orbits of asteroid moons
\citep[see][for instance]{2012AA...543A..68V, 2012AJ....143...24F, 2021A&A...653A..56B}.
Over
long observational baselines, small non-Keplerian influences can be detectable in orbital solutions,
even in relatively
low-resolution datasets.
These non-Keplerian gravitational influences can then be compared to expected perturbations that
would result from the gravitational field of the primary of the system
\citep{2019A&A...623A.132C, fangsylvia, 2022A&A...662A..71F, 2023A&A...677A.189F},
in the case that the system has a totally homogeneous internal structure.
Since the primaries of most binary systems are highly non-spherical \citep{2021A&A...654A..56V}, these influences can be substantial.

Here, we present two orbital models for large, carbonaceous, main belt binary systems (283) Emma and (762) Pulcova.
These systems both have unusually large satellites compared to the diameters of their primary body, and are the smallest known binary systems with primaries that can be resolved through adaptive optics imaging.
In \Cref{sec:observations}, we present the observational datasets included in this study,
where they are sourced, and our data reduction process;
in \Cref{sec:shapes} and \Cref{sec:orbits} we present
the shape models used and developed for this study and
multipole decompositions thereof, and our method
for orbit determination and
the orbital solutions for the two systems.
In \Cref{sec:physicalprops} we present the physical properties derived from the shape and orbital models described in the previous two sections.
In \Cref{sec:discussion}  we discuss the implications of these results.
Finally, \Cref{sec:deadbugs} contextualizes these results in comparison to
the full population of large binary asteroid systems.


\section{Observations and data reduction}
\label{sec:observations}

\subsection{High-angular resolution images}

To construct the most accurate orbit possible, we used all available high-angular
resolution archive data of these objects from 3.5-10\,m class telescopes.
In total, we measured
a total of 68 and 56 positions of the satellites, S/2000 (762) 1 (henceforth nicknamed Pulcamoon) and S/2003 (283) 1 (henceforth nicknamed Emmoon), respectively.
The data for Pulcova (Emma) spans twenty (ten) years in total, from February 22nd, 2000 to September 11th, 2019 (July 14th, 2003 to June 16th 2013).
Included in this set are observations previously published by \cite{merlinepulcova, merlineemma, marchis2008circular, marchis2008eccentric},
for which we have re-reduced the data from the raw images,
and unpublished public archival data.
A summary of these observations can be viewed in \Cref{tab:pulcamoon} and \Cref{tab:emmoon}.
Due to a significant known issue with field orientation in Gemini/Hokupa'a
data\footnote{\url{https://www.gemini.edu/sciops/instruments/uhaos/uhaosIndex.html}},
no astrometric measurements from these observations were included in this study.

Images from the ESO VLT\footnote{ESO programs: 071.C-0669, 074.C-0052, 079.C-0528, 089.C-0944, 091.C-0526}
were acquired with NACO \citep{2003-SPIE-4841-Lenzen, 2003-SPIE-4839-Rousset},
images from Gemini\footnote{Gemini programs: GN-2004B-C-5, GN-2004B-DD-100, GN-2004B-DD-7, GN-2004B-Q-37, GN-2006A-Q-75, GN-2009B-C-7, GN-2010A-C-6, GN-2010B-Q-99, GN-2011A-Q-97}
were acquired with Hokupa'a or NIRI \citep{graves1998HokupaaSPIE, 2003-PASP-115-Hodapp},
images from Keck II\footnote{Keck programs: N32N2, N17N2, N22N2, C29N2, A283N2L, N118, N24N2, and engineering time}
were acquired with NIRC2 \citep{2004-AppOpt-43-vanDam}, and
images from CFHT\footnote{CFHT run IDs: F59, 01AH10B, 01AF35B, 01BF48, 02AH19A, 02AD06, 02AF43, 02AF19, 02AF39} were acquired with PUEO/aobir.
Images in an observational epoch were stacked in groups with the same exposure time and filter, so for example H band images taken consecutively to K band images would be separated into two independent observations. Usual image processing, including sky subtraction, bad pixel removal, flat-field, was performed
using calibration frames \citep{2008AA...478..235C},
and halo subtraction techniques were applied to enhance contrast
\citep{2018Icar..309..134P}.

\begin{figure}
   \centering
   \includegraphics[width=\columnwidth]{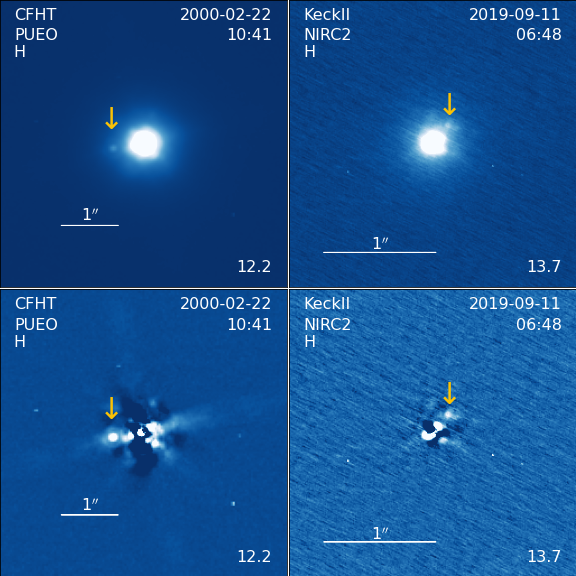}
   \caption{First (left) and most recent (right) images of (762) Pulcova and its Pulcamoon, before (top) and after (bottom) the application of a halo-subtraction algorithm.}
   \label{fig:pulcamoon}
\end{figure}

Relative photometry and astrometry of the satellites were measured using 2D-Gaussian fits on the pre- and post-halo subtracted images. In general, the photometry and astrometry of the primary body were measured on the pre-halo subtracted images, and of the satellite on the corresponding post-halo subtracted images. Measured positions of the satellite change on and order much smaller than the astrometric uncertainties between pre- and post-halo subtracted images, however, in cases where the satellite is at a small angular separation from the primary, it is much easier to identify the position of the satellite post-halo-subtraction. Examples of this processing can be seen in \Cref{fig:pulcamoon}. Formal one-pixel astrometric uncertainties are applied, to account for potential offsets between the primary's photo- and barycenters, however, the quality of the orbital fits suggests that this may be an overestimate.

\subsection{Optical lightcurves}

To construct a shape model for (762) Pulcova, we collected and observed photometric lightcurves of the system. Observations before 2023 were taken from the literature, public databases
\citep{2006A&A...446.1177B, alcdef}\footnote{\url{https://alcdef.org/}}$^{,}$\footnote{\url{https://obswww.unige.ch/~behrend/page_cou.html}} and individual observers, some of which are authors on this manuscript (IS, ES,  MF, EJ, AM, RR, KA, RB). Lightcurves collected from public databases include those observed by A. Waszczak, E. Reina Lorenz, L. P. Strabla, J. Oey, and D. L. Gorshanov. Some of these observations appear in previous publications, including \cite{pulcovabinariespulkovo} and \cite{underoakkevin}. In 2023, observations were made with TRAPPIST-South \citep{2011Msngr.145....2J} on February 16th, and the 60 cm André Peyrot telescope mounted at Les Makes observatory (IAU code 181) on La Réunion island on April 21st-23rd. A summary of these observations appears in \Cref{tab:lc}.


\section{Shape models}
\label{sec:shapes}

To assess potential discrepancies between the shape-based estimates (assuming a homogeneous internal structure) and the orbit-based estimates of the gravitational fields of Pulcova and Emma,
it is necessary to have accurate shape models of the systems' primaries. We discuss the shape models used for Pulcova and Emma in the following section.

\subsection{Pulcova}

\begin{figure}
   \centering
   \includegraphics[width=\columnwidth]{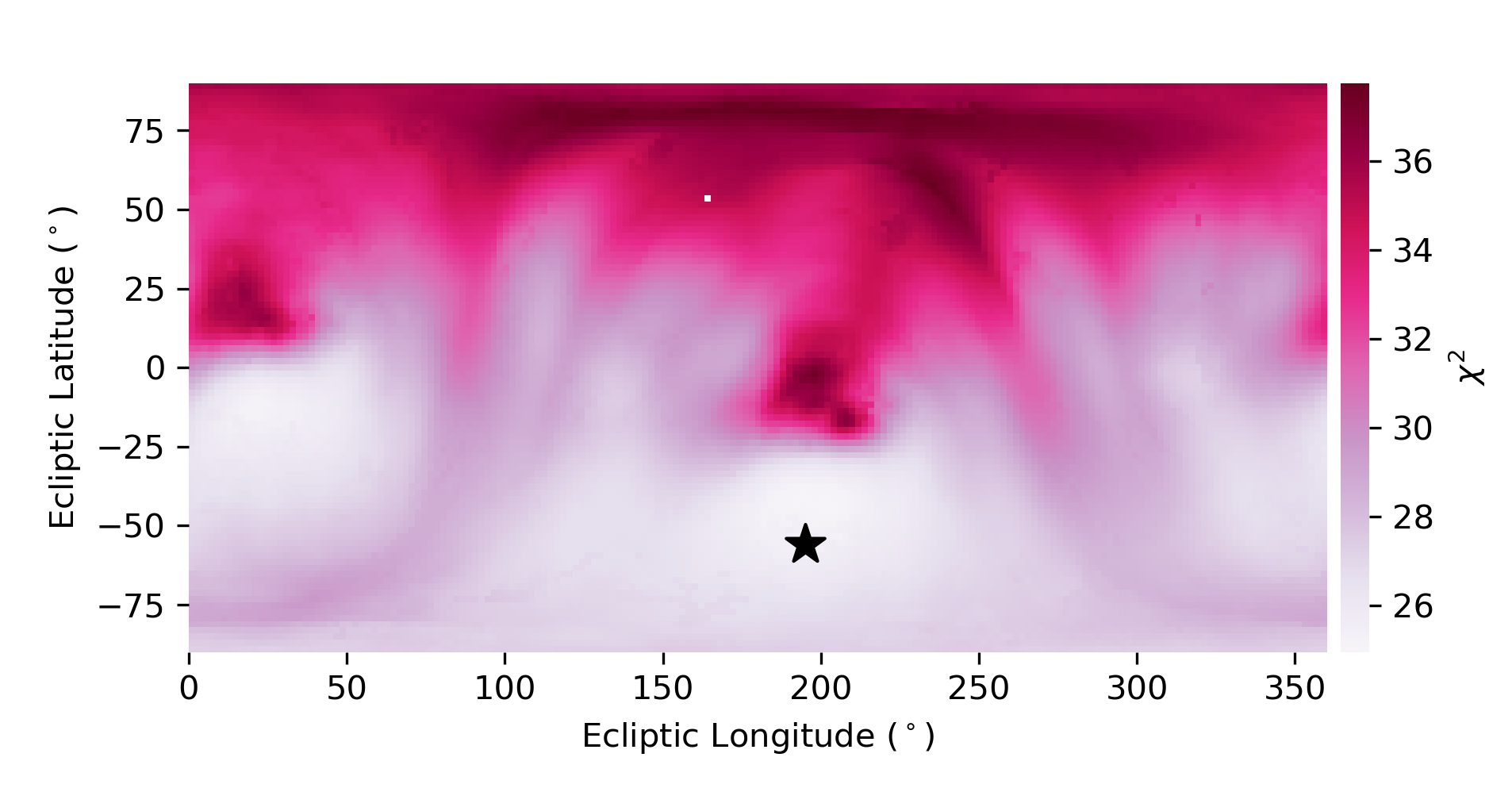}
   \caption{Residuals of spin pole solutions from lightcurve inversion for asteroid (762) Pulcova. Light regions correspond to lower $\chi^2$ values, and therefore better-fit solutions. The coordinates of the best-fit orbital pole are marked with a star.
   }
   \label{fig:spin_pulc}
\end{figure}

For asteroid (762) Pulcova, the shape model is constructed using the standard lightcurve inversion technique
\citep{2001Icar..153...24K, 2001Icar..153...37K},
where the absolute scale of the system is determined with stellar occultations \citep{2011Icar..214..652D}.
A combination of photometric lightcurves and sparse photometry were used to construct this model. No AO images were included as Pulcova's primary was not sufficiently angularly resolved in any of the available archival data.
We found an effective diameter
for Pulcova of \numb{\KM{$136\pm10$} km}, smaller than the previously reported value of \numb{149} km.
A description of the observations used can be found in \Cref{tab:lc}.
The model was adjusted iteratively with the orbital solution,
as the primary spin pole was fitted as part of the orbital model.
Estimations of the spin pole from the lightcurve inversion alone were
imprecise (\Cref{fig:spin_pulc}), with no clear best-fit solution.
Spin estimates from the orbital solution were much more precise,
although uncertainties were still in the range of a few degrees.
The model with the fixed spin pole was found to be a good fit
to the stellar occulations (\Cref{fig:occ_pulc}).
We found the shape model to be very sensitive to small changes in the spin orientation,
and, as such, the spin pole uncertainty introduces a significant source of ambiguity into the shape model.
A topographic map of the shape model of Pulcova can be found in \Cref{fig:shape_pulc}. Future observations, such as high-quality occultations or optical wavelength AO imaging, could be used to improve this model.\KM{Occultation predictions are listed in \Cref{tab:occ_pulc}}. The utility of future lightcurve observations is limited.

\subsection{Emma}

For (283) Emma, we used the shape model available from
DAMIT\footnote{   \url{https://astro.troja.mff.cuni.cz/projects/damit/}} \citep{2010A&A...513A..46D}, previously published in \cite{2017A&A...607A.117V}.
As noted in \Cref{sec:orbits}, a simple Keplerian motion cannot reproduce the observed
positions of the satellite of Emma. We thus studied its orbit by expanding the gravitational
potential to the second order, hence including the
polar oblateness $C_{20} = -J_2$ and the Tesseral and Sectorial coefficients $C_{22}$ and $S_{22}$. \KM{Further analysis of Emma's shape and information about alternative shape models can be found in \Cref{app:shapes}.}

\subsection{Moons}

For both of these systems, the satellites were not disk-resolved in any images (that is, they presented as point sources), and no stellar occultations have been reported to capture the system secondaries. As such, we approximate the shape of the secondaries as a sphere, although there is no evidence to indicate that this is actually the case. The influence of, for example, the $J_2$ of the satellite is assumed to be negligible due to the large size difference and separation between the two components.

We calculate the diameter of the satellites as follows. First, an outlier-rejected set of magnitude differences ($\Delta m$) are derived from the adaptive optics images. There is some level of variability in the $\Delta m$ due to the non-spherical shape of the primary. Then, the relative and absolute size of the satellites is determined from the average value of these magnitude differences, as described in \Cref{eq:moonsizes}, assuming a similar albedo for the primary and the satellite.

\begin{equation}
D_s = D_p\times10^{-0.2\Delta m}
\label{eq:moonsizes}
\end{equation}

For Emmoon, we determine $\numb{D_s=14\pm3}$\,km and for Pulcamoon we determine $\numb{D_s=14\pm4}$\,km, resulting in $D_s/D_p\approx0.1$ for both systems.

\section{Orbit determination}
\label{sec:orbits}
\begin{figure}
    \centering
    \includegraphics[width=\columnwidth]{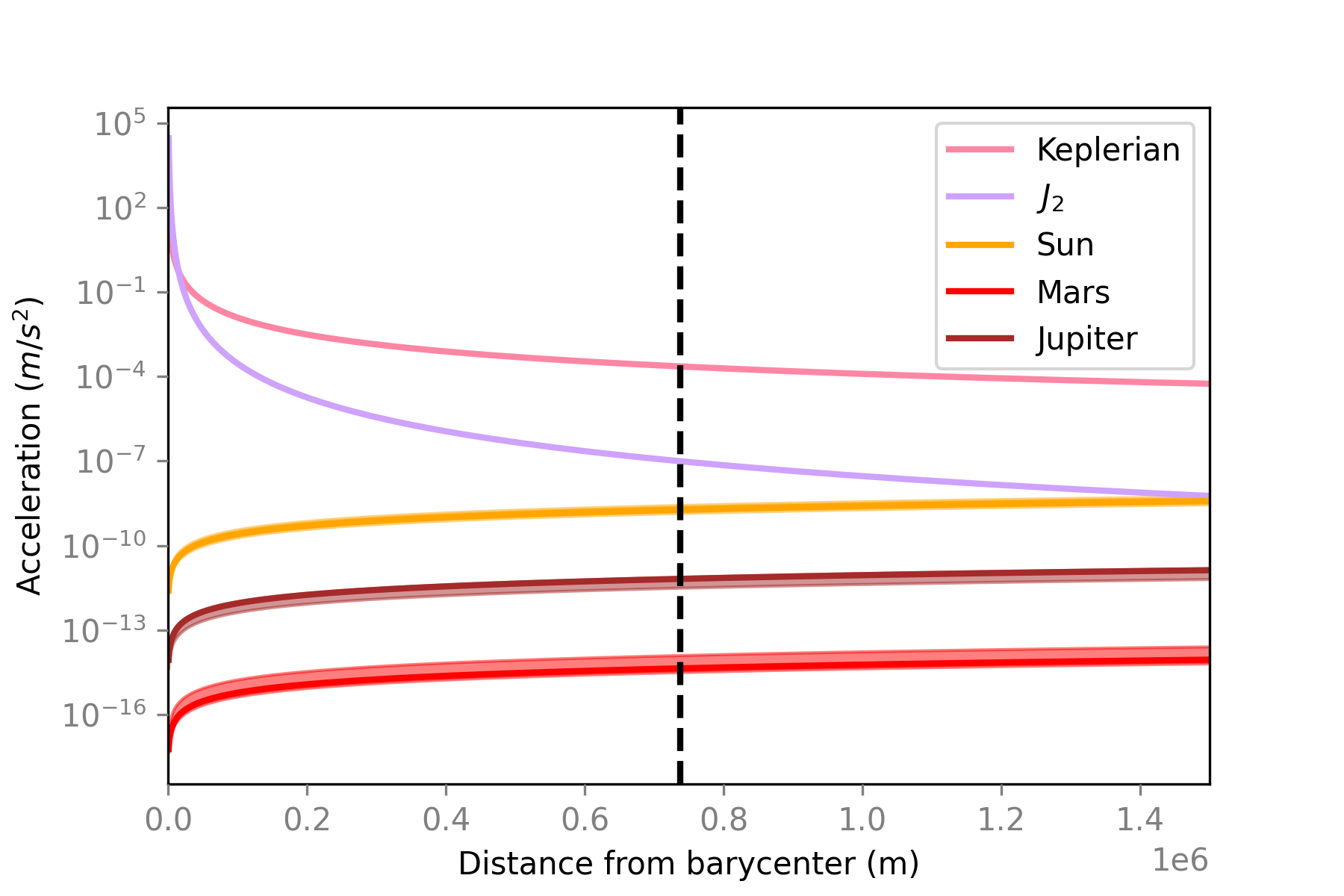}
   \includegraphics[width=\columnwidth]{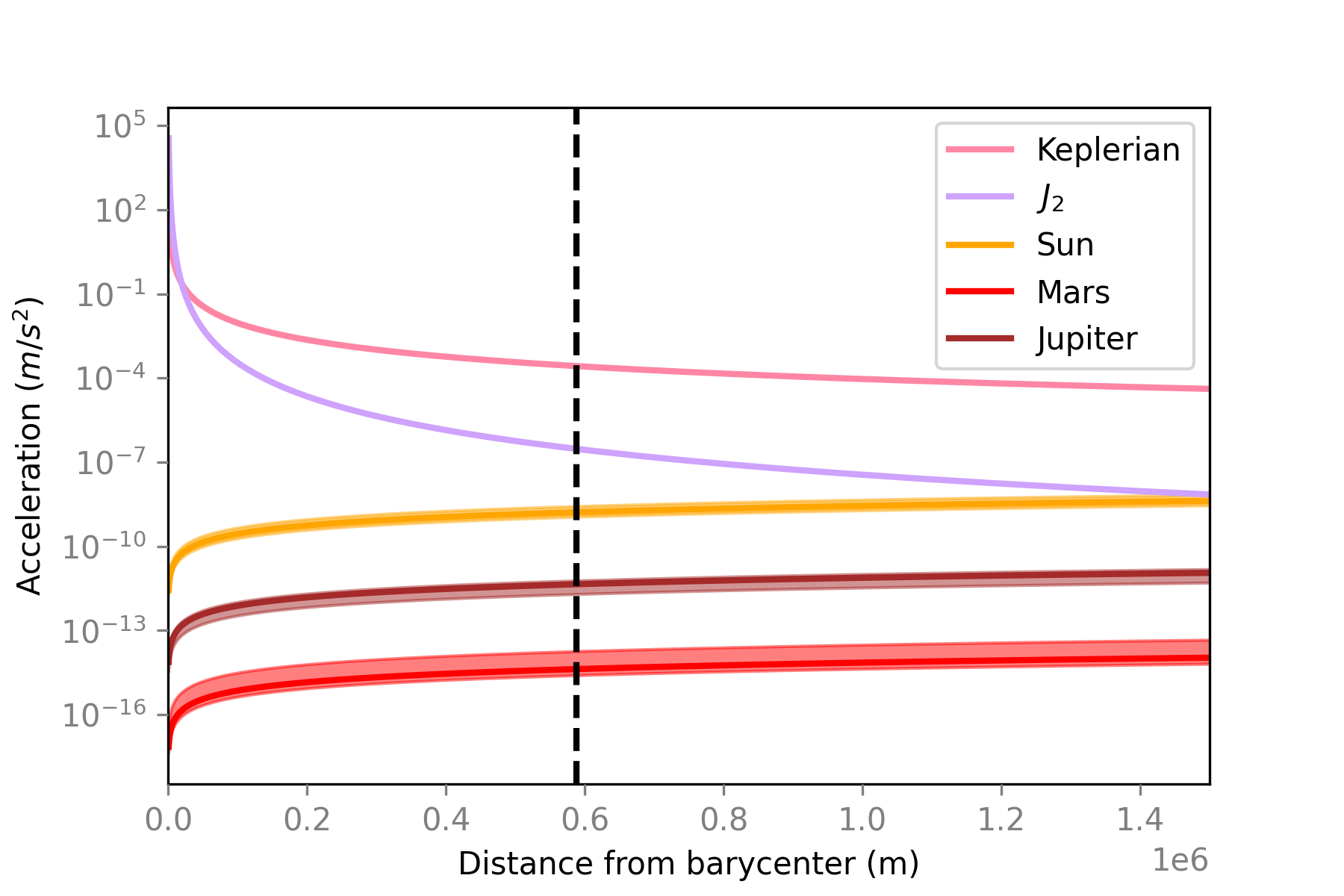}
   \caption{Expected gravitational force due to system gravity and external perturbers, as a function of satellite distance for Pulcova (upper panel) and Emma (lower panel). Shaded regions around planetary forces indicate the range of the force between the objects closest and furthest approach to the asteroid. At the positions of both Pulcamoon and Emmoon (marked with a dashed line), the contribution of external perturbers is negligible.}
   \label{fig:grav}
\end{figure}

We used the \genoid algorithm  \citep{2012AA...543A..68V}
to derive the orbital parameters from
the observed astrometric positions of each system's satellite with respect to the primary body\footnote{\KM{Ephemeris from these orbital solutions can be accessed online at \url{https://ssp.imcce.fr/webservices/miriade/api/ephemsys/}, contact K. Minker or F. Vachier if you have difficulty accessing ephemerides.}}.
\Genoid is a genetic algorithm that searches for a best-fit orbital solution over defined ranges of input parameters. The parameters for the initial generation of solutions are randomly assigned over a large range of values for each parameter (mass, semi-major axis, eccentricity, etc.). For successive generations parameters are pulled from the most successful previous solutions. Depending on the complexity of the problem, differing input parameters can be held fixed or variable (for example, the coordinates of the primary's spin-pole, or higher-order gravitational terms), and external forces (for example, the influences of Solar gravity) can be included or excluded. This process is described in more detail in, e.g.,
\citet{2012AA...543A..68V},
\citet{2014Icar..239..118B},
and
\citet{2022Icar..38215013V}.
For both Pulcova and Emma, we anticipated the gravitational influences of the non-spherical primary ($J_2$) to have a significantly higher influence than any external perturbers (Sun, planets, see \Cref{fig:grav}). As such, Solar and planetary influences were only considered in orbital models where higher-order gravitational terms were also considered. Solutions were tested with terms up to the order 4 included, but ultimately due to the large semi-major axis of these satellites all terms above order two were determined to be negligible.

The uncertainties associated with this technique are statistical, not formal, and encompass the range of values for the parameter found in all solutions compatible with the observed positions at the $1 \sigma$ (or 2 or 3 $\sigma$) level. This provides a full sampling of the parameter space, and accounts for non-linear correlations that may exist. Of course, not all combinations of parameters within the $1 \sigma$ range will be compatible as all parameters are varied simultaneously, and as such outlying solutions can artificially inflate uncertainties on parameters that are generally well constrained.

\subsection{Pulcova}

\begin{table}
\begin{center}
  \caption[Orbital elements of the satellite of Pulcova]{%
    Orbital elements of the satellite of Pulcova,
    expressed in EQJ2000 (equatorial coordinates), obtained with \genoid:
    orbital period $P$, semi-major axis $a$,
    eccentricity $e$, inclination $i$,
    longitude of the ascending node $\Omega$,
    argument of pericenter $\omega$, time of pericenter $t_p$.
    }
  \label{tab:pulcorbj2}
   \begin{tabular}{l ll}
    \hline\hline
    & \multicolumn{2}{c}{S/2000 (762) 1}\\
    \hline
  \noalign{\smallskip}
  \multicolumn{2}{c}{Observing data set} \\
  \noalign{\smallskip}
    Number of observations  & \multicolumn{2}{c}{68} \\
    Time span (days)        & \multicolumn{2}{c}{7141} \\
    RMS (mas)               & \multicolumn{2}{c}{11.04} \\
    \hline
  \noalign{\smallskip}
  \multicolumn{2}{c}{Orbital elements EQJ2000} \\
  \noalign{\smallskip}
    $P$ (day)         & 4.14321 & $\pm$ 0.00038 \\
    $a$ (km)          & 738.4 & $\pm$ 2.6 \\
    $e$               & 0.004 & $_{-0.004}^{+0.018}$ \\
    $i$ (\degr)       & 144.7 & $\pm$ 5.8 \\
    $\Omega$ (\degr)  & 251.1 & $\pm$ 7.5 \\
    $\omega$ (\degr)  & 58.0 & $\pm$ 10.6 \\
    $t_{p}$ (JD)      & 2452841.14505 & $\pm$ 0.16817 \\
    \hline
  \noalign{\smallskip}
  \multicolumn{2}{c}{Derived parameters} \\
  \noalign{\smallskip}
    $M_{\textrm{Pulcova}}$ ($\times 10^{18}$ kg)      & 1.865 & $\pm$ 0.019 \\
    $\lambda_p,\,\beta_p$ (\degr)  & 195, -56 & $\pm$ 9, 5 \\
    $\alpha_p,\,\delta_p$ (\degr)  & 161, -55 & $\pm$ 7, 6 \\
    $\Lambda$ (\degr)              & 0.6 & $\pm$ 0.1 \\
    $\rho$ ( g~cm$^{-3}$) & 1.4 &$\pm0.2$ \\
    \hline
  \end{tabular}
  \end{center}
  \footnotesize{The number of observations and RMS between predicted and
    observed positions are also provided.
    Finally, we report the mass of Pulcova $M_{\textrm{Pulcova}}$,
    the ecliptic J2000 coordinates of the orbital pole
    ($\lambda_p,\,\beta_p$),
    the equatorial J2000 coordinates of the orbital pole
    ($\alpha_p,\,\delta_p$), the
    orbital inclination ($\Lambda$) with respect to the equator of
    Pulcova, and Pulcova's bulk density ($\rho$). Uncertainties are given at 1-$\sigma$.}
\end{table}

The orbital dynamics of Pulcamoon can easily be modeled with a simple Keplerian orbit. Such a model provides a nearly circular orbital geometry in a nearly equatorial plane. This orbital solution is further described in \Cref{tab:pulcorbj2}.

\begin{figure}
    \centering
    \includegraphics[width=1\linewidth]{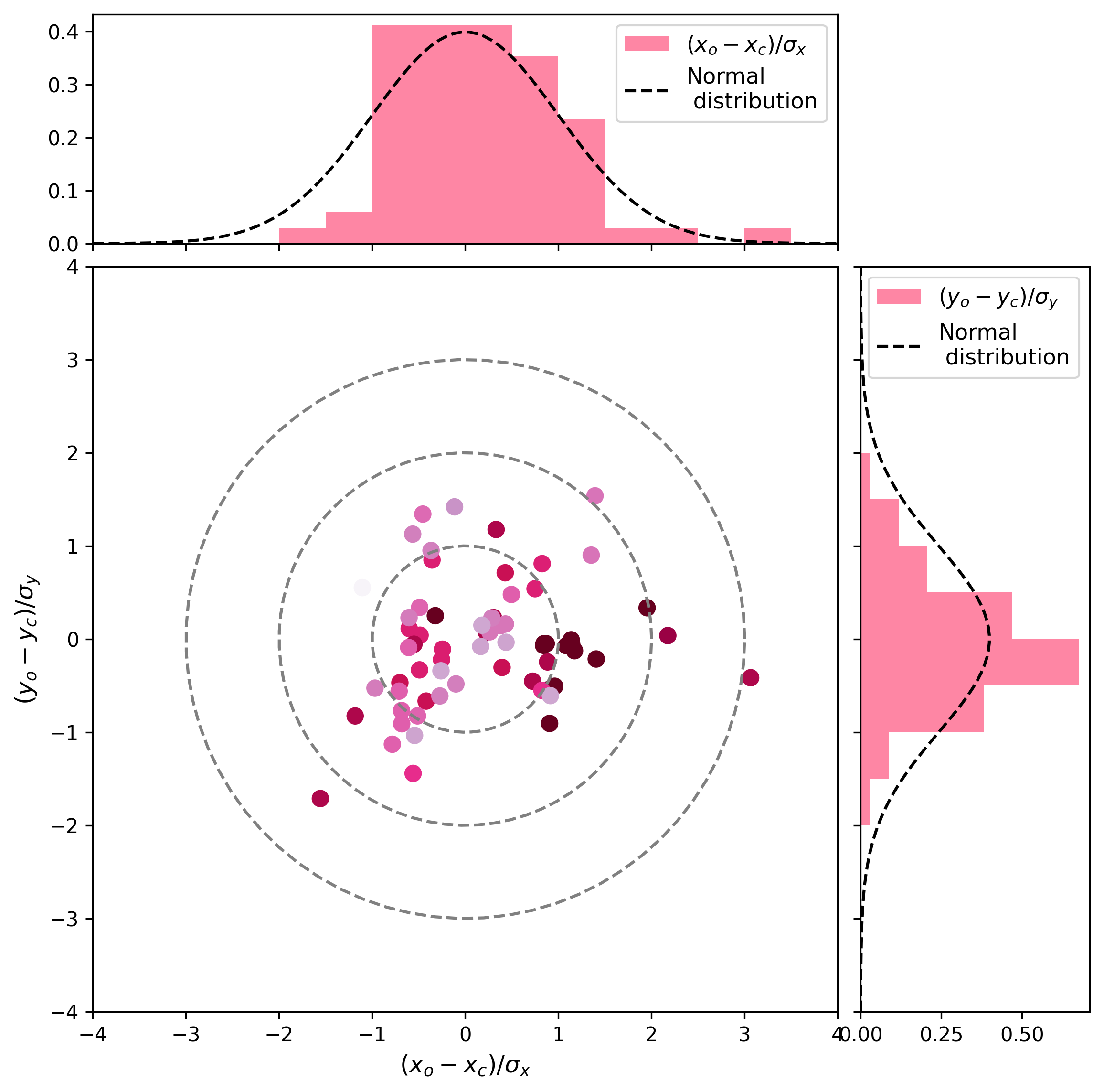}
    \caption{Residuals of Pulcamoon's orbital solution in the x and y directions, scaled to 1 pixel residuals. The residuals follow a roughly gaussian distribution with a standard deviation equivalent to the 1 pixel uncertainties. \KM{Dark red points correspond to older observations, light (lavender) points correspond to recent observations.}}
    \label{fig:pulc_omc}
\end{figure}

This orbit can also be fit equally well using significantly more complicated models with resulting $J_2$ values as high as $J_2=0.13$, and non-zero $C_{22}$ and $S_{22}$ coefficients. However, the severe degeneracy in the higher order terms of the orbital fitting leaves significant ambiguity in Pulcova's internal structure, and it is not worthwhile to pursue resolving this degeneracy until Pulcova's shape model is more clearly defined (see \Cref{sec:shapes} for further discussion on this subject). \KM{Orbital residuals are displayed in \Cref{fig:pulc_omc}.}

We identify significant differences between our orbital solution for Pulcamoon and a previously published solution by \cite{marchis2008circular}. We find a shorter period (4.14 vs. 4.44 days), longer semi-major axis (738 vs. 703 km), and lower eccentricity (0.004 vs. 0.03). We also find Pulcova to be significantly heavier, with a mass of \numb{$1.865\pm0.019\times10^{18}$}\,kg (vs. $1.4\pm0.1\times10^{18}$\,kg), and as a result we find a much higher density ($1.4\pm0.2$\,g~cm$^{-3}$ vs. $0.9\pm0.1$\,g~cm$^{-3}$). Our solution incorporates a longer orbital baseline (2000-2019 vs. 2003-2006). Very early analysis by \cite{asteroidsdohavesatellitesmerline} provide roughly similar results, with an orbital period of 4.0 days, a semi-major axis of 810\,km, and a system density of $1.8\pm0.8$\,g~cm$^{-3}$.

\subsection{Emma}

\begin{figure}
    \centering
    \includegraphics[width=1\linewidth]{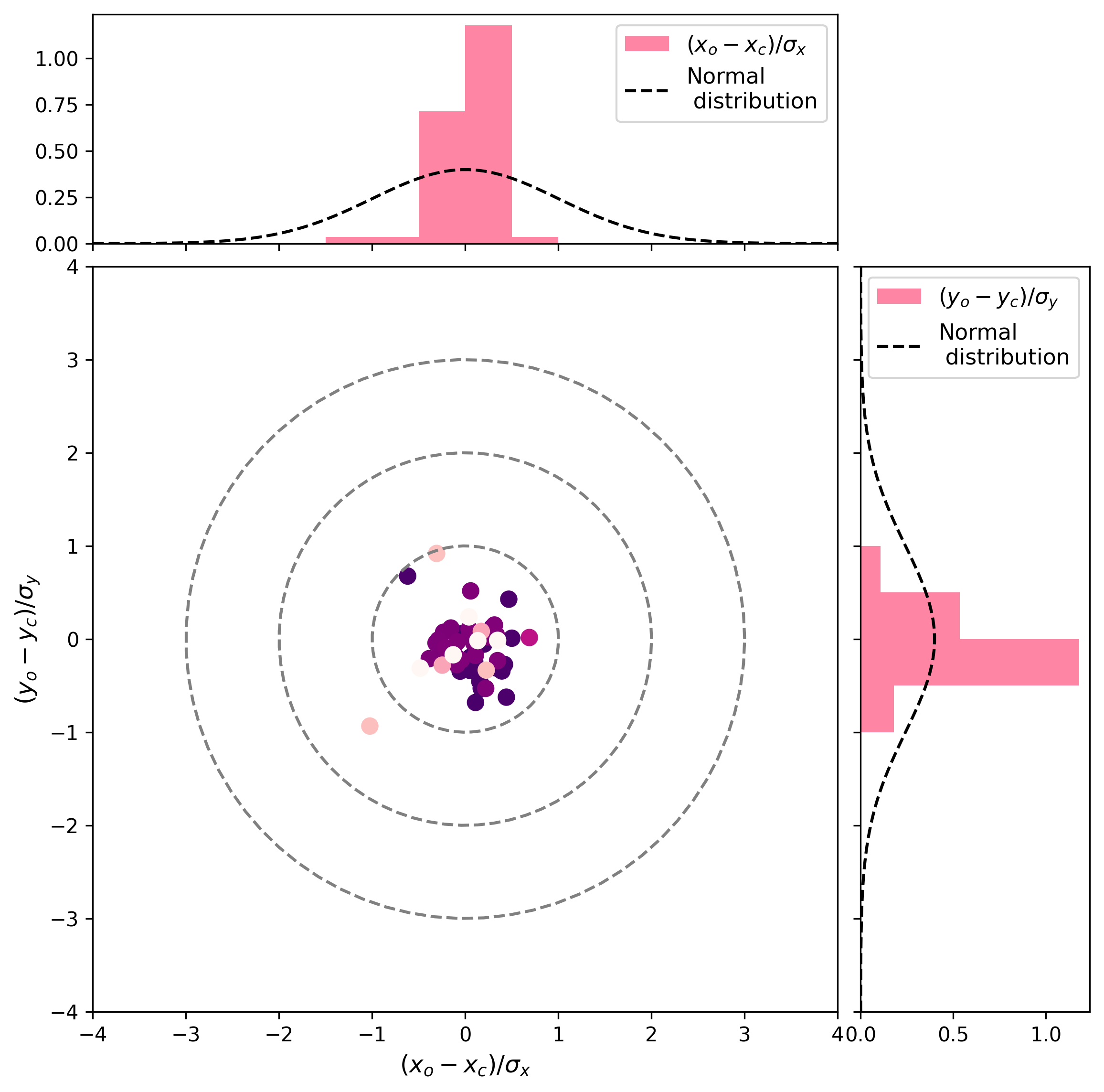}
    \caption{Residuals of Emmoon's orbital solution in the x and y directions, scaled to 1 pixel residuals. The vast majority of points fall within the nominal one-pixel uncertainty level (innermost circle). Dark points correspond to older data and lighter points to newer data.}
    \label{fig:emma_omc}
\end{figure}

The orbital solution for Emma is unambiguous, with a low root-mean-square
(RMS) of the residuals between the observed positions and
those predicted based on the determined orbital solution of
\numb{5.4} mas.
The satellite orbits along a moderately eccentric path (\numb{$e=0.11\pm0.01$})
and a clear detection of \numb{$J_2=0.11\pm0.01$} is observed.
We also note a detection of the Tesseral coefficient $C_{22}=\numb{0.08}\pm0.01$, although the inclusion of this term is not strictly necessary for a reasonable fit.
We were unable to find a strictly Keplerian solution with a reasonable fit to the observational dataset, although portions of the dataset, and no portion of the dataset spanning over one month could be reasonably approximated by a Keplerian orbit.
The details of the orbital solution are presented in \Cref{tab:emmaorb}. One-pixel uncertainties were assigned to all observations;
however this may be an overestimate as the residuals of the orbital fit are on average much smaller than $1 \sigma$. This is illustrated in  \Cref{fig:emma_omc}. Since the model fit so well, uncertainties were reduced to the half pixel level to provide more reasonable uncertainties on resulting physical parameters.

\begin{table}
\begin{center}
  \caption[Orbital elements of the satellite of Emma]{%
    Orbital elements of the satellite of Emma, see \Cref{tab:pulcorbj2} for a description of parameters.
    }
  \label{tab:emmaorb}
   \begin{tabular}{l ll}
    \hline\hline
    & \multicolumn{2}{c}{S/2003 (283) 1}\\
    \hline
  \noalign{\smallskip}
  \multicolumn{2}{c}{Observing data set} \\
  \noalign{\smallskip}
    Number of observations  & \multicolumn{2}{c}{56} \\
    Time span (days)        & \multicolumn{2}{c}{3626} \\
    RMS (mas)               & \multicolumn{2}{c}{5.85} \\
    \hline
  \noalign{\smallskip}
  \multicolumn{2}{c}{Orbital elements EQJ2000} \\
  \noalign{\smallskip}
    $P$ (day)         & 3.41121 & $\pm$ 0.00024 \\
    $a$ (km)          & 588.3 & $\pm$ 0.0 \\
    $e$               & 0.118 & $\pm$ 0.002 \\
    $i$ (\degr)       & 94.2 & $\pm$ 2.5 \\
    $\Omega$ (\degr)  & 347.3 & $\pm$ 1.9 \\
    $\omega$ (\degr)  & 209.1 & $\pm$ 1.1 \\
    $t_{p}$ (JD)      & 2452835.14430 & $\pm$ 0.01320 \\
    $J_2$ & 0.11 & $\pm$ 0.01 \\
    \hline
  \noalign{\smallskip}
  \multicolumn{2}{c}{Derived parameters} \\
  \noalign{\smallskip}
    $M_{\textrm{Emma}}$ ($\times 10^{18}$ kg)      & 1.398 & $\pm$ 0.001 \\
    $\lambda_p,\,\beta_p$ (\degr)  & 257, +18 & $\pm$ 2, 3 \\
    $\alpha_p,\,\delta_p$ (\degr)  & 257, -4 & $\pm$ 2, 3 \\
    $\Lambda$ (\degr)              & 0.3 & $\pm$ 0.2 \\
    $\rho$ ( g~cm$^{-3}$) & 0.9 &$\pm0.3$ \\
    \hline
  \end{tabular}
\end{center}
\end{table}

Orbital solutions were calculated using two values for the radius of Emma, \numb{67}\,km from the updated shape model and the \numb{71}\,km equivalent radius measured from the \cite{2017A&A...607A.117V} shape model (discussed in \Cref{sec:shapes}).
 We find that Emma is only marginally resolved in the AO images used to construct and scale the shape model, and as such it is difficult to determine which value is more reliable.
 We also note that the shape model does not appear to be an exceptionally good fit to the AO observations (see for example the misalignment between the model and the observations in the figure A.17 of \cite{2017A&A...607A.117V}).
Since the determination of $J_2$ is sensitive to Emma's radius, which is held as a fixed parameter, it is illogical to compare an orbital solution based upon one shape model with the physical $J_2$ from another model. As such, we have assumed the value from the shape model to be correct.
 The additional orbital solution is presented \Cref{tab:emma_alt_solution}, using the alternative shape model discussed in \Cref{sec:shapes} with a diameter $D_p=133\pm3$\,km. This model provides a higher density of $1.1\pm0.1$\,g cm$^{-3}$, and a dynamical $J_2=0.13$.
 The literature presents a wide range of estimates (for example \citet{2021PSJ.....2..162M} presents values from 112\,km to 155\,km for Emma's diameter within the same study),
  an effect which likely originates from Emma's highly elongated shape.

When reducing the astrometric uncertainties by a factor of two, we find very similar results, with substantially reduced uncertainties, suggesting that the large uncertainties in the other trials were artificially inflating the 1$\sigma$ fit to an unreasonable level. We remind the reader that the $1 \sigma$ uncertainties are determined statistically, meaning that they represent the range of values found in all solutions that fit the observations at the $1 \sigma$ level. However, occasional outliers can bias these results, artificially inflating the uncertainties.
This solution is broadly similar to the solution reported by \cite{marchis2008eccentric}, although not all parameters are in agreement. We determine a slightly longer orbital period (3.41\,d vs. 3.35\,d) and slightly larger semi-major axis (588\,km vs. 581$\pm3.6$\,km). The eccentricities between the two solutions agree, as do the masses, although ours are determined more precisely. Our solution incorporates observations taken over a period of ten years (compared to three in the Marchis solution), therefore allowing us to constrain Emmoon's precession. Notably, Emmoon's precession due to Emma's irregular shape is observable on timescales much shorter than the 3 year period of observations used in the Marchis solution, but no value of $J_2$ is reported. This may account for some of the discrepancies between the two solutions.

\subsection{Some limitations}

Since the orbital solutions presented in this study come from a large compilation of archival data, there is naturally a level of variance in the quality of these datasets. The quality of an observation is dependent on a number of factors, including the weather, telescope aperture,
the AO correction quality of the instrument, and the distance between the Earth and the asteroid at the time of observation. These parameters are not always optimized, so as such not all observations provide the same quality.

Furthermore, there is a large variation in the pixel scale of different instruments (ranging from 10 mas to 35 mas), which directly translates to the 1-pixel uncertainties used in the calculation of these orbital solutions. This can be seen to be propagated to the RMS residuals of a given orbital solution, and may explain why Pulcova seems to have a poorer-quality fit than Emma, as the mean pixel size was larger for Pulcova than for Emma (20 mas vs. 15 mas).
 This is especially notable when considering CFHT/PUEO observations.
 This is an early-generation AO instrument on a significantly smaller telescope than any of the other instruments considered in this study.
 It also appears that observations by this instrument may have a small
 systematic offset from the true position of the satellite, possibly due to an imperfectly
 calibrated field orientation.
 To minimize the effects of this on our calculations when considering high-order gravitational terms,  we computed the final orbital solution on all observations except for CFHT/PUEO, and then confirmed afterwards that the predicted ephemeris positions matched well to the CFHT observations.

\section{Physical properties}
\label{sec:physicalprops}

\begin{figure}
    \centering
    \includegraphics[width=\columnwidth]{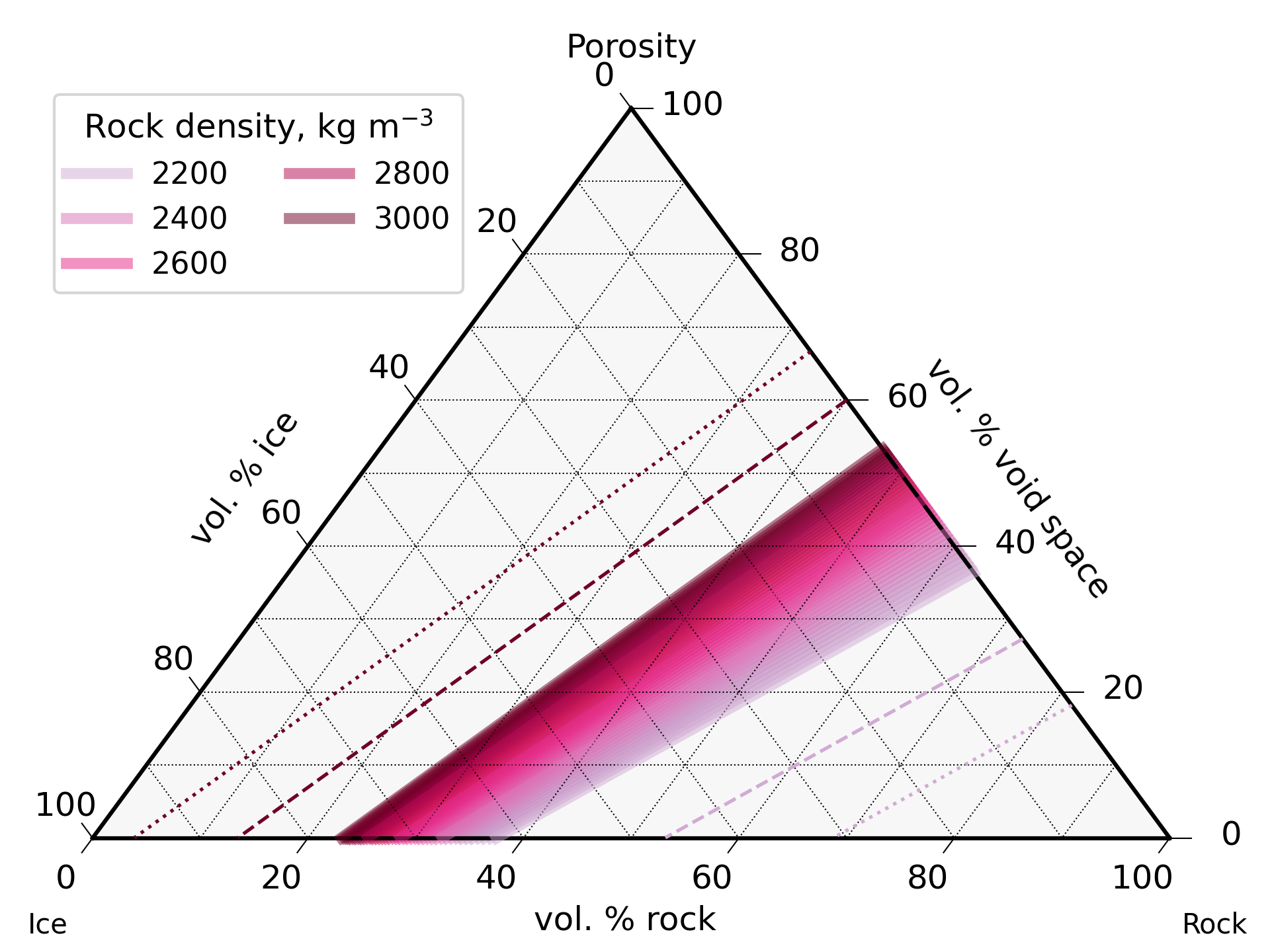}
   \caption{Range of bulk density distribution for Pulcova.
   \label{fig:bulk_pulc}
   }
\end{figure}

From a combined analysis of the shape models and orbital solutions for these two systems we can analyze the internal structure of each of their primary bodies.
For Pulcova, we determine a density of \numb{$1.4\pm0.2$}\,g~cm$^{-3}$. This is similar to the density of other C type asteroids and analogous meteorites \citep{2012P&SS...73...98C}. Beyond this, little information can be derived about the internal structure of Pulcova due to the ambiguity of the orbit. Considering the Keplerian case, a naive interpretation may suggest that this implies a heavily differentiated internal structure for Pulcova to compensate for the substantial difference between the orbital $J_2=0$ and the $J_2=0.10$ from the shape, but this is unlikely to be the case.
In the case of a satellite that orbits on a path that is very nearly circular and coplanar to the primary’s equator, precession cannot be observed,
and therefore many acceptable orbital solutions can be identified, as any arbitrary choice of $J_2$ can produce a reasonable fit to the observational dataset.
\Cref{fig:bulk_pulc} describes potential bulk composition ratios for Pulcova, and demonstrates
that a wide variety of compositions with a mixture of ice, rock, and void are plausible.
The overall fraction of void space is relatively low, most likely on the order of 30\% with an upper limit of 50\%.

For Emma, we determine a mass of $\numb{1.4\pm0.2\times10^{18}}$\,kg, providing a density of \numb{$0.9\pm0.3$}\,g cm$^{-3}$.
We find a difference between the dynamical ($J_2=\numb{0.11\pm0.01}$, determined from the orbital model)
 and physical ($J_2\approx\numb{0.14}$, determined from the shape model) gravitational fields of Emma.
 These differences can be easily reconciled through a two layer (core-crust) internal structure model for Emma
 \citep[similar to Kalliope, see][]{2022A&A...662A..71F}\KM{, further information on such models can be found in \Cref{app:shapes}}.

Notably, the methodology used to produce the shape model of Emma presents some ambiguity. The AO observations included in the shape model computation are low resolution as Emma is poorly resolved and the dataset is limited. As such, it could be possible that there are additional concavities or extensive cratering influencing the low density of Emma's outer layer. Unlike convex surface features, it is impossible to recover concavities with the supplementary lightcurve observations. Some large-scale concavities are already present in the model\KM{, but others could be obscured}. In particular, the possibility of widespread deep cratering could be a plausible explanation for the low-density outer shell \citep[similar to that which can be observed in images of Hyperion with Cassini,][]{hyperiondensity}.

\begin{figure}
    \centering
    \includegraphics[width=\columnwidth]{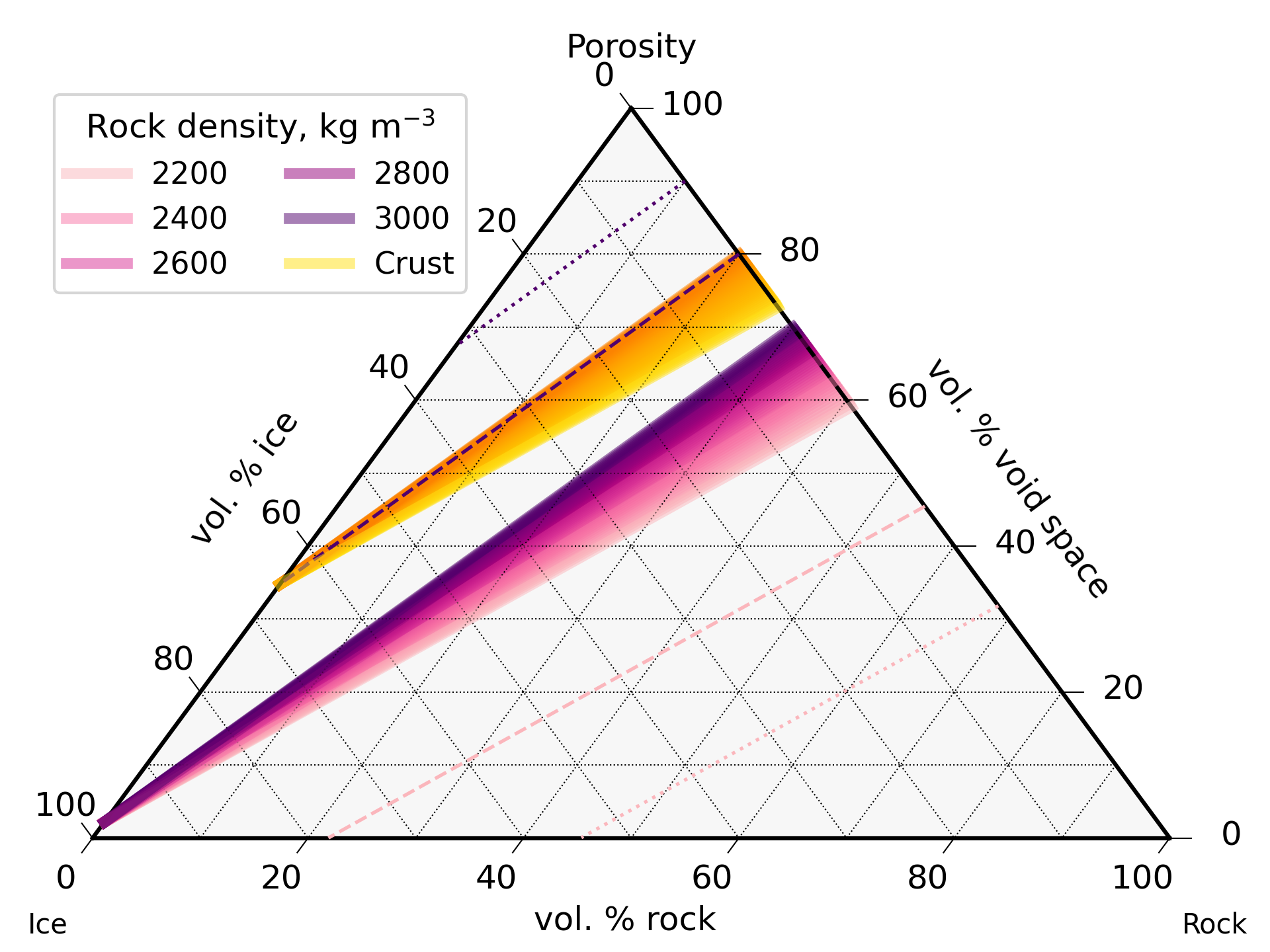}
   \caption{Range of bulk density distribution for Emma, with overall bulk
   density mapped in purple, and crust density in yellow. The same range of rock densities is considered for the composition of both the core and crust material.
   \label{fig:bulk_emma}
}
   \end{figure}

\Cref{fig:bulk_emma} shows potential bulk compositions for Emma overall (in purple) and Emma's crust (in yellow). Emma is likely to be void-dominated, particularly in the crustal layer, which we predict to be at least 30\% void space, but may be up to 80\% void. Although this fraction may seem extreme, similar compositions can be observed in other populations of Solar System objects such as Kuiper Belt objects \citep{arrokothdensity}, comets \citep{67pdensity}, or some giant planet satellites \citep{hyperiondensity}.

\subsection{Constraints on precession}

Due to the uncertainty in the shape models, the value of $J_2$ can only be determined precisely when making assumptions about the primary shape. However, although $J_2$ is the varied parameter, the constrained physical property is actually the satellite's nodal precession $\omega_P$, and this parameter can be precisely and accurately measured. It can be expressed as

\begin{equation}
    \omega_P=-\frac{3 D_p^2J_2\omega cos(\Lambda)}{8(a(1-e^2))^2}
    \label{eq:nodalprecession}
\end{equation}

where $\omega_P$ is expressed in rad/s and $\omega$ is expressed as $2\pi/P$ where $P$ is the period in seconds. Taking a, e to be accurately measured and $\Lambda$ to be $\approx 0$ (as is the case for Emma), one finds that

\begin{equation}
    -\frac{2(a(1-e^2))^2}{3\omega cos(\Lambda)}\omega_P=\frac{ D_p^2J_2}{4}=C
\end{equation}

Where C is a constant. Then, $J_2$ may be expressed as a function of diameter:

\begin{equation}
    J_2=\frac{4C}{D^2_p}
\end{equation}

For Emma, we determine a value of $C=608$\,km$^2$, verified by orbital solutions solved for multiple values of $D_p$.

\section{Discussion}
\label{sec:discussion}

Although the Pulcova and Emma systems share many qualities (size, shape, $D_s/D_p$, ...),
in depth orbital analysis reveals several key differences.
The very low density of (283) Emma ($\rho=0.9\pm0.3$~g~cm$^{-3}$),
alongside the unusual dynamical properties of the satellite
could indicate that the two systems have differing histories, and formed in different ways.

\begin{figure}
   \centering
    \includegraphics[width=\columnwidth]{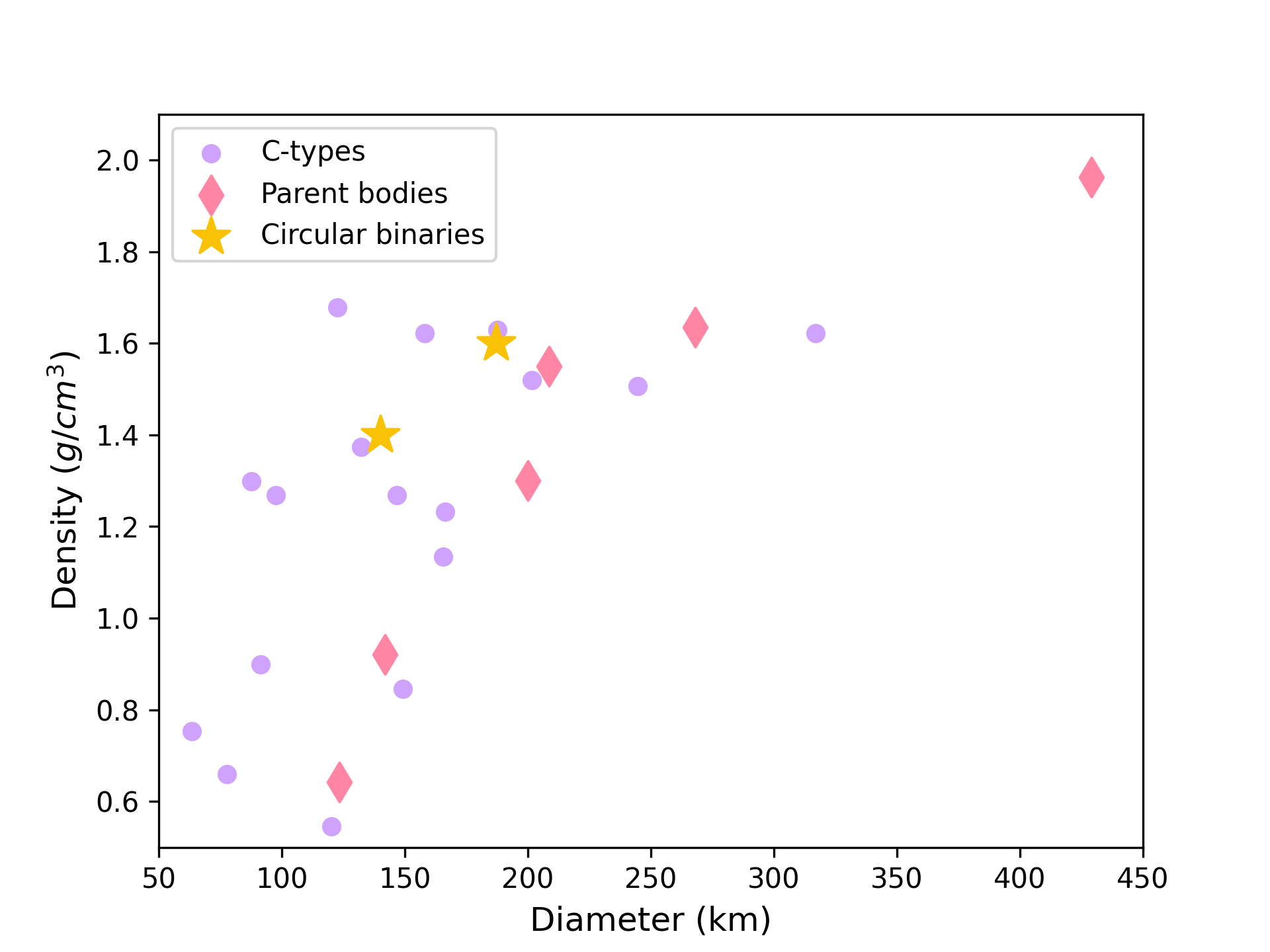}
   \caption{Density distribution for C-type asteroids; the sample is limited to those with uncertainties under 0.5 $g/cm^3$. The parent bodies are (in order of increasing diameter) Tisiphone, Emma, Alauda, Themis, Euphrosyne, Hygiea.
   }
   \label{fig:densities}
\end{figure}

Although the nuanced internal structure of Pulcova remains ambiguous,
due to the imprecise models of its satellite orbit and shape,
the overall bulk density ($\numb{1.4}$~g~cm$^{-3}$) is in line with that which is observed in other C-type asteroids
\citep{2012P&SS...73...98C, 2021A&A...654A..56V, 2023A&A...671A.151B}.
For Emma, we identify an inhomogeneous internal structure, as described in  \Cref{sec:physicalprops}.
 Unlike\KM{, for example,} (4) Vesta,
 this is unlikely to be a result of genuine primordial differentiation. Rather,
we propose that the present day state of Emma is the result of a catastrophic impact
to a larger, primordial parent body, of which Emma is the largest remaining fragment.
The largest solid fragment of the parent body forms Emma's "core", and is surrounded by
a substantial layer of re-accumulated debris with significant voids, forming Emma's low density
outer shell \citep{michel2015reaccumulation}.
Amongst C-type asteroids, we observe that objects that are the parent bodies of large families are typically lower density than those at equivalent diameter which are not, see for example \Cref{fig:densities}.

\begin{figure}
   \centering
   \includegraphics[width=\columnwidth]{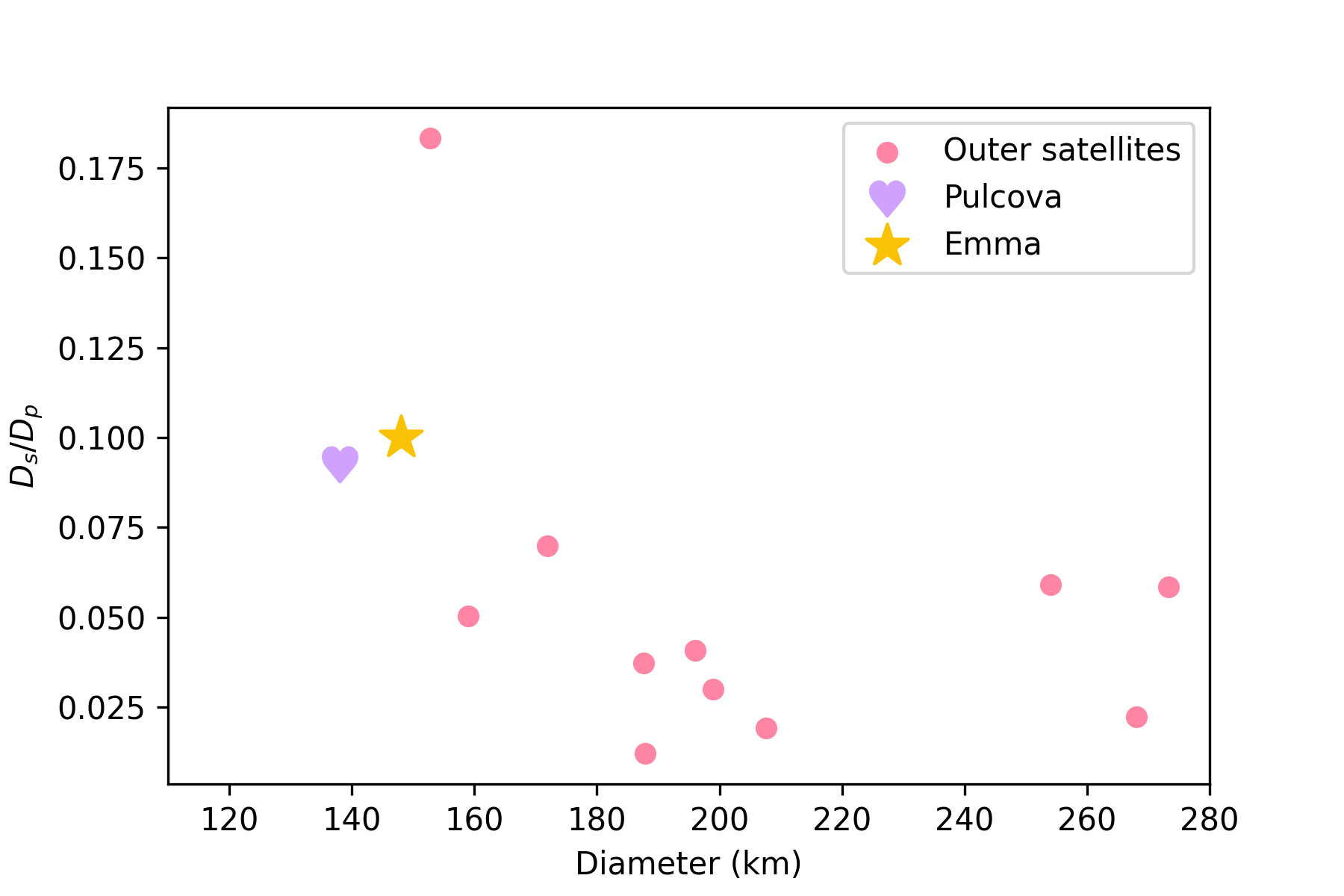}
   \caption{
    Primary-satellite diameter ratio ($D_s/D_p$) vs. system equivalent diameter for known large ($D_{p}>100$\,km) binaries and higher-multiplicity systems. In the case where the system hosts more than one satellite, we consider here the outermost satellite.
   Double asteroids Antiope and Patroclus are excluded, as well as strange asteroid Kleopatra, which likely formed its satellites through a different mechanism \citep{2021A&A...653A..57M}.
   }
   \label{fig:moonsizes}
\end{figure}

In addition to the anticipated density profile of Emma, we find two significant pieces of supporting evidence
to this hypothesis. First, the presence of Emma's large family \citep{2019A&A...622A..47M}, and second, the unusually large size of Emmoon compared to Emma (e.g., \Cref{fig:moonsizes}).

The Emma family, composed of \numb{841} identified members ($D\gtrapprox8$\,km), exhibits compositional homogeneity with nearly all members belonging to a primitive taxonomic type (C,P/X) indicating that it likely originates from an undifferentiated parent body. Spectral analysis indicates that Emma's taxonomic classification varies between C- and P-type like between different measurements\footnote{\url{https://classy.readthedocs.io/}}, which could indicate some large-scale variation in Emma's surface composition. It is typical of carbonaceous dynamical families to exhibit a mixture of C and P type members \citep{2016A&A...586A..15M,2020A&A...643A..38Y}, which may indicate that there is some level of differentiation or compositional mixing within carbonaceous parent bodies \citep{2021A&A...650A.129C, 2021A&A...654A..56V}.

Assuming uniform albedo between components, Emmoon is only a factor of 10 smaller than Emma ($D_s/D_p=0.1$), making it one of the largest known asteroid satellites compared to the system's primary \citep[following that of (22) Kalliope, ][]{2022A&A...662A..71F}. \Cref{fig:moonsizes} shows a comparison between satellite size and system size. Observational biases are likely to affect this distribution, as many smaller (<200km) asteroids fall below the limiting magnitude of new-generation AO instruments like VLT/SPHERE, and very small satellites may not have been detectable with older generation instruments such as Keck/NIRC2 \citep{2016camimoon,2016yangelektraminerva}.

\section{Spins, shapes, and families: Do differing dynamical populations suggest differing formation \KM{pathways}?}
\label{sec:deadbugs}

Over the past several years, the improved characterization of several large binary systems has created the first reasonably reliable physical-dynamical dataset from which to study the population as a whole. Here, we compile binary properties from various studies, including this work,
\citet{2018Icar..309..134P} (Camilla),
\citet{2019A&A...623A.132C} (Daphne),
\citet{2020A&A...641A..80Y} (Euphrosyne),
\citet{2021A&A...650A.129C} (Sylvia),
\citet{2022A&A...662A..71F} (Kalliope),
\citet{asteroidsdohavesatellitesmerline} (Ida, Hermione),
\citet{2021-IMCCEbook-withorbits} and updates thereto (Hektor, Alauda, Eugenia, Minerva, Elektra),
 and discuss observable trends within the population. In addition, when discussing asteroid families, we have compiled family membership from the following studies: \citet{2014Icar..239...46M, 2015PDSS..234.....N, 2019MNRAS.484.3755V, 2017Sci...357.1026D, 2019MNRAS.482.2612T, 2021MNRAS.501..356P, 2019A&A...624A..69D, 2007AJ....134.2160R, 2011MNRAS.412..987R, 2013A&A...551A.117B, 2014Icar..243..111D, 2021A&A...649A.115V,2022AJ....164..167M}.

The prototypical binary system has been well described by many authors over the years,
featuring a singly-synchronous \citep{jacobson2011} satellite on a circular, co-planar orbit \citep{nesvorny2020tidalevolution,2022A&A...662A..71F},
a moderately-elongated and rapidly-rotating primary
\citep{2021A&A...650A.129C, 2021A&A...654A..56V},
and a large family \citep[even if it has yet to be identified][]{broz2022kalliope, SylviafamVokrouhlicky2010}.
Unfortunately, it seems that many of the known binary systems differ from this prototype, with slow spinners, eccentric orbits, and unusual shapes amongst their ranks. However, these atypical binary systems seem to follow their own rules, with predictable correlations between key parameters.
This suggests that these systems are not merely mismatched outliers but rather a distinct population, potentially indicative of a distinct formation mechanism.

\subsection{Shape-eccentricity relation}

\begin{figure}
   \centering
   \includegraphics[width=\columnwidth]{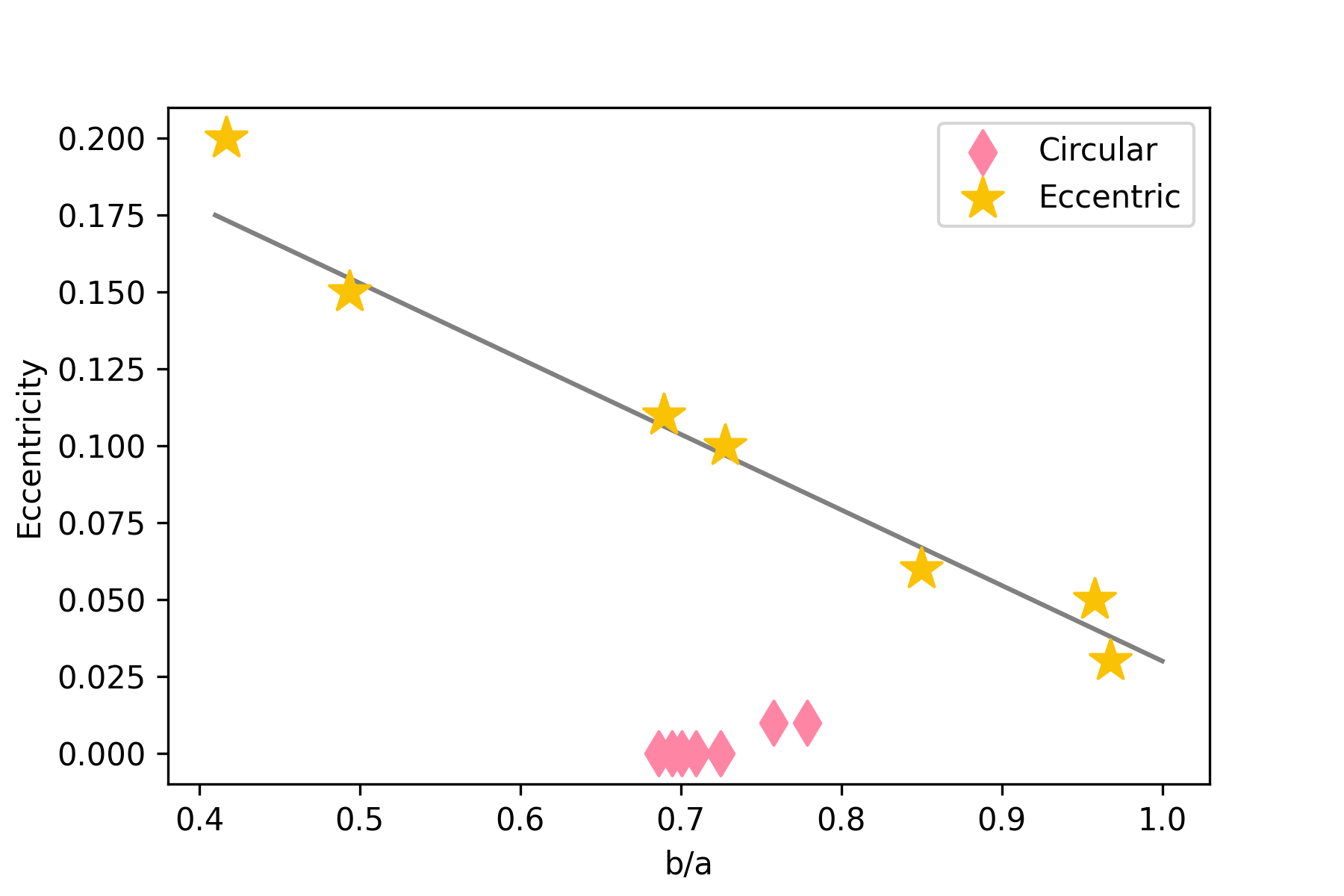}
   \caption{Eccentricity of the (outer) satellite of each binary or multiple system against the elongation of the primary body in the plane orthogonal to the primary spin axis (b/a). The individual asteroids are as follows, from left to right.
   Eccentric series: (243) Ida, (624) Hektor, (283) Emma, (130) Elektra, (702) Alauda, (31) Euphrosyne, (93) Minerva. Circular series: (87) Sylvia, (121) Hermione, (22) Kalliope, (107) Camilla, (762) Pulcova, (45) Eugenia, (41) Daphne.
   }
   \label{fig:deadbugs}
\end{figure}

The first distinction between the prototypical and atypical binary systems can be found in the satellites' orbital dynamics. The prototypical binary (or triple) systems satellites orbit on circular (e<0.01), mostly-coplanar orbits, with a possible additional inner satellite. This is not the case for the atypical binaries, which exhibit a sizeable range of eccentricities (e=0.03-0.20).
The eccentricity of the atypical binary systems is not randomly distributed, rather, it correlates strongly to the shape of the system's primary (see \Cref{fig:deadbugs}). A Pearson Correlation Coefficient (PCC) test of the linearity of this distribution provides a value of -0.98, with a p-value of $10^{-5}$.

Axes ratio values (assuming a tri-axial ellipsoid shape with a>b>c) b/a were calculated by one of the following methods, depending upon the system. For those that were observed by \citet{2021A&A...654A..56V} the b/a were sourced from that study. For the other systems, b/a values were either determined from the dimensions of the shape model listed on DAMIT or from \cite{1996idashape}, or by analysis of the shape model following the method by \citet{1996Icar..124..698D}.
The prototypical binaries occupy a narrow range of elongations (0.67<b/a<0.80), whereas the atypical binaries span a wide range of values (0.4<b/a<1).

Either of these trends can be explained intuitively, but the explanations are difficult to reconcile. First, the prototypical binaries can be explained by tidal circularization \citep{nesvorny2020tidalevolution}, which is effective regardless of the shape of the primary. This mechanism should also be more efficient for tightly bound systems, however, the distance between the primary and the satellite seems to be independent of the system's eccentricity. The correlation between eccentricity and shape in the atypical population can be assumed to be an effect of the non-spherical shape (and, by extension, non-spherical gravity) of the primary on the satellite, either at the time of satellite formation or excited over time due to tides or resonances.
A united explanation for the formation of the entire population is significantly more challenging to justify.

It could be argued that timing of formation is the differentiating factor between the two populations, that one population has not had time to circularize (or, conversely, excite) their satellites to their final orbit. However, this argument is not well founded. Under the assumption that the eccentric orbits are the final state, we would expect to see young families around the typical binaries (which we do not), and likely a wider distribution of eccentricity leading up to the shape-eccentricity sequence, of systems which are partially excited but not yet in their fully excited final geometry. If the circular binaries are the final state, a correlation between eccentricity and family age would be expected (which is not present), with older families having more circular orbits. Furthermore, it fails to explain why all systems with circular orbits occupy a much narrower range of shape than those with eccentric orbits.

\subsection{Binary-Family relationship}

It is often stated that a giant impact is required for the formation of a giant satellite \citep{hartmanndavis1975, broz2022kalliope}, and in the world of asteroids this should also require the formation of a giant family. However, families are conspicuously absent around many large binary systems. This has frequently \citep[e.g.,][]{2010AJ....139.2148V} been justified by the possibility of the families diffusing over time, with the assumption that they once existed and are no longer visible.

\begin{figure}
   \centering
   \includegraphics[width=\columnwidth]{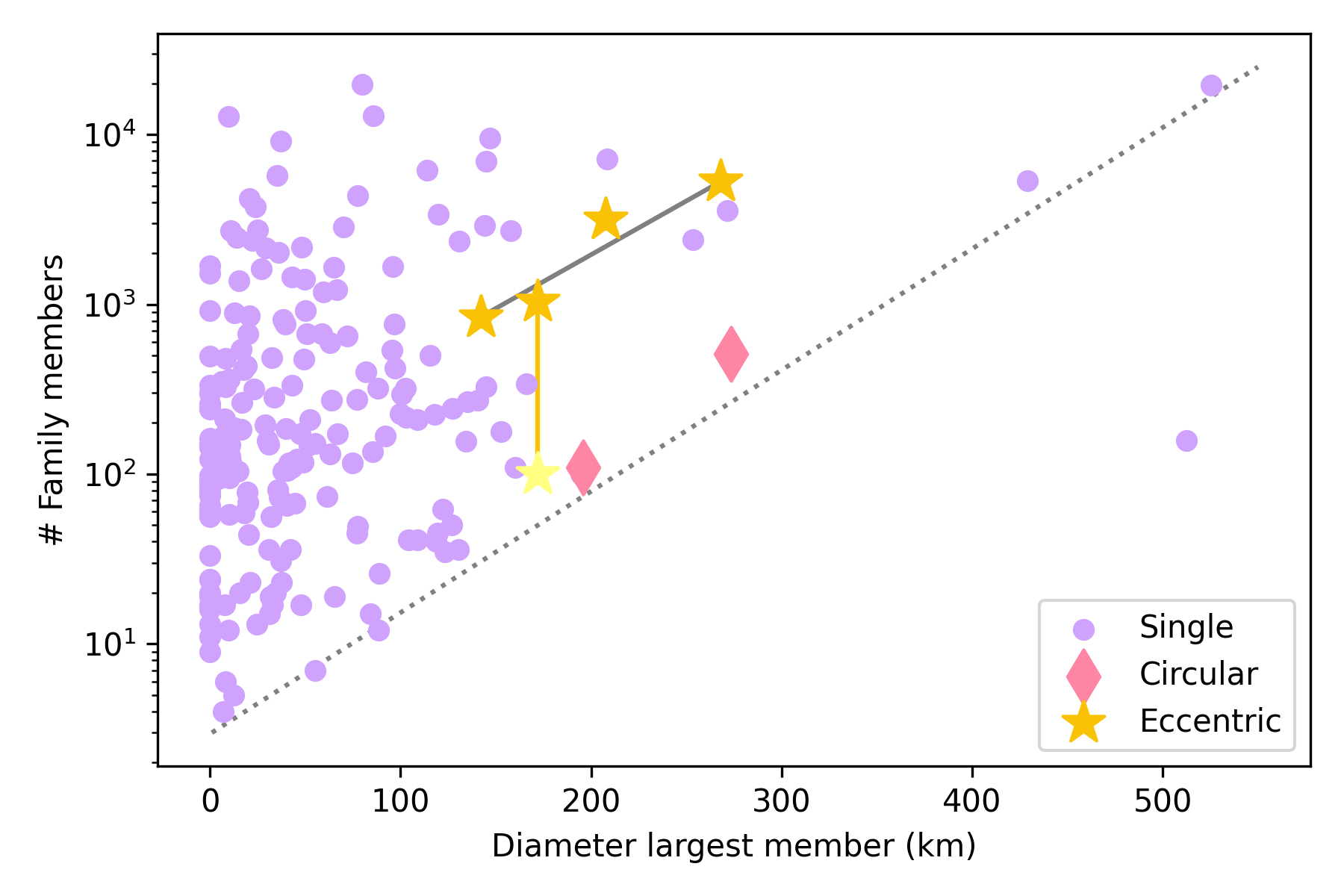}
   \caption{Comparison of binary systems to the general population of parent bodies of asteroid families. Trojan asteroid Hektor is scaled to approximate Main-Belt family size (original family size plotted in light yellow\KM{, and connected to the scaled size with a light yellow bar}).
   }
   \label{fig:deadbug_families}
\end{figure}

When examining the distribution of asteroid family sizes against the diameter of their parent bodies (\Cref{fig:deadbug_families}), it seems that the placement of binary systems is non-random. Two log-linear correlations can be seen. The first represents the
"minimum" detectable size of a family for an asteroid of a given size. This may be an artifact of the methodology used to detect asteroid families \citep{2018NatAs...2..549D}, or it may be a limit on the formation of asteroid families imposed by the parent body's escape velocity, which is linearly correlated to the parent body's diameter.
The second log-linear correlation, parallel to this, represents binary systems with large families. All families around prototypical binary systems fall along the lower-limit line. All families around atypical binaries fall on the "large family" sequence, with the exception of (624) Hektor. Hektor is a Trojan, and its large heliocentric semi-major axis compared to the other Main-Belt binary systems considered here introduces an observational bias against the discovery and detection of Hektor family members. When scaling by a factor of 29 (approximated by comparing the size frequency distributions for objects $D>10$\,km of Hektor, Emma, Alauda, and Euphrosyne families), Hektor falls along the large family line.
Notably, there are both atypical and proto-typical binaries that do not have known families. Only binary systems with primitive (C/P/B/D) compositions are shown in \Cref{fig:deadbug_families}, to minimize influences of the asteroids' composition on the kinematics. This includes the majority of our sample, as only (243) Ida and (22) Kalliope fall outside of this category. The Minerva family is not considered since (93) Minerva itself is a known interloper in that family, but it is possible that this family obfuscates a true family for Minerva \citep{minervafamily}.

\subsubsection{A note about binary-family formation}

\begin{figure}
    \centering
    \includegraphics[width=\linewidth]{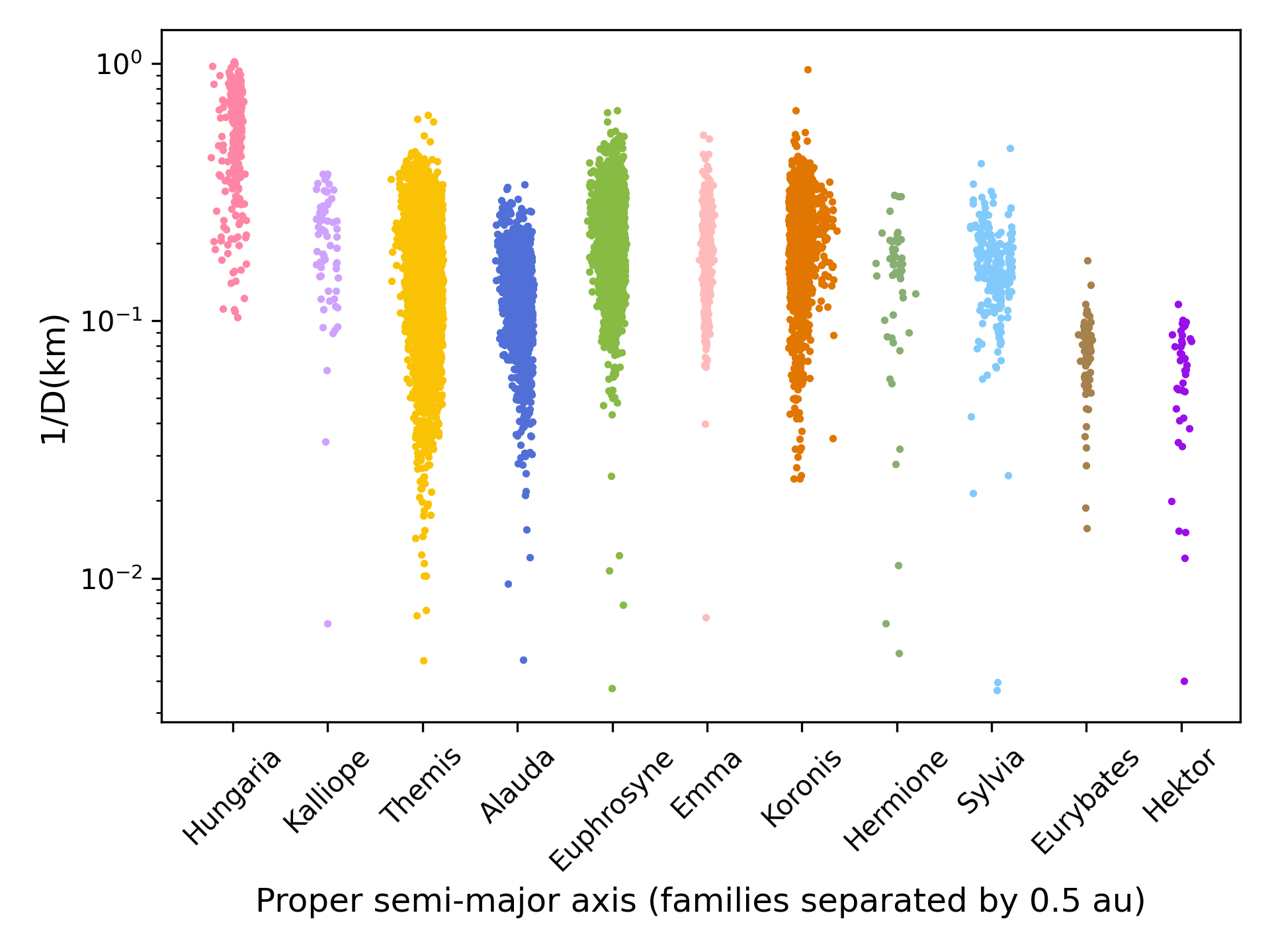}
    \caption{Proper semi-major axis vs. size ($1/D$) for known asteroid families linked to binary formation. \KM{Each family is spaced at an artificial increment equivalent to 0.5\,au for legibility.}}
    \label{fig:EEBs_SMATS}
\end{figure}

Historically, binary systems formed through impacts have been sorted into "Smashed Target Satellites" (SMATS) and "Escaping Ejecta Binaries" (EEB); where the former represents a binary system where the system's primary is the largest remaining fragment of a collisional family, and the latter where the two components are smaller fragments of the collisional family that became gravitationally bound post-impact.
However, in the limit of a fully catastrophic collision without re-accumulation to a singular large body, these definitions become ambiguous. When there are several family members of approximately the same size as the largest member (see for example Koronis family in \Cref{fig:EEBs_SMATS}), it seems inappropriate to proclaim a meaningful physical difference between the largest and nearly-largest remnants of the family. We propose instead a slightly different division, imposing the additional criteria that SMATS must have a parent body which is substantially larger than the next-largest family member, for which we propose a factor of five in mass, approximately equivalent to a factor of 1.7 in diameter, assuming similar densities. This separates systems where there is a singular massive largest member which has either remained mostly intact or disrupted and re-accumulated from those formed through catastrophic collision more meaningfully.

\subsection{Tidal dissipation}
\begin{figure}
   \centering
    \includegraphics[width=\columnwidth]{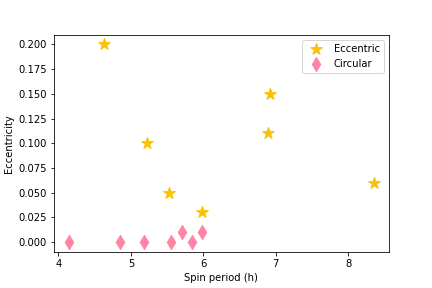}
   \caption{Rotation period vs. satellite eccentricity for large binary systems.
   }
   \label{fig:spins}
\end{figure}

\cite{marchis2008eccentric} suggests that the rapid primary rotation associated with large binary systems may be capable of inducing eccentricity. By extension, this may imply that binary systems with faster-rotating primary bodies are likely to have more-eccentric satellites. This is not globally observed, as all systems with satellites on circular orbits have primary rotation periods $<6$\,h. However, among systems with eccentric satellites and rapidly rotating primary bodies this may explain the negative correlation between spin period and eccentricity observed in the left half of
\Cref{fig:spins}. Our mass ratio for the (283) Emma system places Emmoon slightly above the excitation limit presented in Figure 4 of \cite{marchis2008eccentric}, which may explain why Emma does not follow this trend with its slow rotation period of 6.9\,h. However, this assumes a consistent bulk density between Emma and Emmoon, which may not be the case.
The other two eccentric systems with slower-rotating primaries, Hektor and Alauda, are not well enough understood for meaningful spin-orbit analysis.

Furthermore, the strong linear correlation of eccentricity seen in \Cref{fig:deadbugs} indicates that the eccentric systems are not in random states of eccentricity evolution; more likely they are either in a stable end state, or that all of the systems formed at the same time and have evolved at different speeds. The second possibility can be discarded as there are significantly different age estimates for the systems which also have families ($290\pm67$\,Myr for Emma  vs. $1309\pm312$\,Myr for Euphrosyne \cite{2019A&A...622A..47M}), and it is unlikely that a primordial satellite would survive a family forming impact. Instead, we consider the first possibility, specifically with regards to the tidal evolution of the eccentricity as described in \Cref{eq:ecc ev} \citep{1963MNRAS.126..257Ggoldreich, 1966goldreichsoter}

\begin{equation}
    \frac{de}{dt}=\frac{57 k_p M_s}{8 Q_p M_p} \left(\frac{R_p}{a}\right)^5 n e - \frac{21 k_s M_p}{2 Q_s M_s} \left(\frac{R_s}{a}\right)^5 n e
    \label{eq:ecc ev}
\end{equation}

where $Q_{p/s}$ is the tidal quality factor and $k_{p/s}$ is the tidal Love number for the primary and secondary components, respectively. Setting $\frac{de}{dt}=0$, one finds that either

\begin{equation}
e=0
\end{equation}
or

\begin{equation}
\frac{57 k_p M_s}{8 Q_p M_p} R_p^5 = \frac{21 k_s M_p}{2 Q_s M_s} R_s^5
\end{equation}

Expressed instead as a function of $\rho_{(p/s)}$, and considering the following approximation for the tidal love number of a large rubble pile asteroid \citep{goldreichsarirubblepiles},

\begin{equation}
    k = 10^{-5} \left(\frac{R}{km}\right)
\end{equation}

one finds

\begin{equation}
    \frac{57}{84} \left(\frac{\rho_s}{\rho_p}\right)^2 = \frac{Q_p}{Q_s}
\end{equation}

suggesting that a plausible stable equilibrium point could be reached where $\frac{\rho_s}{\rho_p}$ and $\frac{Q_p}{Q_s}$ are $\approx 1$, with either a slightly less dense primary than satellite, or a slightly larger quality factor for the satellite than for the primary. This could imply that the physical qualities of the systems' components (rigidity, tidal quality factor) are critical to the long-term stability of an eccentric satellite. If one assumes Q to be constant between the components, it can be found that

\begin{equation}
    \frac{\rho_p}{\rho_s}=0.83
\end{equation}

In itself, this does not explain the correlation between an objects shape and eccentricity, as this stability factor is independent of the satellites orbit, meaning that it would be impossible for the satellite to migrate into this favorable position through the tidal evolution of the eccentricity described in \Cref{eq:ecc ev}, or through tidal evolution of the semi-major axis. It could be possible that all satellites originate on this distribution and then some tidally evolve to $e=0$ on a rapid timescale in comparison to the age of the satellites, but then it is not clear what mechanism would cause all of these satellites to form with the ascribed eccentricities in the first place.

\subsection{\KM{Comparison with numerical results}}
\label{sec:kevin}
\KM{
A recent study by \cite{2025arXiv250503325W} demonstrates through SPH and N-body simulations of asteroid impacts the possibility of producing the characteristic elongated shape and rapid rotation associated with large binary asteroid systems. Among the simulations reported in this work, we identify four (numbered 7, 16, 21, 23) that are fully consistent with the narrowly constrained "typical" binary population; defined as trials in which the resulting primary spin period is $<6$\,h and the equatorial elongation is such that $0.67<b/a<0.80$. The simulations end at 38\,h post impact, so no dynamical information about fully formed satellites can be known, prohibiting us from selecting trials by $e$. There exist additional trials that meet the spin requirement but not the elongation requirement; including trials with both more and less elongated primaries. This may be influenced by the fact that some trials induced a fast (4 or 6\,h) pre-impact rotation period. Three out of these four trials lose a significant amount of mass, suggesting that family formation is likely but not strictly necessary during the formation of such an object.
}

\begin{figure}
    \centering
    \includegraphics[width=1\linewidth]{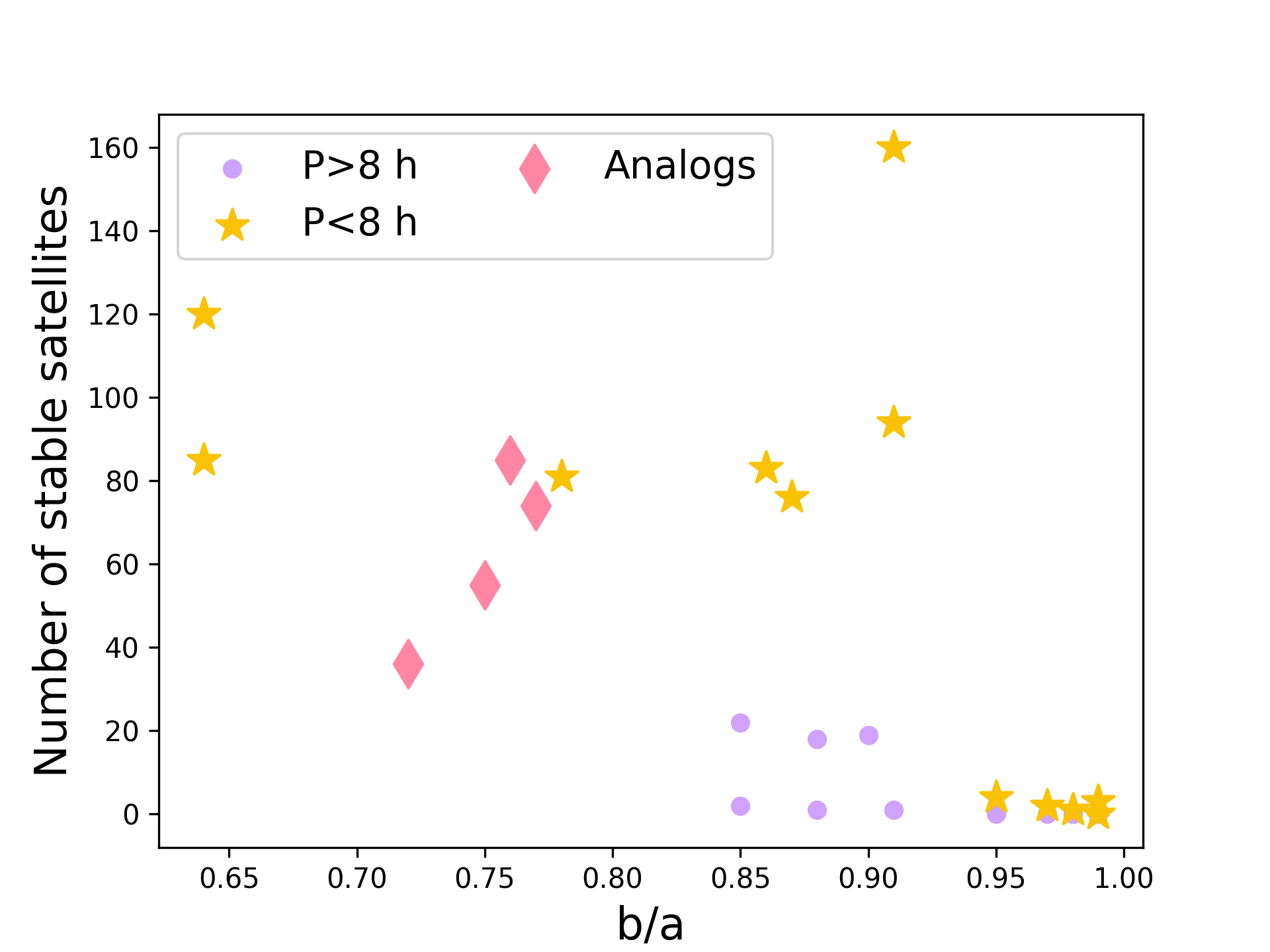}
    \caption{Simulation results from \cite{2025arXiv250503325W}, presented as number of stable satellites vs. $b/a$. Potential analogs of the "typical" population are marked as diamonds.}
    \label{fig:kevin}
\end{figure}

\KM{The results of these simulations are summarized in \Cref{fig:kevin}. A stark contrast can be observed between the elongated remnants (those with $b/a<0.80$) and the relatively spherical remnants; all of the elongated remnants are both fast rotating and surrounded by a large quantity of satellite material (proxied here by "Number of stable satellites", which indicates the number of particles in the simulation on an orbit that was not likely to lead to immediate accretion by the primary). This matches the observed population of binary systems very closely, and may indicate that the formation of an elongated asteroid and production of satellite material are fundamentally linked. In the case of a spherical remnant, cases with and without satellite material are both observed.
}

\subsection{A potential formation \KM{difference}?}

Rather than sampling different evolutionary stages of the same formation process, we propose that the two populations of the so-called "typical" and "atypical" binaries arise from two distinct formation pathways.

\begin{figure}
   \centering
    \includegraphics[width=\columnwidth]{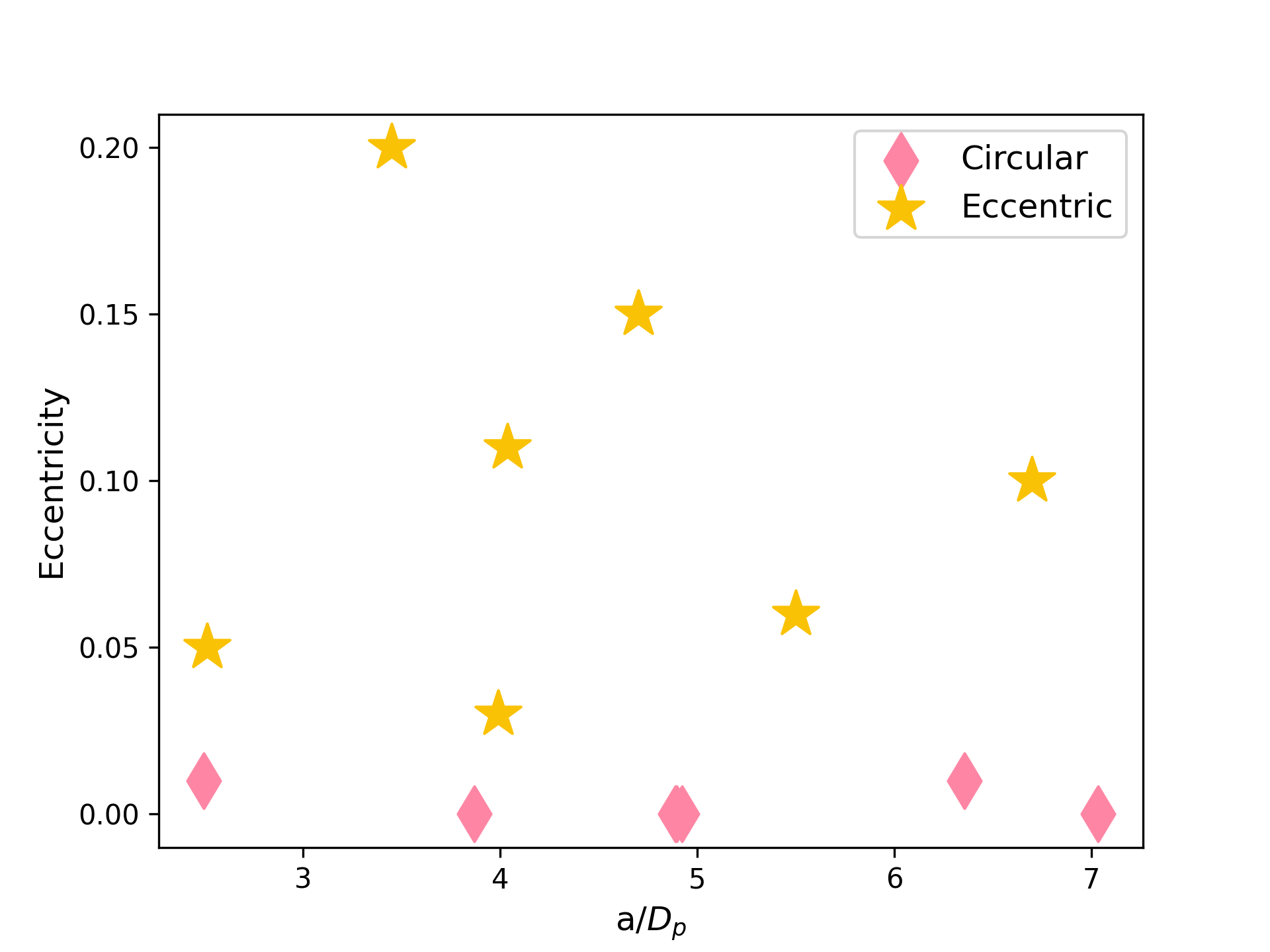}
   \caption{Semi-major axis vs. eccentricity for both populations of binary systems. The parameters seem to be independent.
   }
   \label{fig:a_e}
\end{figure}

First, a population of moderately elongated, very fast rotating asteroids with one or more satellites on circular, mostly co-planar orbits was considered. These asteroids lack families, or have very small families if present.  All identified triples belong to this category, however this may be an observational bias because the majority of well-studied systems belong to this category. Second, there is a population of asteroids with moderately to heavily eccentric satellites, exhibiting a wide but predictable range of shapes, from nearly-spherical ((31) Euphrosyne) to heavily elongated ((624) Hektor, (243) Ida). These systems tend to belong to large families, and are generally the largest remnants of these families. Among these systems, the size of the family may be correlated to the size of the parent asteroid, although the sample size is relatively small. The range of spin periods is wider, extending to over 8 hours. There is little correlation between a binary systems category and the size ratio of its components, or the normalized semi-major axis of the satellite (see for example the lack of correlation between $e$ and $a/D_p$ shown in \Cref{fig:a_e}).

We hypothesize a potential explanation for the formation of these two differing populations; that they stem from formation through catastrophic (eccentric) and sub-catastrophic (circular) impacts. Additional evidence for this is as follows.

An analysis of impact and re-accumulation satellite formation simulations by \cite{durda2004EEBs} suggests that very similar satellites can be formed by both low-energy, sub-catastrophic (largest remaining fragment equates to nearly the same size as the original parent body, generally as the result of a relatively low energy impact) and significantly catastrophic (largest remaining fragment is substantially smaller than the original parent body). Both of these situations can result in large ($D_s/D_p\approx0.10$) and small ($D_s/D_p\approx0.03$) satellites. The significantly catastrophic impacts result in a larger number of family members than the sub-catastrophic impacts, which in some cases generate very few family members. According to the Durda et al. simulations, the sub-catastrophic formation scenario typically results from a small, high-speed impactor, whereas the catastrophic scenario occurs more frequently with larger, slower impactors.

For the systems formed through sub-catastrophic impact, the rapid rotation is likely induced from the same impact that forms the satellite \citep{2025arXiv250503325W}. Restructuring to the interior of the primary body would be minimal, and the ultimate result of the impact would be closer to a major cratering event than total disruption and re-accretion.

For the systems formed through catastrophic impacts a high energy, very catastrophic impact occurs, causing most of the mass of the original parent body to be ejected. Much of this ejecta then re-accumulates onto the parent body, and a small fraction forms the satellite. Remaining ejecta forms the asteroid's family.
There is a relationship between the diameter and elongation of the systems' primaries, in that larger asteroids tend to be more spherical. This may be because the larger, more massive asteroids more efficiently re-accumulate material due to their higher masses.
This has been previously demonstrated for (10) Hygiea \citep{2020NatAs...4..136V}.
If the primary reaccumulates to a non-spherical shape, this could then influence the orbit of the satellite, causing it to settle on an eccentric orbit (it is also possible for the primary to settle into a spherical shape, see once again Hygiea).
Ida is an end-member in this scenario, representing the case where a fully catastrophic impact occurs and there is no remnant representing the majority of the mass of the original parent body, but a collection of several similarly sized "large" fragments. In the case of the Koronis family, these fragments are $D=30-40$ km family members such as Koronis, Ida, and Nassovia.

Several members of the catastrophic group have also been independently linked to a catastrophic impact, including (31) Euphrosyne \citep{2020A&A...641A..80Y}, (243) Ida \citep{durda2004EEBs}, and (283) Emma (this work).
It would be of significant interest to re-study collisional formation of binary asteroids using modern modeling techniques and considering the current observational knowledge of large binary asteroid systems, however, this is beyond the scope of this study. \KM{As discussed in \Cref{sec:kevin}, additional work on this subject has recently been published by \cite{2025arXiv250503325W}, but only considering low-energy collisions where no catastrophic disruption occurs.}

\section{Conclusions}

Although binary asteroid systems (762) Pulcova and (283) Emma may appear very similar, detailed study of their orbital dynamics reveals key differences. For Pulcova, the circular, co-planar satellite geometry limits the possibility for internal structure studies, as well-fitting solutions can be identified across a wide range of $J_2$ values. The possibility of a truly homogeneous internal structure is not excluded. Although a purely Keplerian fit is mathematically feasible, it is not physically realistic.
Alternatively, the (283) Emma system exhibits clear signs of precession, allowing for a precise identification of $J_2=\numb{0.11\pm0.01}$, given a radius of $R=74.5$\,km or $J_2=\numb{0.13}$ given a radius of $R=67$\,km. Assuming the shape model to be accurate, this implies a significant heterogeneity in Emma's internal structure, possibly indicative a catastrophic impact in the system's history. Additionally, when expanding to consider the entire set of large binary asteroid systems, evidence for the presence of two distinct populations can be observed; one presenting a relationship between shape, satellite dynamics, and family formation; and the other a surprising amount of homogeneity between systems.
This is potentially indicative of multiple satellite formation pathways.


\begin{acknowledgements}
\label{sec:acknowledgements}%

Thank you to J. Maia, R. Marschall, H. Agrusa, A. Crida, M. Goldberg and S. Markham for the fruitful discussion which has shaped this manuscript, and to P. Tamblyn, C. Chapman, F. Menard, J. Drummond, W. Owen, J. Christou, B. Enke, L. Sromovsky, N. Siegler, C. Veillet, D. Terrell, G. Duvert, D. Slater
and J. C. Shelton
 who have contributed to the acquisition of the observations used in this study.
TRAPPIST is funded by the Belgian National Fund for Scientific Research (F.R.S.-FNRS) under grant PDR T.0120.21. Based on observations carried out in la Silla Observatory. E. Jehin is a FNRS Director of Research.
This research has made use of the Keck Observatory Archive (KOA),
which is operated by the W. M. Keck Observatory and the NASA
Exoplanet Science Institute (NExScI), under contract with the
National Aeronautics and Space Administration.
Some of the data presented herein were obtained at Keck Observatory, which is a private 501(c)3 non-profit organization operated as a scientific partnership among the California Institute of Technology, the University of California, and the National Aeronautics and Space Administration. The Observatory was made possible by the generous financial support of the W. M. Keck Foundation.
The authors wish to recognize and acknowledge the very significant cultural role and reverence that the summit of Maunakea has always had within the Native Hawaiian community. We are most fortunate to have the opportunity to conduct observations from this mountain.
This publication is based in part on observations obtained at the international Gemini Observatory, a program of NSF NOIRLab, acquired through the Gemini Observatory Archive at NSF NOIRLab, which is managed by the Association of Universities for Research in Astronomy (AURA) under a cooperative agreement with the U.S. National Science Foundation on behalf of the Gemini Observatory partnership: the U.S. National Science Foundation (United States), National Research Council (Canada), Agencia Nacional de Investigaci\'{o}n y Desarrollo (Chile), Ministerio de Ciencia, Tecnolog\'{i}a e Innovaci\'{o}n (Argentina), Minist\'{e}rio da Ci\^{e}ncia, Tecnologia, Inova\c{c}\~{o}es e Comunica\c{c}\~{o}es (Brazil), and Korea Astronomy and Space Science Institute (Republic of Korea).This work was enabled by observations made from the Gemini North telescope, located within the Maunakea Science Reserve and adjacent to the summit of Maunakea. We are grateful for the privilege of observing the Universe from a place that is unique in both its astronomical quality and its cultural significance.
This publication has made use of the Canadian Astronomy Data Centre moving object search tool \citep{CADCmovingobject}.
This research used the
\miriade \citep{2009-EPSC-Berthier},
\ssodnet \citep{2023A&A...671A.151B},
and
\topcat
\citep{2005ASPC..347...29T}
Virtual Observatory tools.
This work used the Minor Planet Physical Properties Catalogue (MP3C)
of the Observatoire de la Côte d'Azur.
It used the
\astropy
\citep{astropy:2013, astropy:2018, astropy:2022} and
\rocks \citep{2023A&A...671A.151B}
python packages.

This paper is based in part on observations obtained with the Adaptive
Optics System Hokupa'a/Quirc, developed and operated by the
University of Hawaii Adaptive Optics Group, with support from the
National Science Foundation.
The Czech Science Foundation has supported the research of JD and JH through grant 22-17783S.
LL was supported by the project GaiaMoons of the Agence Nationale de Recherche (France), grant ANR-22-CE49-0002.
JLM was supported in part by NASA Solar System Working Grant 80NSSC23K0102.

\end{acknowledgements}

\ifx\destination\arxiv
  \bibliographystyle{aux/arxiv}
\fi

\ifx\destination\aanda
 \bibliographystyle{aux/aa}
\fi

\ifx\destination\publisher
  \bibliographystyle{aux/publisher}
  \biboptions{authoryear}
\fi

\bibliography{main}

\begin{thebibliography}{99}
\expandafter\ifx\csname natexlab\endcsname\relax\def\natexlab#1{#1}\fi

\bibitem[{{Alton}(2011)}]{underoakkevin}
{Alton}, K.~B. 2011, Minor Planet Bulletin, 38, 8

\bibitem[{{Astropy Collaboration} {et~al.}(2022){Astropy Collaboration},
  {Price-Whelan}, {Lim}, {Earl}, {Starkman}, {Bradley}, {Shupe}, {Patil},
  {Corrales}, {Brasseur}, {N{"o}the}, {Donath}, {Tollerud}, {Morris},
  {Ginsburg}, {Vaher}, {Weaver}, {Tocknell}, {Jamieson}, {van Kerkwijk},
  {Robitaille}, {Merry}, {Bachetti}, {G{"u}nther}, {Aldcroft},
  {Alvarado-Montes}, {Archibald}, {B{'o}di}, {Bapat}, {Barentsen}, {Baz{'a}n},
  {Biswas}, {Boquien}, {Burke}, {Cara}, {Cara}, {Conroy}, {Conseil}, {Craig},
  {Cross}, {Cruz}, {D'Eugenio}, {Dencheva}, {Devillepoix}, {Dietrich},
  {Eigenbrot}, {Erben}, {Ferreira}, {Foreman-Mackey}, {Fox}, {Freij}, {Garg},
  {Geda}, {Glattly}, {Gondhalekar}, {Gordon}, {Grant}, {Greenfield}, {Groener},
  {Guest}, {Gurovich}, {Handberg}, {Hart}, {Hatfield-Dodds}, {Homeier},
  {Hosseinzadeh}, {Jenness}, {Jones}, {Joseph}, {Kalmbach}, {Karamehmetoglu},
  {Ka{l}uszy{'n}ski}, {Kelley}, {Kern}, {Kerzendorf}, {Koch}, {Kulumani},
  {Lee}, {Ly}, {Ma}, {MacBride}, {Maljaars}, {Muna}, {Murphy}, {Norman},
  {O'Steen}, {Oman}, {Pacifici}, {Pascual}, {Pascual-Granado}, {Patil},
  {Perren}, {Pickering}, {Rastogi}, {Roulston}, {Ryan}, {Rykoff}, {Sabater},
  {Sakurikar}, {Salgado}, {Sanghi}, {Saunders}, {Savchenko}, {Schwardt},
  {Seifert-Eckert}, {Shih}, {Jain}, {Shukla}, {Sick}, {Simpson},
  {Singanamalla}, {Singer}, {Singhal}, {Sinha}, {Sip{H{o}}cz}, {Spitler},
  {Stansby}, {Streicher}, {{ {S}}umak}, {Swinbank}, {Taranu}, {Tewary},
  {Tremblay}, {Val-Borro}, {Van Kooten}, {Vasovi{'c}}, {Verma}, {de Miranda
  Cardoso}, {Williams}, {Wilson}, {Winkel}, {Wood-Vasey}, {Xue}, {Yoachim},
  {Zhang}, {Zonca}, \& {Astropy Project Contributors}}]{astropy:2022}
{Astropy Collaboration}, {Price-Whelan}, A.~M., {Lim}, P.~L., {et~al.} 2022,
  apj, 935, 167

\bibitem[{{Astropy Collaboration} {et~al.}(2018){Astropy Collaboration},
  {Price-Whelan}, {Sip{\H{o}}cz}, {G{\"u}nther}, {Lim}, {Crawford}, {Conseil},
  {Shupe}, {Craig}, {Dencheva}, {Ginsburg}, {VanderPlas}, {Bradley},
  {P{\'e}rez-Su{\'a}rez}, {de Val-Borro}, {Aldcroft}, {Cruz}, {Robitaille},
  {Tollerud}, {Ardelean}, {Babej}, {Bach}, {Bachetti}, {Bakanov}, {Bamford},
  {Barentsen}, {Barmby}, {Baumbach}, {Berry}, {Biscani}, {Boquien}, {Bostroem},
  {Bouma}, {Brammer}, {Bray}, {Breytenbach}, {Buddelmeijer}, {Burke},
  {Calderone}, {Cano Rodr{\'\i}guez}, {Cara}, {Cardoso}, {Cheedella}, {Copin},
  {Corrales}, {Crichton}, {D'Avella}, {Deil}, {Depagne}, {Dietrich}, {Donath},
  {Droettboom}, {Earl}, {Erben}, {Fabbro}, {Ferreira}, {Finethy}, {Fox},
  {Garrison}, {Gibbons}, {Goldstein}, {Gommers}, {Greco}, {Greenfield},
  {Groener}, {Grollier}, {Hagen}, {Hirst}, {Homeier}, {Horton}, {Hosseinzadeh},
  {Hu}, {Hunkeler}, {Ivezi{\'c}}, {Jain}, {Jenness}, {Kanarek}, {Kendrew},
  {Kern}, {Kerzendorf}, {Khvalko}, {King}, {Kirkby}, {Kulkarni}, {Kumar},
  {Lee}, {Lenz}, {Littlefair}, {Ma}, {Macleod}, {Mastropietro}, {McCully},
  {Montagnac}, {Morris}, {Mueller}, {Mumford}, {Muna}, {Murphy}, {Nelson},
  {Nguyen}, {Ninan}, {N{\"o}the}, {Ogaz}, {Oh}, {Parejko}, {Parley}, {Pascual},
  {Patil}, {Patil}, {Plunkett}, {Prochaska}, {Rastogi}, {Reddy Janga},
  {Sabater}, {Sakurikar}, {Seifert}, {Sherbert}, {Sherwood-Taylor}, {Shih},
  {Sick}, {Silbiger}, {Singanamalla}, {Singer}, {Sladen}, {Sooley},
  {Sornarajah}, {Streicher}, {Teuben}, {Thomas}, {Tremblay}, {Turner},
  {Terr{\'o}n}, {van Kerkwijk}, {de la Vega}, {Watkins}, {Weaver}, {Whitmore},
  {Woillez}, {Zabalza}, \& {Astropy Contributors}}]{astropy:2018}
{Astropy Collaboration}, {Price-Whelan}, A.~M., {Sip{\H{o}}cz}, B.~M., {et~al.}
  2018, \aj, 156, 123

\bibitem[{{Astropy Collaboration} {et~al.}(2013){Astropy Collaboration},
  {Robitaille}, {Tollerud}, {Greenfield}, {Droettboom}, {Bray}, {Aldcroft},
  {Davis}, {Ginsburg}, {Price-Whelan}, {Kerzendorf}, {Conley}, {Crighton},
  {Barbary}, {Muna}, {Ferguson}, {Grollier}, {Parikh}, {Nair}, {Unther},
  {Deil}, {Woillez}, {Conseil}, {Kramer}, {Turner}, {Singer}, {Fox}, {Weaver},
  {Zabalza}, {Edwards}, {Azalee Bostroem}, {Burke}, {Casey}, {Crawford},
  {Dencheva}, {Ely}, {Jenness}, {Labrie}, {Lim}, {Pierfederici}, {Pontzen},
  {Ptak}, {Refsdal}, {Servillat}, \& {Streicher}}]{astropy:2013}
{Astropy Collaboration}, {Robitaille}, T.~P., {Tollerud}, E.~J., {et~al.} 2013,
  \aap, 558, A33

\bibitem[{Behrend {et~al.}(2006)Behrend, Bernasconi, Roy, Klotz, Colas,
  Antonini, Aoun, Augustesen, Barbotin, Berger, Berrouachdi, Brochard,
  Cazenave, Cavadore, Coloma, Cotrez, Deconihout, Demeautis, Dorseuil, Dubos,
  Durkee, Frappa, Hormuth, Itkonen, Jacques, Kurtze, Laffont, Lavayssi{\`e}re,
  Lecacheux, Leroy, Manzini, Masi, Matter, Michelsen, Nomen, Oksanen,
  P{\"a}{\"a}kk{\"o}nen, Peyrot, Pimentel, Pray, Rinner, Sanchez, Sonnenberg,
  Sposetti, Starkey, Stoss, Teng, Vignand, \& Waelchli}]{2006A&A...446.1177B}
Behrend, R., Bernasconi, L., Roy, R., {et~al.} 2006, \aap, 446, 1177

\bibitem[{{Berthier} {et~al.}(2023){Berthier}, {Carry}, {Mahlke}, \&
  {Normand}}]{2023A&A...671A.151B}
{Berthier}, J., {Carry}, B., {Mahlke}, M., \& {Normand}, J. 2023, \aap, 671,
  A151

\bibitem[{Berthier {et~al.}(2009)Berthier, Hestroffer, Carry, Vachier, Lainey,
  Emelyanov, Thuillot, Arlot, \& Service}]{2009-EPSC-Berthier}
Berthier, J., Hestroffer, D., Carry, B., {et~al.} 2009, in European Planetary
  Science Congress 2009, 676

\bibitem[{Berthier {et~al.}(2014)Berthier, Vachier, Marchis, {\v D}urech, \&
  Carry}]{2014Icar..239..118B}
Berthier, J., Vachier, F., Marchis, F., {\v D}urech, J., \& Carry, B. 2014,
  \icarus, 239, 118

\bibitem[{Britt {et~al.}(2002)Britt, Yeomans, Housen, \&
  Consolmagno}]{2002-AsteroidsIII-4.2-Britt}
Britt, D.~T., Yeomans, D.~K., Housen, K.~R., \& Consolmagno, G.~J. 2002,
  Asteroids III, 485

\bibitem[{{Bro{\v{z}}} {et~al.}(2022){Bro{\v{z}}}, {Ferrais}, {Vernazza},
  {{\v{S}}eve{\v{c}}ek}, \& {Jutzi}}]{broz2022kalliope}
{Bro{\v{z}}}, M., {Ferrais}, M., {Vernazza}, P., {{\v{S}}eve{\v{c}}ek}, P., \&
  {Jutzi}, M. 2022, \aap, 664, A69

\bibitem[{{Bro{\v{z}}} {et~al.}(2021){Bro{\v{z}}}, {Marchis}, {Jorda},
  {Hanu{\v{s}}}, {Vernazza}, {Ferrais}, {Vachier}, {Rambaux}, {Marsset},
  {Viikinkoski}, {Jehin}, {Benseguane}, {Podlewska-Gaca}, {Carry}, {Drouard},
  {Fauvaud}, {Birlan}, {Berthier}, {Bartczak}, {Dumas}, {Dudzi{\'n}ski},
  {{\v{D}}urech}, {Castillo-Rogez}, {Cipriani}, {Colas}, {Fetick}, {Fusco},
  {Grice}, {Kryszczynska}, {Lamy}, {Marciniak}, {Michalowski}, {Michel},
  {Pajuelo}, {Santana-Ros}, {Tanga}, {Vigan}, {Vokrouhlick{\'y}}, {Witasse}, \&
  {Yang}}]{2021A&A...653A..56B}
{Bro{\v{z}}}, M., {Marchis}, F., {Jorda}, L., {et~al.} 2021, \aap, 653, A56

\bibitem[{{Bro{\v{z}}} {et~al.}(2013){Bro{\v{z}}}, {Morbidelli}, {Bottke},
  {Rozehnal}, {Vokrouhlick{\'y}}, \& {Nesvorn{\'y}}}]{2013A&A...551A.117B}
{Bro{\v{z}}}, M., {Morbidelli}, A., {Bottke}, W.~F., {et~al.} 2013, \aap, 551,
  A117

\bibitem[{Carry(2012)}]{2012P&SS...73...98C}
Carry, B. 2012, \planss, 73, 98

\bibitem[{Carry {et~al.}(2008)Carry, Dumas, Fulchignoni, Merline, Berthier,
  Hestroffer, Fusco, \& Tamblyn}]{2008AA...478..235C}
Carry, B., Dumas, C., Fulchignoni, M., {et~al.} 2008, \aap, 478, 235

\bibitem[{Carry {et~al.}(2010)Carry, Dumas, Kaasalainen, Berthier, Merline,
  Erard, Conrad, Drummond, Hestroffer, Fulchignoni, \&
  Fusco}]{2010Icar..205..460C}
Carry, B., Dumas, C., Kaasalainen, M., {et~al.} 2010, \icarus, 205, 460

\bibitem[{{Carry} {et~al.}(2019){Carry}, {Vachier}, {Berthier}, {Marsset},
  {Vernazza}, {Grice}, {Merline}, {Lagadec}, {Fienga}, {Conrad},
  {Podlewska-Gaca}, {Santana-Ros}, {Viikinkoski}, {Hanu{\v{s}}}, {Dumas},
  {Drummond}, {Tamblyn}, {Chapman}, {Behrend}, {Bernasconi}, {Bartczak},
  {Benkhaldoun}, {Birlan}, {Castillo-Rogez}, {Cipriani}, {Colas}, {Drouard},
  {{\v{D}}urech}, {Enke}, {Fauvaud}, {Ferrais}, {Fetick}, {Fusco}, {Gillon},
  {Jehin}, {Jorda}, {Kaasalainen}, {Keppler}, {Kryszczynska}, {Lamy},
  {Marchis}, {Marciniak}, {Michalowski}, {Michel}, {Pajuelo}, {Tanga}, {Vigan},
  {Warner}, {Witasse}, {Yang}, \& {Zurlo}}]{2019A&A...623A.132C}
{Carry}, B., {Vachier}, F., {Berthier}, J., {et~al.} 2019, \aap, 623, A132

\bibitem[{{Carry} {et~al.}(2021){Carry}, {Vernazza}, {Vachier}, {Neveu},
  {Berthier}, {Hanu{\v{s}}}, {Ferrais}, {Jorda}, {Marsset}, {Viikinkoski},
  {Bartczak}, {Behrend}, {Benkhaldoun}, {Birlan}, {Castillo-Rogez}, {Cipriani},
  {Colas}, {Drouard}, {Dudzi{\'n}ski}, {Desmars}, {Dumas}, {{\v{D}}urech},
  {Fetick}, {Fusco}, {Grice}, {Jehin}, {Kaasalainen}, {Kryszczynska}, {Lamy},
  {Marchis}, {Marciniak}, {Michalowski}, {Michel}, {Pajuelo}, {Podlewska-Gaca},
  {Rambaux}, {Santana-Ros}, {Storrs}, {Tanga}, {Vigan}, {Warner}, {Wieczorek},
  {Witasse}, \& {Yang}}]{2021A&A...650A.129C}
{Carry}, B., {Vernazza}, P., {Vachier}, F., {et~al.} 2021, \aap, 650, A129

\bibitem[{Chapman {et~al.}(1995)Chapman, Veverka, Thomas, Klaasen, Belton,
  Harch, McEwen, Johnson, Helfenstein, Davies, Merline, \&
  Denk}]{1995Natur.374..783C}
Chapman, C.~R., Veverka, J., Thomas, P.~C., {et~al.} 1995, \nat, 374, 783

\bibitem[{{Delbo} {et~al.}(2019){Delbo}, {Avdellidou}, \&
  {Morbidelli}}]{2019A&A...624A..69D}
{Delbo}, M., {Avdellidou}, C., \& {Morbidelli}, A. 2019, \aap, 624, A69

\bibitem[{Delbo {et~al.}(2017)Delbo, Walsh, Bolin, Avdellidou, \&
  Morbidelli}]{2017Sci...357.1026D}
Delbo, M., Walsh, K., Bolin, B., Avdellidou, C., \& Morbidelli, A. 2017,
  {Science}, 357, 1026

\bibitem[{{Dermott} {et~al.}(2018){Dermott}, {Christou}, {Li}, {Kehoe}, \&
  {Robinson}}]{2018NatAs...2..549D}
{Dermott}, S.~F., {Christou}, A.~A., {Li}, D., {Kehoe}, T. J.~J., \&
  {Robinson}, J.~M. 2018, Nature Astronomy, 2, 549

\bibitem[{Dobrovolskis(1996)}]{1996Icar..124..698D}
Dobrovolskis, A.~R. 1996, \icarus, 124, 698

\bibitem[{{Durda} {et~al.}(2004){Durda}, {Bottke}, {Enke}, {Merline},
  {Asphaug}, {Richardson}, \& {Leinhardt}}]{durda2004EEBs}
{Durda}, D.~D., {Bottke}, W.~F., {Enke}, B.~L., {et~al.} 2004, \icarus, 170,
  243

\bibitem[{{\v D}urech {et~al.}(2015){\v D}urech, Carry, Delbo, Kaasalainen, \&
  Viikinkoski}]{2015-AsteroidsIV-Durech}
{\v D}urech, J., Carry, B., Delbo, M., Kaasalainen, M., \& Viikinkoski, M.
  2015, {Asteroid Models from Multiple Data Sources} (Univ. Arizona Press),
  183--202

\bibitem[{{\v D}urech {et~al.}(2011){\v D}urech, Kaasalainen, Herald, Dunham,
  Timerson, Hanu{\v{s}}, Frappa, Talbot, Hayamizu, Warner, Pilcher, \&
  Gal{\'a}d}]{2011Icar..214..652D}
{\v D}urech, J., Kaasalainen, M., Herald, D., {et~al.} 2011, \icarus, 214, 652

\bibitem[{{\v D}urech {et~al.}(2010){\v D}urech, Sidorin, \&
  Kaasalainen}]{2010A&A...513A..46D}
{\v D}urech, J., Sidorin, V., \& Kaasalainen, M. 2010, \aap, 513, A46

\bibitem[{{Dykhuis} {et~al.}(2014){Dykhuis}, {Molnar}, {Van Kooten}, \&
  {Greenberg}}]{2014Icar..243..111D}
{Dykhuis}, M.~J., {Molnar}, L., {Van Kooten}, S.~J., \& {Greenberg}, R. 2014,
  \icarus, 243, 111

\bibitem[{{Fang} \& {Margot}(2012)}]{2012AJ....143...24F}
{Fang}, J. \& {Margot}, J.-L. 2012, \aj, 143, 24

\bibitem[{{Fang} {et~al.}(2012){Fang}, {Margot}, \& {Rojo}}]{fangsylvia}
{Fang}, J., {Margot}, J.-L., \& {Rojo}, P. 2012, \aj, 144, 70

\bibitem[{{Ferrais} {et~al.}(2022){Ferrais}, {Jorda}, {Vernazza}, {Carry},
  {Bro{\v{z}}}, {Rambaux}, {Hanu{\v{s}}}, {Dudzi{\'n}ski}, {Bartczak},
  {Vachier}, {Aristidi}, {Beck}, {Marchis}, {Marsset}, {Viikinkoski}, {Fetick},
  {Drouard}, {Fusco}, {Birlan}, {Podlewska-Gaca}, {Burbine}, {Dyar},
  {Bendjoya}, {Benkhaldoun}, {Berthier}, {Castillo-Rogez}, {Cipriani}, {Colas},
  {Dumas}, {{\v{D}}urech}, {Fauvaud}, {Grice}, {Jehin}, {Kaasalainen},
  {Kryszczynska}, {Lamy}, {Le Coroller}, {Marciniak}, {Michalowski}, {Michel},
  {Prieur}, {Reddy}, {Rivet}, {Santana-Ros}, {Scardia}, {Tanga}, {Vigan},
  {Witasse}, \& {Yang}}]{2022A&A...662A..71F}
{Ferrais}, M., {Jorda}, L., {Vernazza}, P., {et~al.} 2022, \aap, 662, A71

\bibitem[{{Fuksa} {et~al.}(2023){Fuksa}, {Bro{\v{z}}}, {Hanu{\v{s}}},
  {Ferrais}, {Fatka}, \& {Vernazza}}]{2023A&A...677A.189F}
{Fuksa}, M., {Bro{\v{z}}}, M., {Hanu{\v{s}}}, J., {et~al.} 2023, \aap, 677,
  A189

\bibitem[{{Goldreich}(1963)}]{1963MNRAS.126..257Ggoldreich}
{Goldreich}, P. 1963, \mnras, 126, 257

\bibitem[{{Goldreich} \& {Sari}(2009)}]{goldreichsarirubblepiles}
{Goldreich}, P. \& {Sari}, R. 2009, \apj, 691, 54

\bibitem[{{Goldreich} \& {Soter}(1966)}]{1966goldreichsoter}
{Goldreich}, P. \& {Soter}, S. 1966, \icarus, 5, 375

\bibitem[{Gomes-Júnior {et~al.}(2022)Gomes-Júnior, Morgado, Benedetti-Rossi,
  Boufleur, Rommel, Banda-Huarca, Kilic, Braga-Ribas, \& Sicardy}]{sora_pred}
Gomes-Júnior, A.~R., Morgado, B.~E., Benedetti-Rossi, G., {et~al.} 2022,
  Monthly Notices of the Royal Astronomical Society, 511, 1167

\bibitem[{{Graves} {et~al.}(1998){Graves}, {Northcott}, {Roddier}, {Roddier},
  \& {Close}}]{graves1998HokupaaSPIE}
{Graves}, J.~E., {Northcott}, M.~J., {Roddier}, F.~J., {Roddier}, C.~A., \&
  {Close}, L.~M. 1998, in Society of Photo-Optical Instrumentation Engineers
  (SPIE) Conference Series, Vol. 3353, Adaptive Optical System Technologies,
  ed. D.~{Bonaccini} \& R.~K. {Tyson}, 34--43

\bibitem[{{Gwyn} {et~al.}(2012){Gwyn}, {Hill}, \&
  {Kavelaars}}]{CADCmovingobject}
{Gwyn}, S. D.~J., {Hill}, N., \& {Kavelaars}, J.~J. 2012, \pasp, 124, 579

\bibitem[{{Hartmann} \& {Davis}(1975)}]{hartmanndavis1975}
{Hartmann}, W.~K. \& {Davis}, D.~R. 1975, \icarus, 24, 504

\bibitem[{{Herald} {et~al.}(2024){Herald}, {Gault}, {Carlson}, {Guhl},
  {Frappa}, {Giacchini}, {Hayamizu}, {Kerr}, \&
  {Moore}}]{heraldPDSocculatation}
{Herald}, D., {Gault}, D., {Carlson}, N., {et~al.} 2024

\bibitem[{{Hirabayashi} {et~al.}(2020){Hirabayashi}, {Trowbridge}, \&
  {Bodewits}}]{arrokothdensity}
{Hirabayashi}, M., {Trowbridge}, A.~J., \& {Bodewits}, D. 2020, \apjl, 891, L12

\bibitem[{Hodapp {et~al.}(2003)Hodapp, Jensen, Irwin, Yamada, Chung, Fletcher,
  Robertson, Hora, Simons, Mays, Nolan, Bec, Merrill, \&
  Fowler}]{2003-PASP-115-Hodapp}
Hodapp, K.~W., Jensen, J.~B., Irwin, E.~M., {et~al.} 2003, Publications of the
  Astronomical Society of the Pacific, 115, 1388

\bibitem[{IMCCE(2021)}]{2021-IMCCEbook-withorbits}
IMCCE. 2021, {Introduction aux \'{e}ph\'{e}m\'{e}rides et ph\'{e}nom\`{e}nes
  astronomiques}, ed. J.~{Berthier}, P.~{Descamps}, \& F.~{Mignard} (edp
  sciences)

\bibitem[{{Jacobson} \& {Scheeres}(2011)}]{jacobson2011}
{Jacobson}, S.~A. \& {Scheeres}, D.~J. 2011, \apjl, 736, L19

\bibitem[{{Jehin} {et~al.}(2011){Jehin}, {Gillon}, {Queloz}, {Magain},
  {Manfroid}, {Chantry}, {Lendl}, {Hutsem{\'e}kers}, \&
  {Udry}}]{2011Msngr.145....2J}
{Jehin}, E., {Gillon}, M., {Queloz}, D., {et~al.} 2011, The Messenger, 145, 2

\bibitem[{{Jorda} {et~al.}(2016){Jorda}, {Gaskell}, {Capanna}, {Hviid}, {Lamy},
  {{\v{D}}urech}, {Faury}, {Groussin}, {Guti{\'e}rrez}, {Jackman}, {Keihm},
  {Keller}, {Knollenberg}, {K{\"u}hrt}, {Marchi}, {Mottola}, {Palmer},
  {Schloerb}, {Sierks}, {Vincent}, {A'Hearn}, {Barbieri}, {Rodrigo}, {Koschny},
  {Rickman}, {Barucci}, {Bertaux}, {Bertini}, {Cremonese}, {Da Deppo},
  {Davidsson}, {Debei}, {De Cecco}, {Fornasier}, {Fulle}, {G{\"u}ttler}, {Ip},
  {Kramm}, {K{\"u}ppers}, {Lara}, {Lazzarin}, {Lopez Moreno}, {Marzari},
  {Naletto}, {Oklay}, {Thomas}, {Tubiana}, \& {Wenzel}}]{67pdensity}
{Jorda}, L., {Gaskell}, R., {Capanna}, C., {et~al.} 2016, \icarus, 277, 257

\bibitem[{Kaasalainen(2011)}]{2011-IPI-5-Kaasalainen}
Kaasalainen, M. 2011, Inverse Problems and Imaging, 5, 37

\bibitem[{Kaasalainen \& Torppa(2001)}]{2001Icar..153...24K}
Kaasalainen, M. \& Torppa, J. 2001, \icarus, 153, 24

\bibitem[{Kaasalainen {et~al.}(2001)Kaasalainen, Torppa, \&
  Muinonen}]{2001Icar..153...37K}
Kaasalainen, M., Torppa, J., \& Muinonen, K. 2001, \icarus, 153, 37

\bibitem[{Lenzen {et~al.}(2003)Lenzen, Hartung, Brandner, Finger, Hubin,
  Lacombe, Lagrange, Lehnert, Moorwood, \& Mouillet}]{2003-SPIE-4841-Lenzen}
Lenzen, R., Hartung, M., Brandner, W., {et~al.} 2003, SPIE, 4841, 944

\bibitem[{{Marchis} {et~al.}(2008{\natexlab{a}}){Marchis}, {Descamps}, {Baek},
  {Harris}, {Kaasalainen}, {Berthier}, {Hestroffer}, \&
  {Vachier}}]{marchis2008circular}
{Marchis}, F., {Descamps}, P., {Baek}, M., {et~al.} 2008{\natexlab{a}},
  \icarus, 196, 97

\bibitem[{{Marchis} {et~al.}(2008{\natexlab{b}}){Marchis}, {Descamps},
  {Berthier}, {Hestroffer}, {Vachier}, {Baek}, {Harris}, \&
  {Nesvorn{\'y}}}]{marchis2008eccentric}
{Marchis}, F., {Descamps}, P., {Berthier}, J., {et~al.} 2008{\natexlab{b}},
  \icarus, 195, 295

\bibitem[{Marchis {et~al.}(2012)Marchis, Enriquez, Emery, Mueller, Baek,
  Pollock, Assafin, Vieira~Martins, Berthier, Vachier, Cruikshank, Lim,
  Reichart, Ivarsen, Haislip, \& LaCluyze}]{2012-Icarus-221-Marchis}
Marchis, F., Enriquez, J.~E., Emery, J.~P., {et~al.} 2012, \icarus, 221, 1130

\bibitem[{{Marchis} {et~al.}(2021){Marchis}, {Jorda}, {Vernazza}, {Bro{\v{z}}},
  {Hanu{\v{s}}}, {Ferrais}, {Vachier}, {Rambaux}, {Marsset}, {Viikinkoski},
  {Jehin}, {Benseguane}, {Podlewska-Gaca}, {Carry}, {Drouard}, {Fauvaud},
  {Birlan}, {Berthier}, {Bartczak}, {Dumas}, {Dudzi{\'n}ski}, {{\v{D}}urech},
  {Castillo-Rogez}, {Cipriani}, {Colas}, {Fetick}, {Fusco}, {Grice},
  {Kryszczynska}, {Lamy}, {Marciniak}, {Michalowski}, {Michel}, {Pajuelo},
  {Santana-Ros}, {Tanga}, {Vigan}, {Witasse}, \& {Yang}}]{2021A&A...653A..57M}
{Marchis}, F., {Jorda}, L., {Vernazza}, P., {et~al.} 2021, \aap, 653, A57

\bibitem[{Margot {et~al.}(2015)Margot, Pravec, Taylor, Carry, \&
  Jacobson}]{2015-AsteroidsIV-Margot}
Margot, J.-L., Pravec, P., Taylor, P., Carry, B., \& Jacobson, S. 2015,
  {Asteroid Systems: Binaries, Triples, and Pairs}, ed. P.~Michel, F.~DeMeo, \&
  W.~F. Bottke (Univ. Arizona Press), 355--374

\bibitem[{{Marschall} {et~al.}(2022){Marschall}, {Nesvorn{\'y}}, {Deienno},
  {Wong}, {Levison}, \& {Bottke}}]{2022AJ....164..167M}
{Marschall}, R., {Nesvorn{\'y}}, D., {Deienno}, R., {et~al.} 2022, \aj, 164,
  167

\bibitem[{{Marsset} {et~al.}(2016{\natexlab{a}}){Marsset}, {Carry}, {Yang},
  {Marchis}, {Vernazza}, {Dumas}, {Berthier}, \& {Vachier}}]{2016camimoon}
{Marsset}, M., {Carry}, B., {Yang}, B., {et~al.} 2016{\natexlab{a}}, \iaucirc,
  9282, 1

\bibitem[{{Marsset} {et~al.}(2016{\natexlab{b}}){Marsset}, {Vernazza},
  {Birlan}, {DeMeo}, {Binzel}, {Dumas}, {Milli}, \&
  {Popescu}}]{2016A&A...586A..15M}
{Marsset}, M., {Vernazza}, P., {Birlan}, M., {et~al.} 2016{\natexlab{b}}, \aap,
  586, A15

\bibitem[{{Masiero} {et~al.}(2021){Masiero}, {Mainzer}, {Bauer}, {Cutri},
  {Grav}, {Kramer}, {Pittichov{\'a}}, \& {Wright}}]{2021PSJ.....2..162M}
{Masiero}, J.~R., {Mainzer}, A.~K., {Bauer}, J.~M., {et~al.} 2021, PSJ, 2, 162

\bibitem[{{Merline} {et~al.}(2000){Merline}, {Close}, {Dumas}, {Shelton},
  {Menard}, {Chapman}, \& {Slater}}]{merlinepulcova}
{Merline}, W.~J., {Close}, L.~M., {Dumas}, C., {et~al.} 2000, in AAS/Division
  for Planetary Sciences Meeting Abstracts, Vol.~32, AAS/Division for Planetary
  Sciences Meeting Abstracts \#32, 13.06

\bibitem[{{Merline} {et~al.}(2003){Merline}, {Dumas}, {Siegler}, {Close},
  {Chapman}, {Tamblyn}, {Terrell}, {Conrad}, {Menard}, \&
  {Duvert}}]{merlineemma}
{Merline}, W.~J., {Dumas}, C., {Siegler}, N., {et~al.} 2003, \iaucirc, 8165, 1

\bibitem[{{Merline} {et~al.}(2002){Merline}, {Weidenschilling}, {Durda},
  {Margot}, {Pravec}, \& {Storrs}}]{asteroidsdohavesatellitesmerline}
{Merline}, W.~J., {Weidenschilling}, S.~J., {Durda}, D.~D., {et~al.} 2002, in
  Asteroids III, ed. J.~{Bottke}, W.~F., A.~{Cellino}, P.~{Paolicchi}, \& R.~P.
  {Binzel}, 289--312

\bibitem[{{Michel} {et~al.}(2015){Michel}, {Jutzi}, {Richardson}, {Goodrich},
  {Hartmann}, \& {O`Brien}}]{michel2015reaccumulation}
{Michel}, P., {Jutzi}, M., {Richardson}, D.~C., {et~al.} 2015, \planss, 107, 24

\bibitem[{Milani {et~al.}(2014)Milani, Cellino, Kne{\v z}evi{\'c},
  Novakovi{\'c}, Spoto, \& Paolicchi}]{2014Icar..239...46M}
Milani, A., Cellino, A., Kne{\v z}evi{\'c}, Z., {et~al.} 2014, \icarus, 239, 46

\bibitem[{{Milani} {et~al.}(2019){Milani}, {Kne{\v{z}}evi{\'c}}, {Spoto}, \&
  {Paolicchi}}]{2019A&A...622A..47M}
{Milani}, A., {Kne{\v{z}}evi{\'c}}, Z., {Spoto}, F., \& {Paolicchi}, P. 2019,
  \aap, 622, A47

\bibitem[{{Moth{\'e}-Diniz} {et~al.}(2005){Moth{\'e}-Diniz}, {Roig}, \&
  {Carvano}}]{minervafamily}
{Moth{\'e}-Diniz}, T., {Roig}, F., \& {Carvano}, J.~M. 2005, \icarus, 174, 54

\bibitem[{{Nesvorny}(2015)}]{2015PDSS..234.....N}
{Nesvorny}, D. 2015, {Nesvorny HCM Asteroid Families V3.0}, NASA Planetary Data
  System, id. EAR-A-VARGBDET-5-NESVORNYFAM-V3.0

\bibitem[{Nesvorn{\'y} {et~al.}(2018)Nesvorn{\'y}, Vokrouhlick{\'y}, Bottke, \&
  Levison}]{2018NatAs...2..878N}
Nesvorn{\'y}, D., Vokrouhlick{\'y}, D., Bottke, W.~F., \& Levison, H.~F. 2018,
  Nature Astronomy, 2, 878

\bibitem[{{Nesvorn{\'y}} {et~al.}(2020){Nesvorn{\'y}}, {Vokrouhlick{\'y}},
  {Bottke}, {Levison}, \& {Grundy}}]{nesvorny2020tidalevolution}
{Nesvorn{\'y}}, D., {Vokrouhlick{\'y}}, D., {Bottke}, W.~F., {Levison}, H.~F.,
  \& {Grundy}, W.~M. 2020, \apjl, 893, L16

\bibitem[{Pajuelo {et~al.}(2018)Pajuelo, Carry, Vachier, Marsset, Berthier,
  Descamps, Merline, Tamblyn, Grice, Conrad, Storrs, Timerson, Dunham, Preston,
  Vigan, Yang, Vernazza, Fauvaud, Bernasconi, Romeuf, Behrend, Dumas, Drummond,
  Margot, Kervella, Marchis, \& Girard}]{2018Icar..309..134P}
Pajuelo, M., Carry, B., Vachier, F., {et~al.} 2018, \icarus, 309, 134

\bibitem[{Park {et~al.}(2021)Park, Folkner, Williams, \& Boggs}]{park2021jpl}
Park, R.~S., Folkner, W.~M., Williams, J.~G., \& Boggs, D.~H. 2021, The
  Astronomical Journal, 161, 105

\bibitem[{{Pavela} {et~al.}(2021){Pavela}, {Novakovi{\'c}}, {Carruba}, \&
  {Radovi{\'c}}}]{2021MNRAS.501..356P}
{Pavela}, D., {Novakovi{\'c}}, B., {Carruba}, V., \& {Radovi{\'c}}, V. 2021,
  \mnras, 501, 356

\bibitem[{{Ragozzine} \& {Brown}(2007)}]{2007AJ....134.2160R}
{Ragozzine}, D. \& {Brown}, M.~E. 2007, \aj, 134, 2160

\bibitem[{{Rivkin} {et~al.}(2021){Rivkin}, {Chabot}, {Stickle}, {Thomas},
  {Richardson}, {Barnouin}, {Fahnestock}, {Ernst}, {Cheng}, {Chesley}, {Naidu},
  {Statler}, {Barbee}, {Agrusa}, {Moskovitz}, {Terik Daly}, {Pravec},
  {Scheirich}, {Dotto}, {Della Corte}, {Michel}, {K{\"u}ppers}, {Atchison}, \&
  {Hirabayashi}}]{2021PSJ.....2..173R}
{Rivkin}, A.~S., {Chabot}, N.~L., {Stickle}, A.~M., {et~al.} 2021, PSJ, 2, 173

\bibitem[{Rousset {et~al.}(2003)Rousset, Lacombe, Puget, Hubin, Gendron, Fusco,
  Arsenault, Charton, Feautrier, Gigan, Kern, Lagrange, Madec, Mouillet,
  Rabaud, Rabou, Stadler, \& Zins}]{2003-SPIE-4839-Rousset}
Rousset, G., Lacombe, F., Puget, P., {et~al.} 2003, SPIE, 4839, 140

\bibitem[{{Ro{\.z}ek} {et~al.}(2011){Ro{\.z}ek}, {Breiter}, \&
  {Jopek}}]{2011MNRAS.412..987R}
{Ro{\.z}ek}, A., {Breiter}, S., \& {Jopek}, T.~J. 2011, \mnras, 412, 987

\bibitem[{Scheeres {et~al.}(2015)Scheeres, Britt, Carry, \&
  Holsapple}]{2015-AsteroidsIV-Scheeres}
Scheeres, D.~J., Britt, D., Carry, B., \& Holsapple, K.~A. 2015, {Asteroid
  Interiors and Morphology}, ed. P.~Michel, F.~DeMeo, \& W.~F. Bottke (Univ.
  Arizona Press), 745--766

\bibitem[{Taylor(2005)}]{2005ASPC..347...29T}
Taylor, M.~B. 2005, in Astronomical Society of the Pacific Conference Series,
  Vol. 347, Astronomical Data Analysis Software and Systems XIV, ed.
  P.~{Shopbell}, M.~{Britton}, \& R.~{Ebert}, 29

\bibitem[{{Thomas} {et~al.}(2007){Thomas}, {Armstrong}, {Asmar}, {Burns},
  {Denk}, {Giese}, {Helfenstein}, {Iess}, {Johnson}, {McEwen}, {Nicolaisen},
  {Porco}, {Rappaport}, {Richardson}, {Somenzi}, {Tortora}, {Turtle}, \&
  {Veverka}}]{hyperiondensity}
{Thomas}, P.~C., {Armstrong}, J.~W., {Asmar}, S.~W., {et~al.} 2007, \nat, 448,
  50

\bibitem[{{Thomas} {et~al.}(1996){Thomas}, {Belton}, {Carcich}, {Chapman},
  {Davies}, {Sullivan}, \& {Veverka}}]{1996idashape}
{Thomas}, P.~C., {Belton}, M.~J.~S., {Carcich}, B., {et~al.} 1996, \icarus,
  120, 20

\bibitem[{{Tsirvoulis}(2019)}]{2019MNRAS.482.2612T}
{Tsirvoulis}, G. 2019, \mnras, 482, 2612

\bibitem[{Vachier {et~al.}(2012)Vachier, Berthier, \&
  Marchis}]{2012AA...543A..68V}
Vachier, F., Berthier, J., \& Marchis, F. 2012, \aap, 543, A68

\bibitem[{{Vachier} {et~al.}(2022){Vachier}, {Carry}, \&
  {Berthier}}]{2022Icar..38215013V}
{Vachier}, F., {Carry}, B., \& {Berthier}, J. 2022, \icarus, 382, 115013

\bibitem[{van Dam {et~al.}(2004)van Dam, Le~Mignant, \&
  Macintosh}]{2004-AppOpt-43-vanDam}
van Dam, M.~A., Le~Mignant, D., \& Macintosh, B. 2004, Applied Optics, 43, 5458

\bibitem[{{Vereshchagina}(2011)}]{pulcovabinariespulkovo}
{Vereshchagina}, I.~A. 2011, arXiv e-prints, arXiv:1102.0152

\bibitem[{{Vernazza} {et~al.}(2021){Vernazza}, {Ferrais}, {Jorda},
  {Hanu{\v{s}}}, {Carry}, {Marsset}, {Bro{\v{z}}}, {Fetick}, {Viikinkoski},
  {Marchis}, {Vachier}, {Drouard}, {Fusco}, {Birlan}, {Podlewska-Gaca},
  {Rambaux}, {Neveu}, {Bartczak}, {Dudzi{\'n}ski}, {Jehin}, {Beck}, {Berthier},
  {Castillo-Rogez}, {Cipriani}, {Colas}, {Dumas}, {{\v{D}}urech}, {Grice},
  {Kaasalainen}, {Kryszczynska}, {Lamy}, {Le Coroller}, {Marciniak},
  {Michalowski}, {Michel}, {Santana-Ros}, {Tanga}, {Vigan}, {Witasse}, {Yang},
  {Antonini}, {Audejean}, {Aurard}, {Behrend}, {Benkhaldoun}, {Bosch},
  {Chapman}, {Dalmon}, {Fauvaud}, {Hamanowa}, {Hamanowa}, {His}, {Jones},
  {Kim}, {Kim}, {Krajewski}, {Labrevoir}, {Leroy}, {Livet}, {Molina},
  {Montaigut}, {Oey}, {Payre}, {Reddy}, {Sabin}, {Sanchez}, \&
  {Socha}}]{2021A&A...654A..56V}
{Vernazza}, P., {Ferrais}, M., {Jorda}, L., {et~al.} 2021, \aap, 654, A56

\bibitem[{{Vernazza} {et~al.}(2020){Vernazza}, {Jorda}, {{\v{S}}eve{\v{c}}ek},
  {Bro{\v{z}}}, {Viikinkoski}, {Hanu{\v{s}}}, {Carry}, {Drouard}, {Ferrais},
  {Marsset}, {Marchis}, {Birlan}, {Podlewska-Gaca}, {Jehin}, {Bartczak},
  {Dudzinski}, {Berthier}, {Castillo-Rogez}, {Cipriani}, {Colas}, {DeMeo},
  {Dumas}, {Durech}, {Fetick}, {Fusco}, {Grice}, {Kaasalainen}, {Kryszczynska},
  {Lamy}, {Le Coroller}, {Marciniak}, {Michalowski}, {Michel}, {Rambaux},
  {Santana-Ros}, {Tanga}, {Vachier}, {Vigan}, {Witasse}, {Yang}, {Gillon},
  {Benkhaldoun}, {Szakats}, {Hirsch}, {Duffard}, {Chapman}, \&
  {Maestre}}]{2020NatAs...4..136V}
{Vernazza}, P., {Jorda}, L., {{\v{S}}eve{\v{c}}ek}, P., {et~al.} 2020, Nature
  Astronomy, 4, 136

\bibitem[{Viikinkoski {et~al.}(2017)Viikinkoski, Hanu{\v{s}}, Kaasalainen,
  Marchis, \& {\v{D}}urech}]{2017A&A...607A.117V}
Viikinkoski, M., Hanu{\v{s}}, J., Kaasalainen, M., Marchis, F., \&
  {\v{D}}urech, J. 2017, \aap, 607, A117

\bibitem[{Viikinkoski {et~al.}(2015)Viikinkoski, Kaasalainen, \&
  Durech}]{2015A&A...576A...8V}
Viikinkoski, M., Kaasalainen, M., \& Durech, J. 2015, \aap, 576, A8

\bibitem[{{Vinogradova}(2019)}]{2019MNRAS.484.3755V}
{Vinogradova}, T.~A. 2019, \mnras, 484, 3755

\bibitem[{{Vokrouhlick{\'y}} {et~al.}(2010{\natexlab{a}}){Vokrouhlick{\'y}},
  {Nesvorn{\'y}}, {Bottke}, \& {Morbidelli}}]{SylviafamVokrouhlicky2010}
{Vokrouhlick{\'y}}, D., {Nesvorn{\'y}}, D., {Bottke}, W.~F., \& {Morbidelli},
  A. 2010{\natexlab{a}}, \aj, 139, 2148

\bibitem[{{Vokrouhlick{\'y}} {et~al.}(2010{\natexlab{b}}){Vokrouhlick{\'y}},
  {Nesvorn{\'y}}, {Bottke}, \& {Morbidelli}}]{2010AJ....139.2148V}
{Vokrouhlick{\'y}}, D., {Nesvorn{\'y}}, D., {Bottke}, W.~F., \& {Morbidelli},
  A. 2010{\natexlab{b}}, \aj, 139, 2148

\bibitem[{{Vokrouhlick{\'y}} {et~al.}(2021){Vokrouhlick{\'y}}, {Novakovi{\'c}},
  \& {Nesvorn{\'y}}}]{2021A&A...649A.115V}
{Vokrouhlick{\'y}}, D., {Novakovi{\'c}}, B., \& {Nesvorn{\'y}}, D. 2021, \aap,
  649, A115

\bibitem[{{Walsh} {et~al.}(2025){Walsh}, {Ballouz}, {Agrusa}, {Hanus}, {Jutzi},
  \& {Michel}}]{2025arXiv250503325W}
{Walsh}, K.~J., {Ballouz}, R.-L., {Agrusa}, H.~F., {et~al.} 2025, arXiv
  e-prints, arXiv:2505.03325

\bibitem[{Walsh \& Jacobson(2015)}]{2015-AsteroidsIV-Walsh}
Walsh, K.~J. \& Jacobson, S.~A. 2015, {Formation and Evolution of Binary
  Asteroids}, ed. P.~Michel, F.~DeMeo, \& W.~F. Bottke, 375--393

\bibitem[{{Warner} {et~al.}(2011){Warner}, {Stephens}, \& {Harris}}]{alcdef}
{Warner}, B.~D., {Stephens}, R.~D., \& {Harris}, A.~W. 2011, Minor Planet
  Bulletin, 38, 172

\bibitem[{{Wieczorek} \& {Meschede}(2018)}]{shtools}
{Wieczorek}, M.~A. \& {Meschede}, M. 2018, Geochemistry, Geophysics,
  Geosystems, 19, 2574

\bibitem[{{Yang} {et~al.}(2020{\natexlab{a}}){Yang}, {Hanu{\v{s}}},
  {Bro{\v{z}}}, {Chrenko}, {Willman}, {{\v{S}}eve{\v{c}}ek}, {Masiero}, \&
  {Kaluna}}]{2020A&A...643A..38Y}
{Yang}, B., {Hanu{\v{s}}}, J., {Bro{\v{z}}}, M., {et~al.} 2020{\natexlab{a}},
  \aap, 643, A38

\bibitem[{{Yang} {et~al.}(2020{\natexlab{b}}){Yang}, {Hanu{\v{s}}}, {Carry},
  {Vernazza}, {Bro{\v{z}}}, {Vachier}, {Rambaux}, {Marsset}, {Chrenko},
  {{\v{S}}eve{\v{c}}ek}, {Viikinkoski}, {Jehin}, {Ferrais}, {Podlewska-Gaca},
  {Drouard}, {Marchis}, {Birlan}, {Benkhaldoun}, {Berthier}, {Bartczak},
  {Dumas}, {Dudzi{\'n}ski}, {{\v{D}}urech}, {Castillo-Rogez}, {Cipriani},
  {Colas}, {Fetick}, {Fusco}, {Grice}, {Jorda}, {Kaasalainen}, {Kryszczynska},
  {Lamy}, {Marciniak}, {Michalowski}, {Michel}, {Pajuelo}, {Santana-Ros},
  {Tanga}, {Vigan}, \& {Witasse}}]{2020A&A...641A..80Y}
{Yang}, B., {Hanu{\v{s}}}, J., {Carry}, B., {et~al.} 2020{\natexlab{b}}, \aap,
  641, A80

\bibitem[{{Yang} {et~al.}(2016){Yang}, {Wahhaj}, {Beauvalet}, {Marchis},
  {Dumas}, {Marsset}, {Nielsen}, \& {Vachier}}]{2016yangelektraminerva}
{Yang}, B., {Wahhaj}, Z., {Beauvalet}, L., {et~al.} 2016, \apjl, 820, L35

\end{thebibliography}

\appendix

\section{\KM{Supplementary analysis of shape models}}
\label{app:shapes}

\subsection{\KM{(762) Pulcova}}

\begin{figure*}[h!]
    \centering
    \includegraphics[width=1\textwidth]{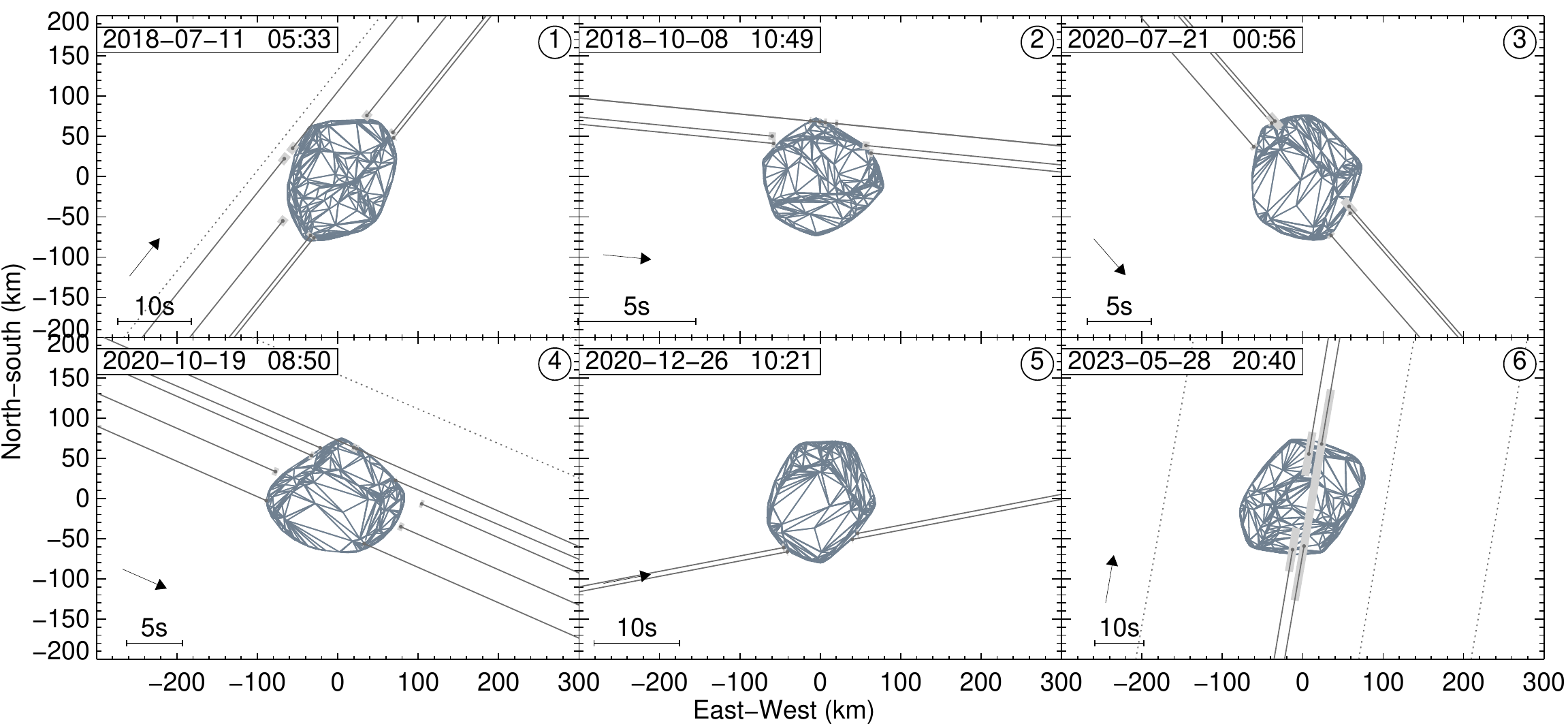}
   \caption{Shape model of (762) Pulcova developed by lightcurve inversion compared to stellar occulations from the literature. Occultations are sourced from \cite{heraldPDSocculatation}.}
   \label{fig:occ_pulc}
\end{figure*}

\KM{This section contains supplementary material regarding the shape model of asteroid (762) Pulcova. A topographic map of our shape model for (762) Pulcova can be found in \Cref{fig:shape_pulc}, and a comparison of this model with the stellar occultations used to scale the model can be found in \Cref{fig:occ_pulc}.}

\subsection{\KM{(283) Emma}}
\subsubsection{Shape}

\begin{figure}[h!]
   \centering
   \includegraphics[width=\columnwidth]{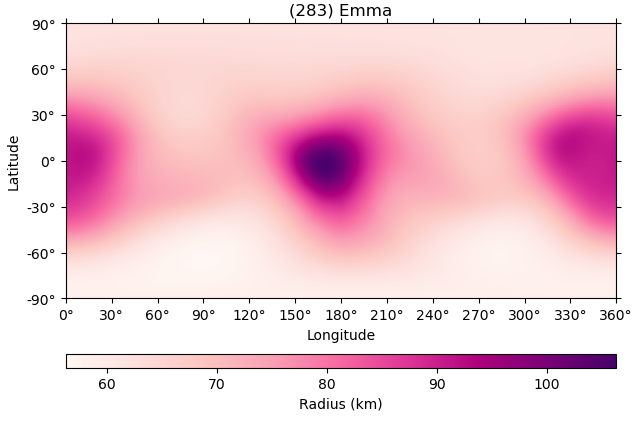}
   \includegraphics[width=\columnwidth]{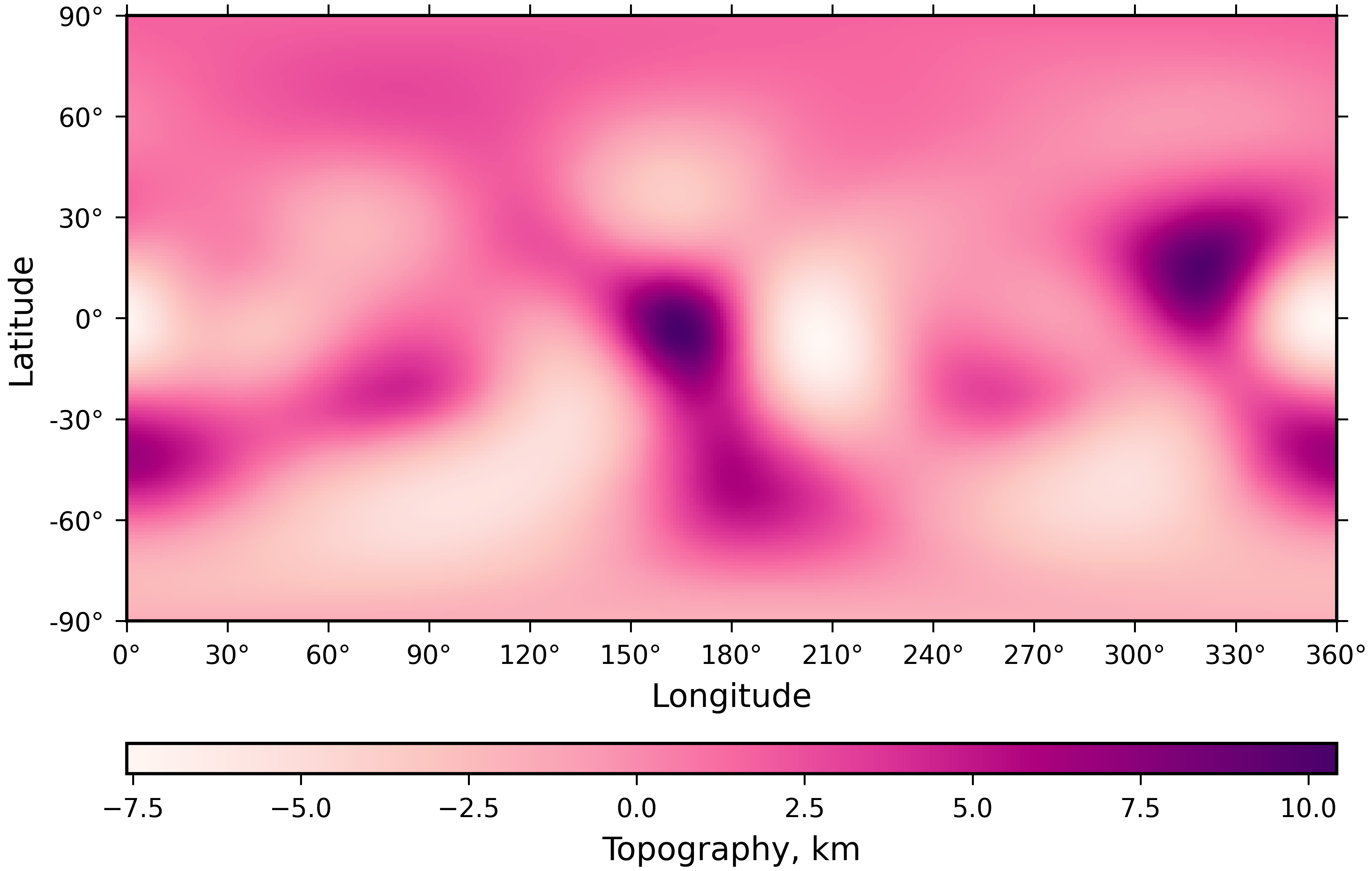}
   \caption{Topographic maps of asteroid (283) Emma, before (upper panel) and after (lower panel) the subtraction of a tri-axial ellipsoid of the same dimensions. Clear elongation can be observed in the equatorial plane. }

   \label{fig:shape_emma}
\end{figure}

To compare the orbital gravitational coefficients to the shape model, we decompose the
gravitational field of Emma, assuming a homogeneous interior, using
SHTOOLS\footnote{\url{https://shtools.github.io/SHTOOLS/}} \citep{shtools} to process the decomposition of topographical maps of the asteroid into spherical harmonics coefficients. The maps were determined from the shape models of the asteroid, as illustrated in \Cref{fig:shape_emma}.

The spherical harmonics coefficients $C_{\ell m}$ associated with the gravitational field were determined as described in \Cref{eq:SHcoeff}

\begin{equation}
U(r)=\frac{GM}{r} \sum_{\ell=0}^{\infty} \sum_{m=-\ell}^{\ell} \left(\frac{R_0}{r}\right)^\ell C_{\ell m} Y_{\ell m}(\theta, \phi)
\label{eq:SHcoeff}
\end{equation}

 and $Y_{\ell\textit{m}}(\theta, \phi)$ are the spherical harmonics, defined as
\begin{equation}
    Y_{\ell\textit{m}}(\theta, \phi) =
    \begin{cases}
        \bar{P}_{\ell m}(\cos{\theta}) \cos{m \phi}, & \text{if } m\geq 0\\
        \bar{P}_{\ell |m|}(\cos{\theta}) \sin{|m|\phi}, & \text{if } m < 0
    \end{cases}
\end{equation}
where $\bar{P}_{\ell m}$ are the normalized associated Legendre polynomials.
Because unnormalized coefficients are used

\begin{equation}
    \bar{P}_{\ell m}=P_{\ell m}
\end{equation}

in the case where $m$ is negative, $S_{\ell m}$ are defined as
\begin{equation}
S_{\ell -m}= C_{\ell m}
\end{equation}

For example, $S_{22}=C_{2 -2}$. We use unnormalized coefficients for both the orbital solution and field determined from the shape model (i.e., as opposed to $4\pi$ or orthonormal normalizations frequently used in other fields of planetary science).

Although higher order coefficients (e.g., up to $\ell>10$) can be estimated from the shape model, only quadrupole coefficients (at most) can realistically be determined from the gravitational field with the present dataset, and as such only the low order terms realistically need to be considered. Falsely assuming static values for components of the gravitational field that do not match physical values may have adverse effects on the convergence of the orbit, and may bias results even more than neglecting the higher-order terms entirely.

Similarly, if the spin-pole of the primary body is incorrectly determined, even by a very small amount, it may have adverse effects on determining the orbital solution.
In the first case, this is likely to lead to false determinations of the orbital geometry, overly circularizing the orbital path in order to minimize the visibility of the expected precession of the orbit.
This also applies in the second case, and also higher order gravitational terms will trend to zero, giving the impression of a falsely-Keplerian orbital fit that is not representative of the true gravitational forces acting upon the satellite. Correctly identifying the primary spin-pole is non-trivial, as uncertainties on this value from shape models are much larger than necessary to adversely affect the orbit.

\begin{table}
    \centering
    \caption{Size estimates derived from different shape models of (283) Emma.}
    \begin{tabular}{ll}
    \hline\hline
    Model     & Diameter (km) \\
    \hline
     Literature    & 142$\pm$14 \\
     High rugosity    & $\approx$125 \\
     Low rugosity & 133$\pm$3\\
      \hline
    \end{tabular}
    \label{tab:emma_diameters}
\end{table}

\begin{figure}[h!]
    \centering
    \includegraphics[width=0.7\linewidth]{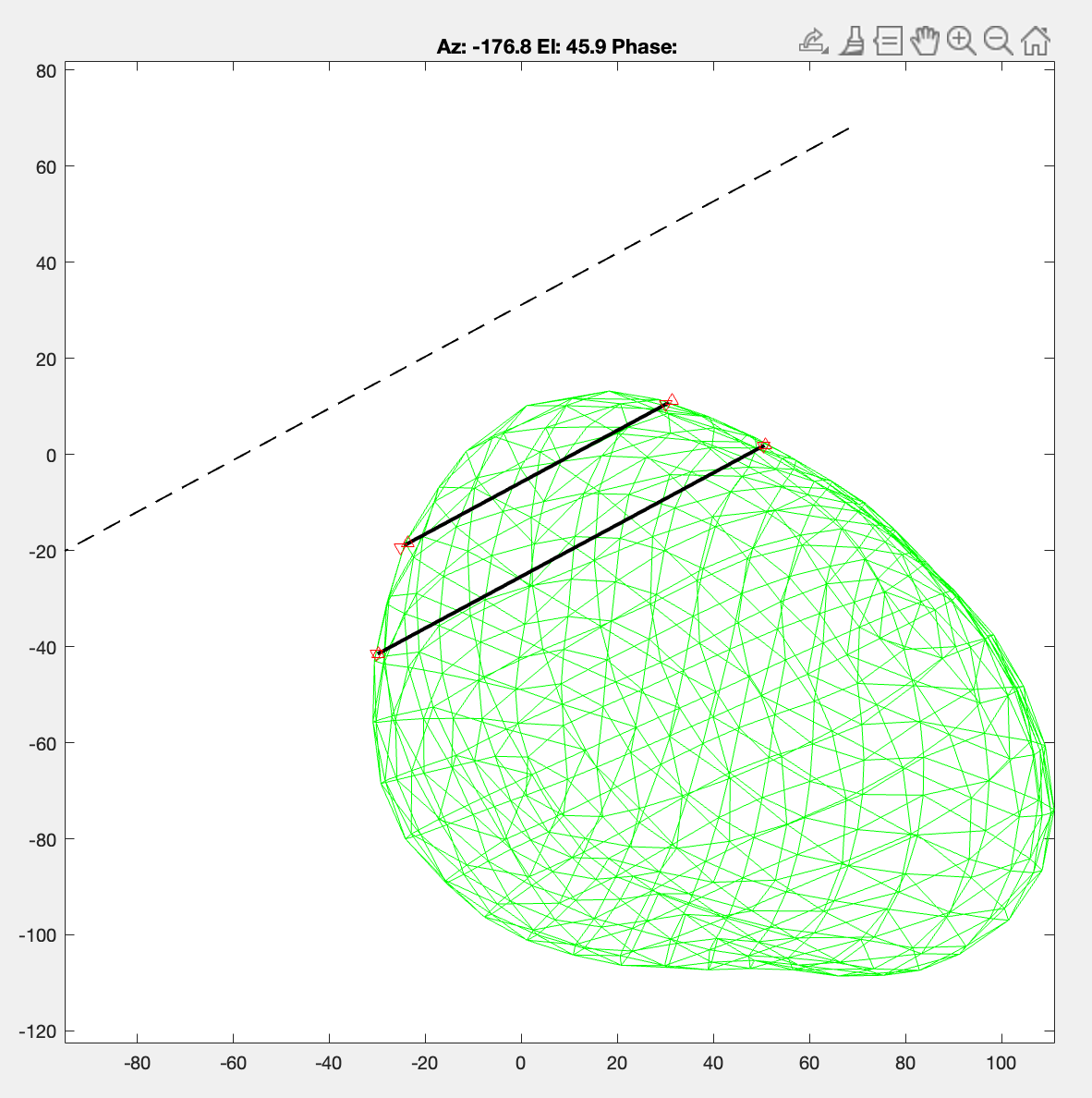}
    \includegraphics[width=0.7\linewidth]{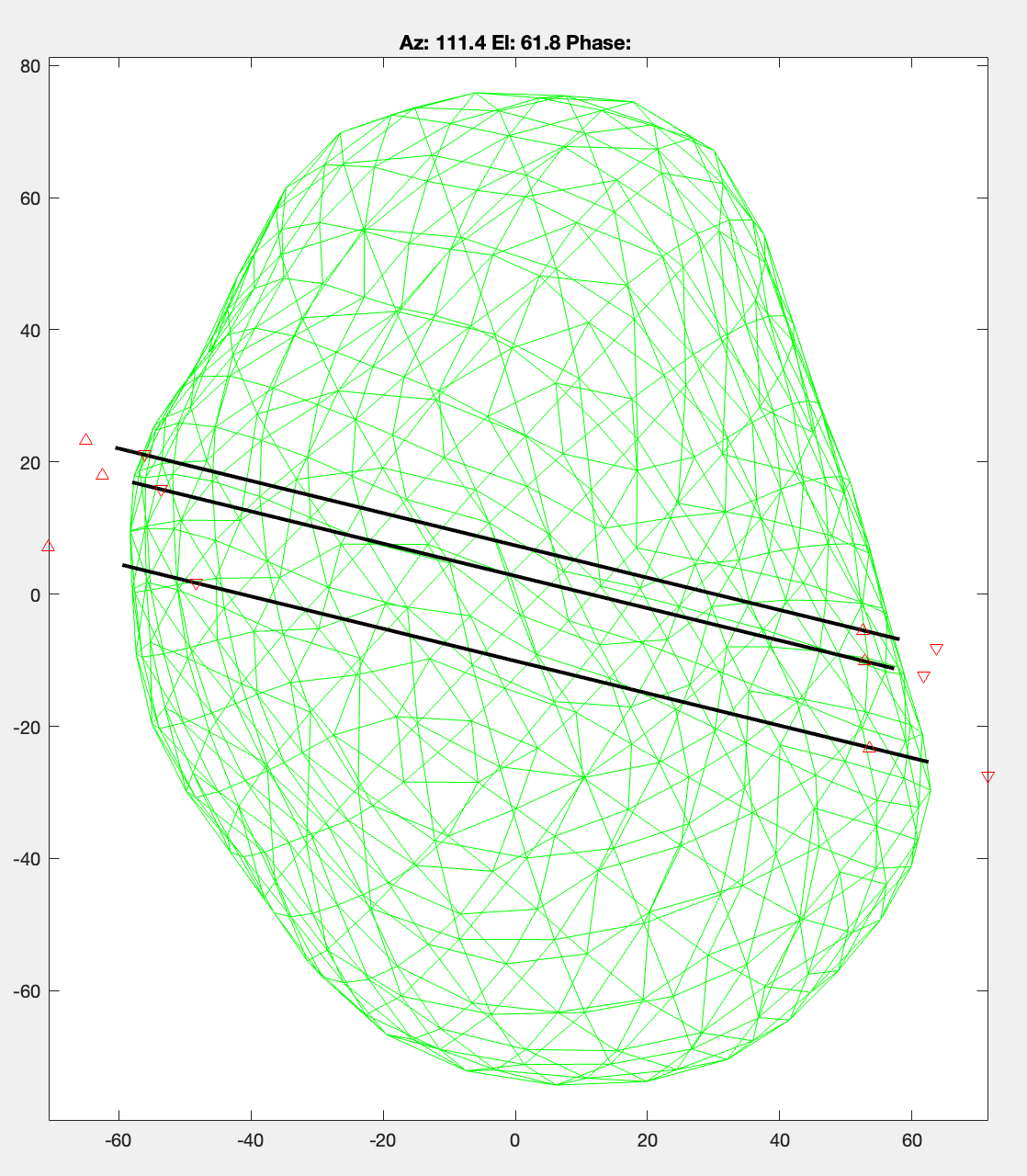}
    \caption{Match of Emma's shape model to occultation data. The model is a good fit, but the limited chords leave some ambiguity in the scaling of the model. Black lines represent a detected chord, orange triangles represented the uncertainties on the start and stop time of the occultation.}
    \label{fig:emma_occ}
\end{figure}

Noting the limitations of the \cite{2017A&A...607A.117V, 2010A&A...513A..46D} model, we constructed a new shape model for Emma based on new and archival lightcurves, occultations, and AO images. We found limited utility from the AO images, and as such the sizing of the model was constrained primarily by the occultations. Unfortunately, the available occultation data is limited, with only two usable events and a low number of chords per event (both events match well to the model but do not constrain it effectively, see \Cref{fig:emma_occ}. As such, the size of the model is not particularly well constrained. When varying the rugosity of the model, we found solutions with sizes between $D_p\approx125$\,km and $D_p=133$\,km. Comparing with independently calculated diameter estimates \citep[e.g., thermal physical modeling,][]{2012-Icarus-221-Marchis}, we adopt the value of $D_p=133\pm3$\,km, noting that the uncertainties are formal, and the nature of the solution leaves some ambiguity. These size estimates are summarized in \Cref{tab:emma_diameters}.

This model is primarily constrained by photometric lightcurves, and as such is subject to some biases. In particular, lightcurve inversion models fail to accurately constrain the oblateness (that is, the extension of the model in the direction parallel to the spin-pole). Unfortunately, this can interfere significantly with attempts to analyze the internal structure of an object, as the lowest-order term ($J_2$) is directly related to the object's oblateness. Although topographical features of this model and the equatorial elongation are likely accurate, the current best-fit solution presents an oblateness which is incompatible with our orbital solution, as discussed in \Cref{sec:physicalprops}. This could not be resolved by internal structure modeling as very large under-dense regions are required to reconcile the physical and dynamical $J_2$ values. As such, we return to the \cite{2017A&A...607A.117V} model. Future disk-resolved observations or stellar occultations could appropriately constrain the size and shape of this object. Opportunities for disk-resolved imaging are limited, as it is rarely possible to resolve the primary in near-infrared AO images with existing instruments, and Emma is rarely bright enough to be targeted with most visible wavelength AO cameras. As such, we propose that stellar occultations are essential to future characterization of Emma's shape. In \Cref{app:occ} we present occultation predictions for Emma over the next several years, restricted to events with a high probability of success. We also include predictions for Pulcova, whose shape is subject to similar uncertainties.

\subsubsection{Internal structure}

 \begin{figure}[ht]
    \centering
    \includegraphics[width=\columnwidth]{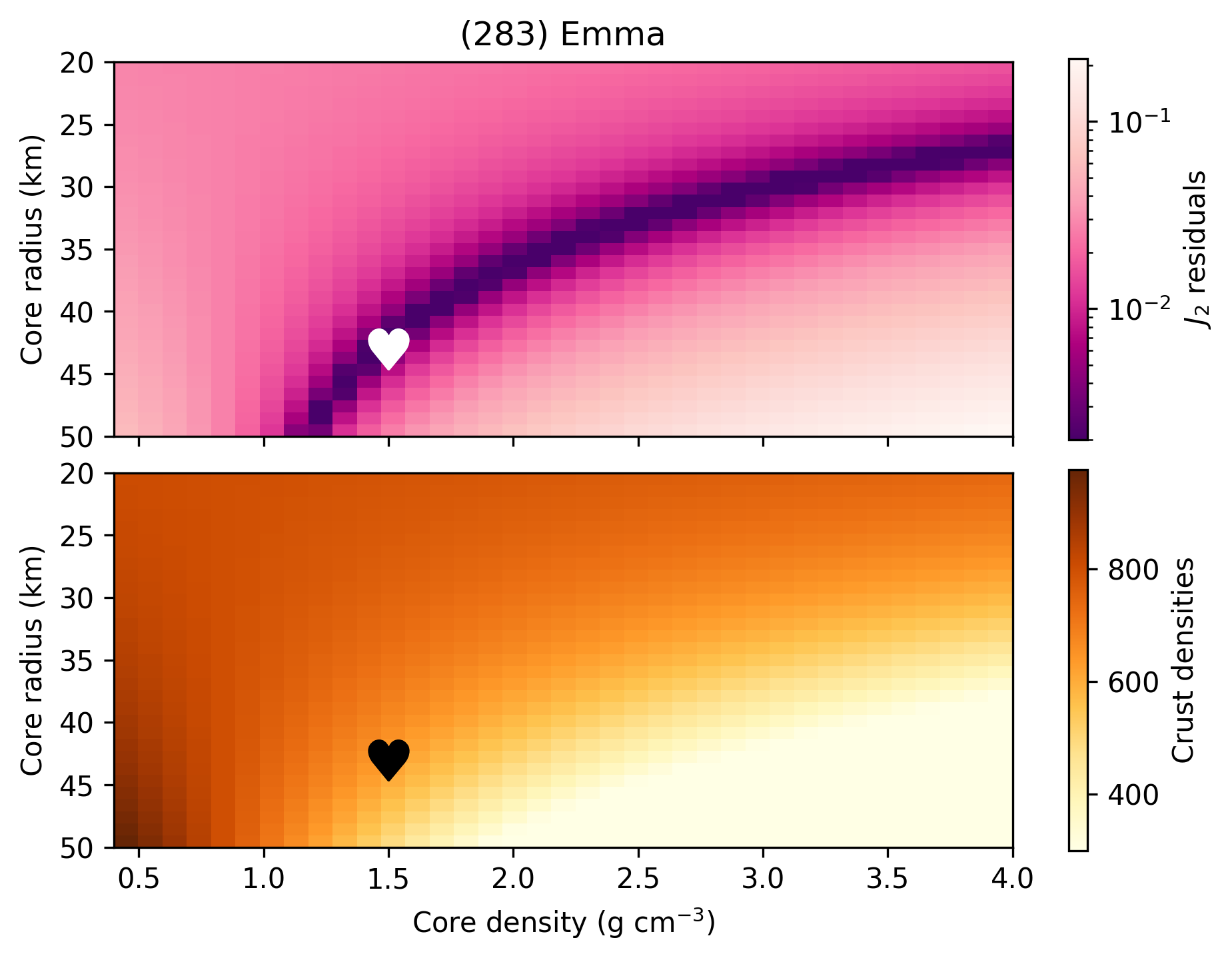}
    \caption{Potential internal structure solutions for (283) Emma, assuming a spherical core. A range of potential core densities 0.5-4~g~cm$^{-3}$ has been proposed, in line with the range of densities of observed rocky bodies in the Solar System. Upper panel: The colorbar represents the residuals between the observed and modeled $J_2$ values, with a darker color corresponding to a better fit. Lower panel: Corresponding crust densities for different cores. Not all solutions are physical. Scenarios with no core and smaller cores than pictured were also considered, but the residuals were unrealistically high. \KM{A proposed reasonable solution corresponding to a core of similar density to known C- and P-type asteroids (1.5 g~cm$^{-3}$) is marked with a heart.}
     }
    \label{fig:emma_j2}
\end{figure}

Since the total mass of Emma must be conserved within the model, it is necessary to keep the density of the core low enough to maintain a plausible density for the outer layer.
Reasonable models have core densities ranging from 1.2 and 4 g~cm$^{-3}$ for spherical cores ranging
between 56 and 100\,km in diameter (see \Cref{fig:emma_j2}).
The corresponding crust densities are approximately 0.6~g~cm$^{3}$.
This allows for an interior (core) layer consistent with the densities of other known C/P type asteroids \citep{2012P&SS...73...98C},
 surrounded by a very low density outer shell. Although these models use spherical cores, similar results can be achieved with non-spherical differentiated layers.

\begin{figure}[h!]
    \centering
    \includegraphics[width=1\columnwidth]{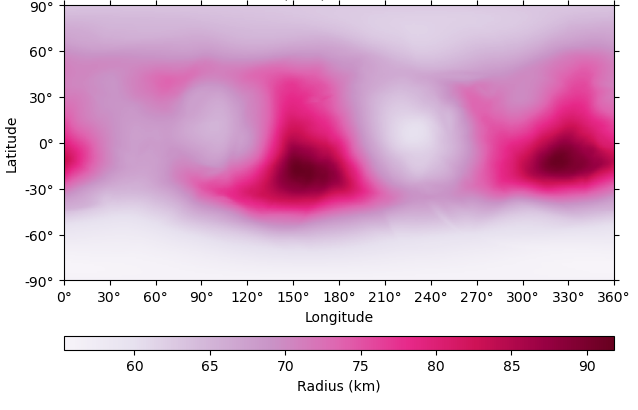}
   \caption{Topographic map of Pulcova.
   }
   \label{fig:shape_pulc}
\end{figure}

\onecolumn
\section{Photometric lightcurves \label{app:lc}}

  \begin{longtable}{crrrr}
    \caption[Photometric lightcurves of (762) Pulcova.]{
      Date, duration ($\mathcal{L}$, in hours), number of points ($\mathcal{N}_p$), phase angle ($\alpha$),
      filter, residual (against the shape model),
      for each
      lightcurve. \label{tab:lc}
    }\\

    \hline\hline
    Date & \multicolumn{1}{c}{$\mathcal{L}$} & \multicolumn{1}{c}{$\mathcal{N}_p$} &
    \multicolumn{1}{c}{$\alpha$} & \multicolumn{1}{c}{RMS} \\
    & \multicolumn{1}{c}{(h)} && \multicolumn{1}{c}{(\degr)}& \multicolumn{1}{c}{(mag)} \\
    \hline
    \endfirsthead

    \multicolumn{4}{c}{{\tablename\ \thetable{} -- continued from previous page}} \\
    \hline\hline
    Date & \multicolumn{1}{c}{$\mathcal{L}$} & \multicolumn{1}{c}{$\mathcal{N}_p$} &
    \multicolumn{1}{c}{$\alpha$} &  \multicolumn{1}{c}{RMS} \\
    & \multicolumn{1}{c}{(h)} && \multicolumn{1}{c}{(\degr)}& \multicolumn{1}{c}{(mag)} \\
    \hline
    \endhead

    \hline \multicolumn{4}{r}{{Continued on next page}} \\ \hline
    \endfoot

    \hline
    \endlastfoot

    2006-02-17 &  1.9 &  52 &  14.5  &  0.010 \\
    2006-03-09 &  3.1 &  88 &   9.3  &  0.026 \\
    2006-03-12 &  2.7 &  71 &   8.5  &  0.013 \\
    2006-03-26 &  2.1 &  60 &   5.9  &  0.026 \\
    2006-03-27 &  4.4 & 112 &   5.8  &  0.028 \\
    2008-08-30 &  3.5 &  50 &   3.5  &  0.011 \\
    2008-10-08 &  4.5 &  69 &  13.3  &  0.031 \\
    2008-10-15 &  3.3 &  37 &  14.5  &  0.028 \\
    2008-11-02 &  0.6 &   7 &  16.3  &  0.004 \\
     2008-11-05 &  3.3 &  62 &  16.5  &  0.019 \\
     2009-11-29 &  1.9 & 152 &  10.9  &  0.018 \\
     2009-12-12 &  5.0 & 378 &  13.9  &  0.028 \\
     2009-12-17 &  3.0 & 224 &  14.9  &  0.033 \\
     2020-07-19 &  1.7 &  68 &  17.8 &  0.011 \\
     2020-07-26 &  1.6 &  63 &  17.7 &  0.010 \\
     2020-07-29 &  1.8 &  57 &  17.6 &  0.018 \\
     2020-07-31 &  2.8 &  26 &  17.5 &  0.010 \\
     2020-07-31 &  1.4 &  24 &  17.5 &  0.005 \\
     2020-08-05 &  3.1 &  46 &  17.3 &  0.010 \\
     2020-08-06 &  3.3 &  53 &  17.2 &  0.011 \\
     2020-08-08 &  3.2 & 128 &  17.0 &  0.020 \\
     2020-08-12 &  2.7 &  94 &  16.6 &  0.016 \\
     2020-08-17 &  3.6 & 135 &  16.1 &  0.009 \\
     2020-08-18 &  2.9 & 102 &  16.0 &  0.013 \\
     2020-08-19 &  4.6 & 171 &  15.9 &  0.010 \\
     2020-08-20 &  4.0 & 140 &  15.7 &  0.013 \\
     2020-08-24 &  5.5 & 207 &  15.2 &  0.013 \\
     2020-08-25 &  1.5 &  52 &  15.0 &  0.021 \\
     2020-08-31 &  1.5 &  59 &  14.0 &  0.014 \\
     2020-09-01 &  6.2 & 267 &  13.8 &  0.015 \\
     2020-09-02 &  5.8 & 265 &  13.6 &  0.020 \\
     2020-09-03 &  4.9 & 256 &  13.4 &  0.025 \\
     2020-09-12 &  5.6 & 245 &  11.6 &  0.016 \\
     2020-09-13 &  1.7 &  77 &  11.3 &  0.008 \\
     2020-09-14 &  4.2 & 162 &  11.1 &  0.006 \\
     2020-09-19 &  4.3 & 183 &  10.0 &  0.020 \\
     2020-09-20 &  2.9 & 119 &   9.7 &  0.010 \\
     2020-09-25 &  2.5 & 104 &   8.5 &  0.018 \\
     2020-09-26 &  3.3 & 132 &   8.3 &  0.012 \\
     2020-09-28 &  2.8 & 104 &   7.8 &  0.008 \\
     2023-02-16 &  2.0 &  99 &  17.8 &  0.022 \\
     2023-02-17 &  5.9 & 272 &  17.6 &  0.020 \\
     2023-04-21 &  7.8 & 145 &   6.6 &  0.022 \\
     2023-04-22 &  8.7 & 247 &   6.7 &  0.023 \\
     2023-04-23 &  1.4 &  47 &   6.8 &  0.008 \\
\hline
  \end{longtable}


\section{Astrometric and photometric measurements of Pulcamoon and Emmoon}

Here we report the astrometry and photometry measured for Pulcamoon (\Cref{tab:pulcamoon}) and Emmoon (\Cref{tab:emmoon}). Positions are relative to the photocenter of the primary. Since Pulcova and Emma are rarely resolved in these observations, the photocenter can be considered to be equivalent to the systems' barycenter with negligible uncertainties.

\begin{center}
  \begin{longtable}{cclllrrrrrrr}
  \caption[Astrometry of Pulcova]{Astrometry of Pulcova's satellite S/2000 (762) 1.
    \label{tab:pulcamoon}
  }\\

    \hline\hline
     Date & UTC & Tel. & Cam. & Filter &
     \multicolumn{1}{c}{$X_o$} &
     \multicolumn{1}{c}{$Y_o$} &
     \multicolumn{1}{c}{$X_{o-c}$} &
     \multicolumn{1}{c}{$Y_{o-c}$} &
     \multicolumn{1}{c}{$\sigma$} &
     \multicolumn{1}{c}{$\Delta M$} &
     \multicolumn{1}{c}{$\delta M$} \\
    &&&&&
     \multicolumn{1}{c}{(mas)} & \multicolumn{1}{c}{(mas)} &
     \multicolumn{1}{c}{(mas)} & \multicolumn{1}{c}{(mas)} &
     \multicolumn{1}{c}{(mas)} &
     \multicolumn{1}{c}{(mag)} & \multicolumn{1}{c}{(mag)}  \\
    \hline
    \endfirsthead

    \multicolumn{11}{c}{{\tablename\ \thetable{} -- continued from previous page}} \\
    \hline\hline
     Date & UTC & Tel. & Cam. & Filter &
     \multicolumn{1}{c}{$X_o$} &
     \multicolumn{1}{c}{$Y_o$} &
     \multicolumn{1}{c}{$X_{o-c}$} &
     \multicolumn{1}{c}{$Y_{o-c}$} &
     \multicolumn{1}{c}{$\sigma$} &
     \multicolumn{1}{c}{$\Delta M$} &
     \multicolumn{1}{c}{$\delta M$} \\
    &&&&&
     \multicolumn{1}{c}{(mas)} & \multicolumn{1}{c}{(mas)} &
     \multicolumn{1}{c}{(mas)} & \multicolumn{1}{c}{(mas)} &
     \multicolumn{1}{c}{(mas)} &
     \multicolumn{1}{c}{(mag)} & \multicolumn{1}{c}{(mag)}  \\
    \hline
    \endhead

    \hline \multicolumn{11}{r}{{Continued on next page}} \\ \hline
    \endfoot

    \hline
    \endlastfoot
2000-02-22 & 10:16:57.26 & CFHT & PUEO & Kp & 563.6 & -60.7 & 68.3 & 11.7 & 35.0 & 5.5 & 0.4  \\
2000-02-22 & 10:41:38.92 & CFHT & PUEO & H & 533.7 & -82.6 & 33.6 & -17.7 & 35.0 & 5.7 & 0.2  \\
2000-02-22 & 11:09:44.65 & CFHT & PUEO & H & 537.3 & -74.8 & 32.1 & -18.6 & 35.0 & 5.7 & 0.3  \\
2000-02-22 & 11:36:51.89 & CFHT & PUEO & J & 541.4 & -79.5 & 31.7 & -31.7 & 35.0 & 5.5 & 0.8  \\
2000-02-23 & 06:59:14.60 & CFHT & PUEO & H & 354.5 & 256.3 & 41.1 & -4.4 & 35.0 & 5.0 & 0.5  \\
2000-02-23 & 07:05:41.06 & CFHT & PUEO & J & 350.0 & 259.0 & 39.5 & -2.6 & 35.0 & 4.7 & 1.1  \\
2000-02-23 & 07:09:24.54 & CFHT & PUEO & J & 348.6 & 261.9 & 39.8 & -0.3 & 35.0 & 4.8 & 1.1  \\
2000-02-23 & 07:16:34.03 & CFHT & PUEO & Kp & 334.9 & 260.9 & 29.4 & -2.3 & 35.0 & 5.0 & 0.2  \\
2000-02-23 & 07:20:29.00 & CFHT & PUEO & Kp & 333.3 & 262.1 & 29.6 & -1.7 & 35.0 & 4.9 & 0.3  \\
2000-02-23 & 07:24:50.44 & CFHT & PUEO & IR & 332.1 & 262.7 & 30.4 & -1.7 & 35.0 & 5.4 & 0.2  \\
2000-02-23 & 07:29:24.51 & CFHT & PUEO & IR & 337.6 & 262.6 & 38.0 & -2.5 & 35.0 & 5.0 & 0.7  \\
2000-02-23 & 07:41:48.00 & CFHT & PUEO & H2 & 333.8 & 264.8 & 40.0 & -2.0 & 35.0 & 5.4 & 0.3  \\
2000-02-23 & 07:44:47.37 & CFHT & PUEO & H  & 281.1 & 276.0 & -11.3 & 8.8 & 35.0 & 5.2 & 0.2  \\
2000-02-23 & 08:02:30.00 & CFHT & PUEO & Jc & 333.2 & 262.1 & 49.2 & -7.5 & 35.0 & 4.9 & 1.1  \\
2002-09-27 & 07:51:10.15 & Keck II & NIRC2 & Kp & 302.1 & -81.1 & 21.8 & 0.4 & 10.0 & 5.3 & 0.1  \\
2003-07-14 & 14:11:03.34 & Keck II & NIRC2 & H & -227.7 & -209.6 & -5.5 & -0.6 & 10.0 & 5.1 & 0.6  \\
2003-08-10 & 15:01:24.87 & Keck II & NIRC2 & Kp & 325.9 & 185.5 & 8.8 & -2.5 & 10.0 & 5.0 & 0.1  \\
2003-08-10 & 15:04:56.89 & Keck II & NIRC2 & Kp & 324.9 & 182.6 & 7.2 & -4.5 & 10.0 & 5.0 & 0.0  \\
2003-08-14 & 12:10:45.05 & Keck II & NIRC2 & H & 242.9 & 274.4 & 3.3 & 11.8 & 10.0 & 5.8 & 0.1  \\
2003-08-14 & 13:12:09.16 & Keck II & NIRC2 & H & 287.3 & 248.7 & 30.6 & -4.2 & 10.0 & 5.7 & 0.1  \\
2003-08-17 & 14:32:00.46 & Keck II & NIRC2 & Kp & -284.8 & 142.8 & -11.8 & -8.3 & 10.0 & 5.1 & 0.1  \\
2003-08-18 & 14:00:41.12 & Keck II & NIRC2 & Kp & 222.5 & 280.0 & 2.3 & 0.8 & 10.0 & 5.3 & 0.2  \\
2003-08-18 & 14:01:46.28 & Keck II & NIRC2 & Kp & 223.6 & 281.4 & 3.0 & 2.3 & 10.0 & 5.2 & 0.0  \\
2003-08-18 & 15:29:17.68 & Keck II & NIRC2 & Kp & 231.3 & 249.3 & -15.6 & -17.1 & 10.0 & 4.9 & 0.3  \\
2004-10-30 & 14:37:44.88 & Gem-N & NIRI & Kp & -364.4 & -236.8 & -9.2 & -14.6 & 21.9 & 4.9 & 0.3  \\
2004-11-02 & 12:06:44.30 & Gem-N & NIRI & Kp & 224.3 & 14.4 & 8.6 & -6.7 & 21.9 & 4.8 & 0.5  \\
2004-11-03 & 13:45:17.53 & Gem-N & NIRI & Kp & -338.6 & -241.2 & -15.4 & -10.3 & 21.9 & 4.6 & 0.2  \\
2004-11-05 & 14:26:03.90 & Gem-N & NIRI & Kp & 326.3 & 246.7 & 9.4 & 15.6 & 21.9 & 4.4 & 0.0  \\
2006-03-18 & 12:35:45.87 & Gem-N & NIRI & Kp & -204.7 & -347.5 & -10.8 & -7.2 & 21.9 & 4.9 & 0.3  \\
2006-03-18 & 12:43:50.10 & Gem-N & NIRI & Kp & -195.5 & -344.5 & -5.4 & -2.4 & 21.9 & 4.6 & 0.0  \\
2006-03-18 & 12:48:05.20 & Gem-N & NIRI & Kp & -193.7 & -347.9 & -5.6 & -4.8 & 21.9 & 4.8 & 0.1  \\
2006-04-28 & 08:26:40.73 & Gem-N & NIRI & Kp & -450.1 & -83.2 & -10.6 & 0.9 & 21.9 & 4.7 & 0.0  \\
2006-04-28 & 08:31:36.40 & Gem-N & NIRI & Kp & -451.8 & -83.5 & -13.2 & 2.5 & 21.9 & 4.3 & 0.0  \\
2006-05-16 & 06:04:56.40 & Gem-N & NIRI & Kp & 334.8 & -246.9 & 5.8 & 3.8 & 21.9 & 5.3 & 0.6  \\
2006-05-18 & 07:17:31.00 & Gem-N & NIRI & Kp & -334.1 & 265.8 & -7.8 & 18.6 & 21.9 & 5.1 & 0.1  \\
2006-06-23 & 06:00:02.53 & Gem-N & NIRI & Kp & 304.0 & 136.8 & 18.1 & 17.7 & 21.9 & 5.4 & 0.2  \\
2006-06-23 & 06:18:51.70 & Gem-N & NIRI & Kp & 298.1 & 135.8 & 16.4 & 11.8 & 21.9 & 5.5 & 0.3  \\
2007-07-26 & 04:30:42.27 & ESO/VLT & NACO & Ks & -80.1 & -132.6 & -7.5 & -19.2 & 13.3 & 7.0 & 0.6  \\
2007-07-27 & 02:53:10.53 & ESO/VLT & NACO & H & 311.6 & -261.0 & 10.9 & -7.3 & 13.3 & 4.9 & 0.0  \\
2009-09-02 & 13:45:51.49 & Keck II & NIRC2 & H & -263.4 & -281.5 & -6.8 & -7.7 & 10.0 & 4.6 & 0.0  \\
2009-09-02 & 13:50:27.59 & Keck II & NIRC2 & J & -264.4 & -282.1 & -6.8 & -9.1 & 10.0 & 4.6 & 0.0  \\
2009-09-02 & 13:53:23.28 & Keck II & NIRC2 & FeII & -264.3 & -273.4 & -6.1 & -0.9 & 10.0 & 4.8 & 0.1  \\
2009-09-02 & 13:55:31.40 & Keck II & NIRC2 & FeII & -265.8 & -277.7 & -7.1 & -5.6 & 10.0 & 4.8 & 0.0  \\
2009-09-02 & 14:20:44.71 & Keck II & NIRC2 & J & -259.2 & -262.7 & 4.9 & 4.8 & 10.0 & 4.5 & 0.4  \\
2009-10-01 & 14:18:23.50 & Keck II & NIRC2 & Kp & -334.0 & -288.0 & -7.8 & -11.3 & 10.0 & 4.7 & 0.0  \\
2009-11-24 & 04:52:40.75 & Gem-N & NIRI & Kp & -289.1 & -360.4 & -11.2 & -18.1 & 21.9 & 4.4 & 0.0  \\
2009-11-25 & 05:36:00.71 & Gem-N & NIRI & Kp & -318.1 & 191.6 & -10.7 & 7.5 & 21.9 & 4.5 & 0.0  \\
2010-03-02 & 05:13:40.74 & Gem-N & NIRI & Kp & 164.0 & -76.7 & -10.0 & 29.4 & 21.9 & 4.5 & 1.0  \\
2010-09-30 & 15:10:34.20 & Gem-N & NIRI & Kp & 210.8 & 61.8 & 29.6 & 19.7 & 21.9 & 1.3 & 0.0  \\
2010-09-30 & 15:15:40.08 & Gem-N & NIRI & Kp & 213.0 & 76.3 & 30.5 & 33.7 & 21.9 & -4.7 & 0.0  \\
2010-10-09 & 15:19:07.45 & Gem-N & NIRI & Kp & 314.4 & 115.1 & 8.2 & 3.1 & 21.9 & 4.7 & 0.0  \\
2010-10-09 & 15:21:43.45 & Gem-N & NIRI & Kp & 311.7 & 113.7 & 5.8 & 1.7 & 21.9 & 4.5 & 0.0  \\
2010-10-09 & 15:26:12.33 & Gem-N & NIRI & Kp & 315.0 & 115.6 & 9.4 & 3.5 & 21.9 & 4.6 & 0.1  \\
2011-02-12 & 08:10:55.80 & Gem-N & NIRI & Kp & -519.5 & -142.2 & -13.2 & 5.0 & 21.9 & 5.1 & 0.1  \\
2011-02-12 & 08:15:41.55 & Gem-N & NIRI & Kp & -527.8 & -159.5 & -21.2 & -11.5 & 21.9 & 4.8 & 0.1  \\
2011-04-15 & 05:49:37.20 & Gem-N & NIRI & Kp & -367.1 & -98.1 & 6.2 & 4.9 & 21.9 & 4.5 & 0.1  \\
2011-04-15 & 05:52:32.20 & Gem-N & NIRI & Kp & -375.8 & -114.0 & -2.2 & -10.6 & 21.9 & 4.8 & 0.3  \\
2011-04-15 & 05:57:24.40 & Gem-N & NIRI & Kp & -380.0 & -117.5 & -6.0 & -13.4 & 21.9 & 4.8 & 0.3  \\
2011-04-18 & 05:19:16.70 & Gem-N & NIRI & Kp & 136.4 & 168.2 & -8.1 & 20.8 & 21.9 & 4.2 & 0.4  \\
2011-04-18 & 05:23:35.58 & Gem-N & NIRI & Kp & 130.5 & 171.7 & -12.4 & 24.7 & 21.9 & 5.0 & 0.0  \\
2012-05-01 & 05:54:05.80 & ESO/VLT & NACO & Ks & 140.8 & 343.6 & -1.5 & 18.9 & 13.3 & 4.8 & 0.1  \\
2013-06-03 & 09:29:42.28 & ESO/VLT & NACO & Ks & 280.4 & -82.1 & 5.8 & -0.5 & 13.3 & 5.1 & 0.0  \\
2013-06-04 & 08:48:34.43 & ESO/VLT & NACO & Ks & 261.4 & -110.1 & 2.2 & -1.1 & 13.3 & 4.9 & 0.2  \\
2013-06-05 & 09:46:59.86 & ESO/VLT & NACO & Ks & -257.9 & 58.3 & -7.2 & -13.8 & 13.3 & 5.5 & 0.1  \\
2013-08-26 & 08:00:27.36 & Keck II & NIRC2 & Kp & 155.3 & -72.0 & 1.8 & 1.5 & 10.0 & 5.2 & 0.1  \\
2013-08-26 & 08:03:35.80 & Keck II & NIRC2 & H & 149.9 & -76.4 & -2.6 & -3.4 & 10.0 & 5.3 & 0.8  \\
2013-08-26 & 08:08:27.03 & Keck II & NIRC2 & Kp & 160.1 & -78.3 & 9.1 & -6.1 & 10.0 & 5.0 & 0.5  \\
2019-09-11 & 06:48:18.14 & Keck II & NIRC2 & H & -134.6 & 152.3 & -11.1 & 5.5 & 10.0 & 5.2 & 0.2  \\
  \end{longtable}
\footnotesize{
    Date, mid-observing time (UTC), telescope, camera, filter,
    astrometry ($X$ is aligned with Right Ascension, and $Y$ with Declination, and
    $o$ and $c$ indices stand for observed and computed positions),
    and photometry (magnitude difference $\Delta M$ with uncertainty $\delta M$).}
\end{center}

\begin{center}
  \begin{longtable}{cclllrrrrrrr}
  \caption[Astrometry of Emma]{Astrometry of Emma's satellite S/2003 (283) 1, see \Cref{tab:pulcamoon} for a description of columns.
    \label{tab:emmoon}
  }\\

    \hline\hline
     Date & UTC & Tel. & Cam. & Filter &
     \multicolumn{1}{c}{$X_o$} &
     \multicolumn{1}{c}{$Y_o$} &
     \multicolumn{1}{c}{$X_{o-c}$} &
     \multicolumn{1}{c}{$Y_{o-c}$} &
     \multicolumn{1}{c}{$\sigma$} &
     \multicolumn{1}{c}{$\Delta M$} &
     \multicolumn{1}{c}{$\delta M$} \\
    &&&&
     \multicolumn{1}{c}{(mas)} & \multicolumn{1}{c}{(mas)} &
     \multicolumn{1}{c}{(mas)} & \multicolumn{1}{c}{(mas)} &
     \multicolumn{1}{c}{(mas)} &
     \multicolumn{1}{c}{(mag)} & \multicolumn{1}{c}{(mag)}  \\
    \hline
    \endfirsthead

    \multicolumn{11}{c}{{\tablename\ \thetable{} -- continued from previous page}} \\
    \hline\hline
     Date & UTC & Tel. & Cam. & Filter &
     \multicolumn{1}{c}{$X_o$} &
     \multicolumn{1}{c}{$Y_o$} &
     \multicolumn{1}{c}{$X_{o-c}$} &
     \multicolumn{1}{c}{$Y_{o-c}$} &
     \multicolumn{1}{c}{$\sigma$} &
     \multicolumn{1}{c}{$\Delta M$} &
     \multicolumn{1}{c}{$\delta M$} \\
    &&&&&
     \multicolumn{1}{c}{(mas)} & \multicolumn{1}{c}{(mas)} &
     \multicolumn{1}{c}{(mas)} & \multicolumn{1}{c}{(mas)} &
     \multicolumn{1}{c}{(mas)} &
     \multicolumn{1}{c}{(mag)} & \multicolumn{1}{c}{(mag)}  \\
    \hline
    \endhead

    \hline \multicolumn{11}{r}{{Continued on next page}} \\ \hline
    \endfoot

    \hline
    \endlastfoot
2003-07-14 & 13:27:33.79 & Keck II & NIRC2 & Kp & -162.2 & -204.6 & 2.8 & -1.3 & 5.0 & 5.4 & 0.1  \\
2003-07-14 & 13:43:51.23 & Keck II & NIRC2 & H & -162.4 & -213.2 & 0.4 & -0.5 & 5.0 & 5.0 & 0.1  \\
2003-07-14 & 14:22:52.32 & Keck II & NIRC2 & H & -156.1 & -237.8 & 1.0 & -3.0 & 5.0 & 5.2 & 0.1  \\
2003-07-15 & 06:55:23.68 & ESO/VLT & NACO & H & 94.2 & -386.4 & -0.3 & -0.5 & 27.1 & 4.5 & 0.1  \\
2003-07-15 & 07:13:30.66 & ESO/VLT & NACO & H & 99.4 & -384.5 & 0.3 & -3.6 & 27.1 & 4.7 & 0.1  \\
2003-07-15 & 07:17:01.87 & ESO/VLT & NACO & Ks & 99.0 & -382.9 & -1.0 & -3.0 & 27.1 & 4.5 & 0.0  \\
2003-07-15 & 07:20:20.00 & ESO/VLT & NACO & J & 100.4 & -383.4 & -0.4 & -4.4 & 27.1 & 4.6 & 0.1  \\
2003-07-15 & 10:18:18.05 & ESO/VLT & NACO & H & 151.6 & -323.2 & 9.1 & -5.4 & 27.1 & 5.7 & 0.6  \\
2003-07-16 & 10:27:24.63 & ESO/VLT & NACO & Ks & 168.1 & 374.8 & 4.7 & -24.1 & 27.1 & 4.9 & 0.5  \\
2003-08-10 & 11:15:37.42 & Keck II & NIRC2 & H & -235.0 & -27.0 & -0.4 & 0.1 & 5.0 & 4.8 & 0.1  \\
2003-08-10 & 11:21:37.99 & Keck II & NIRC2 & H & -232.5 & -32.2 & 2.0 & -0.9 & 5.0 & 5.0 & 0.1  \\
2003-08-14 & 09:56:33.69 & Keck II & NIRC2 & H & -69.6 & -423.1 & 3.6 & 9.6 & 5.0 & 5.3 & 0.0  \\
2003-08-14 & 10:01:07.46 & Keck II & NIRC2 & H & -64.6 & -427.1 & 7.0 & 6.4 & 5.0 & 5.0 & 0.1  \\
2003-08-14 & 12:02:42.55 & Keck II & NIRC2 & H & -22.7 & -440.1 & 5.1 & 8.5 & 5.0 & 4.9 & 0.0  \\
2003-08-15 & 08:43:56.41 & Keck II & NIRC2 & Kp & 285.4 & 25.8 & 3.0 & 0.4 & 5.0 & 5.0 & 0.0  \\
2003-08-15 & 08:46:06.17 & Keck II & NIRC2 & Kp & 287.6 & 25.9 & 5.1 & -0.8 & 5.0 & 5.2 & 0.1  \\
2003-08-16 & 08:50:51.79 & Keck II & NIRC2 & Kp & 31.8 & 521.2 & 6.2 & -13.6 & 5.0 & 5.1 & 0.1  \\
2003-08-16 & 08:54:09.34 & Keck II & NIRC2 & Kp & 27.8 & 527.3 & 3.2 & -7.3 & 5.0 & 5.3 & 0.1  \\
2003-08-17 & 11:25:49.17 & Keck II & NIRC2 & Kp & -205.7 & -242.8 & 3.7 & -0.2 & 5.0 & 5.3 & 0.3  \\
2003-08-17 & 11:29:47.37 & Keck II & NIRC2 & Kp & -207.6 & -246.9 & 1.0 & -2.0 & 5.0 & 4.8 & 1.2  \\
2003-08-18 & 11:16:21.70 & Keck II & NIRC2 & Kp & 235.2 & -208.1 & -1.1 & -1.2 & 5.0 & 5.0 & 0.2  \\
2003-08-18 & 11:19:44.20 & Keck II & NIRC2 & Kp & 237.3 & -205.8 & 0.3 & -0.7 & 5.0 & 5.2 & 0.0  \\
2004-10-30 & 12:16:14.56 & Gem-N & NIRI & Kp & 449.9 & -13.7 & -9.5 & 3.4 & 10.9 & 4.9 & 0.0  \\
2004-10-30 & 12:20:25.00 & Gem-N & NIRI & Kp & 451.2 & -22.5 & -8.0 & -3.7 & 10.9 & 4.7 & 0.1  \\
2004-10-30 & 15:23:38.50 & Gem-N & NIRI & Kp & 432.3 & -98.0 & -8.4 & -3.8 & 10.9 & 5.4 & 0.1  \\
2004-11-02 & 15:20:25.30 & Gem-N & NIRI & Kp & 432.6 & 135.1 & -11.4 & -1.6 & 10.9 & 4.9 & 0.0  \\
2004-11-05 & 10:30:54.30 & Gem-N & NIRI & Kp & 143.2 & 372.0 & -7.5 & 2.4 & 10.9 & 4.4 & 0.2  \\
2004-11-14 & 06:31:27.98 & ESO/VLT & NACO & Ks & -251.1 & -265.0 & 4.4 & -0.6 & 6.6 & 4.7 & 0.0  \\
2004-11-15 & 05:42:44.13 & ESO/VLT & NACO & Ks & -126.1 & 350.0 & -4.2 & -2.2 & 6.6 & 4.9 & 0.2  \\
2004-11-16 & 04:58:43.32 & ESO/VLT & NACO & Ks & 456.2 & 133.5 & -10.3 & 0.5 & 6.6 & 5.3 & 0.0  \\
2004-11-16 & 05:56:33.99 & ESO/VLT & NACO & Ks & 463.1 & 109.0 & -10.7 & 0.0 & 6.6 & 4.8 & 0.0  \\
2004-11-17 & 05:08:26.55 & ESO/VLT & NACO & Ks & 93.9 & -359.5 & -0.5 & 0.7 & 6.6 & 4.8 & 0.1  \\
2004-11-18 & 06:19:14.29 & ESO/VLT & NACO & Ks & -339.9 & 162.3 & 3.1 & -2.3 & 6.6 & 5.2 & 0.1  \\
2004-12-07 & 03:38:52.48 & ESO/VLT & NACO & Ks & 352.6 & -235.6 & -5.5 & -1.7 & 6.6 & 5.1 & 0.0  \\
2004-12-07 & 03:55:46.41 & ESO/VLT & NACO & H & 346.9 & -243.9 & -4.9 & -4.1 & 6.6 & 5.1 & 0.0  \\
2004-12-07 & 04:11:37.00 & ESO/VLT & NACO & J & 343.6 & -248.1 & -2.2 & -2.8 & 6.6 & 5.2 & 0.1  \\
2004-12-08 & 04:17:32.34 & ESO/VLT & NACO & Ks & -342.7 & -163.5 & 9.0 & 2.8 & 6.6 & 5.1 & 0.1  \\
2004-12-10 & 05:46:42.04 & ESO/VLT & NACO & Ks & 458.1 & -61.4 & -6.3 & -3.4 & 6.6 & 4.9 & 0.1  \\
2004-12-14 & 03:55:27.72 & ESO/VLT & NACO & Ks & 205.3 & -333.3 & 3.4 & -3.4 & 6.6 & 5.1 & 0.1  \\
2004-12-14 & 04:12:40.24 & ESO/VLT & NACO & H & 192.7 & -332.4 & -0.6 & 1.1 & 6.6 & 5.2 & 0.0  \\
2004-12-14 & 04:32:17.00 & ESO/VLT & NACO & J & 183.1 & -330.9 & -0.2 & 6.6 & 6.6 & 5.8 & 0.2  \\
2004-12-19 & 03:29:18.70 & ESO/VLT & NACO & Ks & -83.8 & 345.1 & 2.0 & -6.6 & 6.6 & 4.9 & 0.1  \\
2004-12-20 & 01:42:23.80 & ESO/VLT & NACO & Ks & 441.4 & 147.3 & -9.3 & -1.7 & 6.6 & 5.1 & 0.0  \\
2004-12-20 & 04:33:50.87 & ESO/VLT & NACO & Ks & 457.2 & 76.5 & -7.4 & -1.0 & 6.6 & 4.9 & 0.0  \\
2004-12-28 & 02:44:40.39 & ESO/VLT & NACO & Ks & -101.8 & -346.4 & 6.2 & 0.9 & 6.6 & 4.8 & 0.1  \\
2004-12-28 & 04:52:34.74 & ESO/VLT & NACO & Ks & -162.6 & -323.1 & 8.0 & 1.4 & 6.6 & 5.0 & 0.1  \\
2006-06-07 & 05:58:15.80 & Gem-N & NIRI & Kp & -4.2 & 202.1 & 15.8 & -2.2 & 10.9 & 4.5 & 0.9  \\
2009-11-24 & 06:06:17.75 & Gem-N & NIRI & Kp & -297.5 & 33.0 & -5.2 & -5.1 & 10.9 & 4.8 & 0.0  \\
2009-11-25 & 05:28:36.52 & Gem-N & NIRI & Kp & 189.5 & 335.3 & 4.9 & -3.5 & 10.9 & 4.5 & 0.0  \\
2010-09-30 & 14:55:08.20 & Gem-N & NIRI & Kp & -68.0 & -208.5 & -21.9 & -18.0 & 10.9 & -2.1 & 0.0  \\
2010-10-13 & 14:35:29.70 & Gem-N & NIRI & Kp & 113.6 & -164.7 & 5.3 & -5.1 & 10.9 & 6.6 & 1.6  \\
2010-10-21 & 15:35:12.15 & Gem-N & NIRI & Kp & -164.4 & 91.8 & -7.1 & 18.8 & 10.9 & 100.0 & 0.0  \\
2013-06-15 & 05:10:30.83 & ESO/VLT & NACO & Ks & -5.9 & -370.5 & 0.8 & 8.7 & 6.6 & 6.4 & 0.2  \\
2013-06-15 & 05:57:23.71 & ESO/VLT & NACO & Ks & 20.1 & -376.6 & 4.6 & 5.5 & 6.6 & 7.3 & 0.2  \\
2013-06-15 & 07:07:05.40 & ESO/VLT & NACO & Ks & 41.5 & -382.6 & -6.8 & 1.7 & 6.6 & 5.9 & 0.4  \\
2013-06-17 & 04:59:48.19 & ESO/VLT & NACO & H & -77.7 & 312.4 & 2.6 & -5.0 & 6.6 & 4.9 & 0.3  \\
2013-06-17 & 06:26:58.65 & ESO/VLT & NACO & H & -126.6 & 297.9 & -0.4 & -7.1 & 6.6 & 5.0 & 0.2  \\
  \end{longtable}
\end{center}

\section{Occultation predictions}
\label{app:occ}

The stellar occultation predictions presented in \Cref{tab:occ_emma} and \Cref{tab:occ_pulc} were obtained with the SORA python package \citep{sora_pred} using JPL planetary and lunar ephemeris DE440 \citep{park2021jpl}.
\begin{table}[h!]
\begin{centering}
\caption{List of stellar occultation predictions for (283) Emma. }
\begin{tabular}{lllllll}
\hline\hline
Epoch UTC & ICRS Star Coord at Epoch & C/A & P/A & Vel & Dist & G \\
yyyy-mm-dd hh:mm:ss & h m s \quad\quad\quad\quad\quad$\mathrm{{}^{\circ}}$ $\mathrm{{}^{\prime}}$ $\mathrm{{}^{\prime\prime}}$ & $\mathrm{{}^{\prime\prime}}$ & $\mathrm{{}^{\circ}}$ & $\mathrm{km\ s^{-1}}$ & $\mathrm{AU}$ & $\mathrm{mag}$  \\
\hline\\
2025-08-15 19:26:34.520 & 02 34 51.35569 +24 39 25.00063 & 2.085 & 144.73 & 13.53 & 2.283 & 15.495 \\
2025-10-08 12:06:10.860 & 02 37 41.14978 +28 05 08.21590 & 1.809 & 358.49 & -7.55 & 1.774 & 13.496 \\
2025-10-24 22:34:21.760 & 02 24 53.85482 +27 35 11.49668 & 1.972 & 343.77 & -10.46 & 1.717 & 13.636 \\
2025-11-01 03:50:15.260 & 02 18 31.35750 +27 06 24.81280 & 2.995 & 338.9 & -10.9 & 1.714 & 14.036 \\
2025-12-12 14:46:27.460 & 01 55 26.85157 +23 17 09.95675 & 3.129 & 100.64 & -4.63 & 1.959 & 12.204 \\
2025-12-14 13:35:41.100 & 01 55 22.59585 +23 08 23.39293 & 2.886 & 91.83 & -4.37 & 1.98 & 14.898 \\
2027-01-03 03:20:55.820 & 09 02 15.36344 +16 49 37.16820 & 1.808 & 4.9 & -11.01 & 2.367 & 13.981 \\
2027-01-27 16:43:17.240 & 08 42 30.65272 +17 20 52.71581 & 2.679 & 186.56 & -14.84 & 2.294 & 15.785 \\
2027-02-26 01:40:19.860 & 08 19 41.00071 +17 52 01.98768 & 1.84 & 3.11 & -9.88 & 2.442 & 15.033 \\
2028-03-16 11:54:21.420 & 12 30 13.81997 -14 02 04.13225 & 1.757 & 194.43 & -14.04 & 2.538 & 14.338 \\
2028-06-05 10:00:38.480 & 11 56 46.83882 -09 19 11.22410 & 0.729 & 169.88 & 5.61 & 3.018 & 14.441 \\
2029-04-03 07:21:10.200 & 17 26 08.48697 -31 13 14.49667 & 0.287 & 22.87 & 6.28 & 2.7 & 15.233 \\
2029-04-04 10:09:47.820 & 17 26 29.80911 -31 15 16.00401 & 2.48 & 24.26 & 5.83 & 2.684 & 15.797 \\
2029-04-13 08:33:23.600 & 17 28 17.33700 -31 30 04.90703 & 0.513 & 52.71 & 2.5 & 2.558 & 14.983 \\
2029-05-09 16:17:00.660 & 17 21 59.38875 -31 55 43.16345 & 0.329 & 358.45 & -8.13 & 2.254 & 15.219 \\
2029-05-27 04:17:17.720 & 17 09 10.18469 -31 45 26.34116 & 0.693 & 7.45 & -11.98 & 2.131 & 14.837 \\
2029-06-12 23:30:02.300 & 16 53 45.78689 -31 06 44.64872 & 4.067 & 14.77 & -12.86 & 2.09 & 15.757 \\
\hline\\
\end{tabular}\\
\footnotesize{The table contains the date and instant of the closest approach (UTC), the ICRS (J2000) star coordinates at occultation, the closest apparent distance between the
star and the target ($C/A$), the position angle of the shadow across the Earth ($P/A$) (counter-clockwise, zero at South), the velocity in $km/s$, the distance ($D$) to the Earth ($AU$), and the Gaia average magnitude ($G$). }
\label{tab:occ_emma}
\end{centering}
\end{table}

\begin{table}[h!]
\begin{centering}
\caption{List of stellar occultation predictions for (762) Pulcova,  \Cref{tab:occ_emma} for a description of columns. }
\begin{tabular}{lllllll}
\hline\hline
Epoch UTC & ICRS Star Coord at Epoch & C/A & P/A & Vel & Dist & G \\
yyyy-mm-dd hh:mm:ss & h m s \quad\quad\quad\quad\quad$\mathrm{{}^{\circ}}$ $\mathrm{{}^{\prime}}$ $\mathrm{{}^{\prime\prime}}$ & $\mathrm{{}^{\prime\prime}}$ & $\mathrm{{}^{\circ}}$ & $\mathrm{km\ s^{-1}}$ & $\mathrm{AU}$ & $\mathrm{mag}$  \\
\hline\\
2025-08-26 07:46:57.460 & 22 41 15.23344 +03 05 49.41220 & 1.89 & 352.01 & -14.53 & 2.461 & 11.17 \\
2025-08-28 20:21:42.620 & 22 39 17.17723 +03 01 23.62798 & 2.719 & 350.96 & -14.69 & 2.455 & 15.836 \\
2026-09-18 04:21:17.200 & 03 36 28.45370 +35 36 05.24633 & 0.378 & 110.43 & 6.65 & 2.583 & 15.92 \\
2026-10-03 02:54:51.100 & 03 35 48.94628 +36 37 41.14428 & 0.157 & 49.18 & -5.35 & 2.401 & 13.574 \\
2026-10-04 06:26:35.160 & 03 35 31.47424 +36 41 28.02676 & 1.746 & 224.55 & -5.5 & 2.388 & 14.999 \\
2026-10-20 11:25:09.140 & 03 27 44.30484 +37 14 26.07831 & 0.43 & 5.03 & -9.1 & 2.232 & 15.323 \\
2026-10-22 08:43:00.100 & 03 26 25.55525 +37 15 27.25459 & 0.291 & 2.45 & -9.54 & 2.217 & 15.732 \\
2026-10-31 04:40:25.420 & 03 19 23.84902 +37 11 12.79817 & 3.5 & 172.49 & -11.36 & 2.16 & 14.287 \\
2026-11-05 22:10:08.060 & 03 14 14.49107 +36 59 57.67113 & 0.56 & 347.14 & -12.25 & 2.133 & 14.056 \\
2026-11-20 00:57:03.040 & 03 00 55.36704 +36 04 46.02633 & 0.466 & 335.15 & -13.04 & 2.106 & 15.049 \\
2026-12-04 22:17:00.360 & 02 48 41.63083 +34 33 43.73080 & 0.415 & 321.36 & -11.75 & 2.138 & 15.059 \\
2026-12-14 01:52:26.520 & 02 43 21.48768 +33 29 49.05517 & 1.504 & 309.79 & -10.1 & 2.186 & 15.753 \\
2027-01-03 12:23:28.960 & 02 39 25.49566 +31 15 21.39431 & 0.654 & 79.21 & 7.01 & 2.359 & 15.973 \\
2027-01-07 23:23:52.180 & 02 40 04.48920 +30 50 34.42370 & 0.347 & 243.68 & 7.19 & 2.405 & 15.029 \\
2029-04-24 10:59:18.640 & 15 54 03.70038 -37 53 36.00516 & 0.788 & 178.59 & -9.69 & 2.299 & 15.465 \\
2029-05-11 00:12:56.220 & 15 40 18.62508 -37 33 09.56181 & 0.329 & 193.95 & -12.62 & 2.233 & 15.843 \\
2029-05-11 17:07:27.540 & 15 39 39.81542 -37 31 13.29291 & 0.471 & 14.5 & -12.7 & 2.232 & 14.434 \\
2029-05-23 16:01:32.000 & 15 28 40.39682 -36 45 37.17132 & 0.785 & 23.7 & -13.19 & 2.231 & 13.471 \\
2029-06-04 04:36:28.600 & 15 19 06.10678 -35 43 34.37645 & 3.174 & 33.07 & -12.42 & 2.266 & 15.774 \\
\hline\\
\end{tabular}\\
\label{tab:occ_pulc}
\end{centering}
\end{table}
\FloatBarrier

\section{Alternate orbital solution for (283) Emma}

\begin{table}[h!]
\begin{center}
  \caption[Orbital elements of the satellite of Emma]{%
    Orbital elements of the satellite of Emma,
    expressed in EQJ2000, obtained with \genoid; see \Cref{tab:pulcorbj2} for a description of parameters.}
  \label{tab:emma_alt_solution}
   \begin{tabular}{l ll}
    \hline\hline
    & \multicolumn{2}{c}{S2003-283-1}\\
    \hline
  \noalign{\smallskip}
  \multicolumn{2}{c}{Observing data set} \\
  \noalign{\smallskip}
    Number of observations  & \multicolumn{2}{c}{56} \\
    Time span (days)        & \multicolumn{2}{c}{3626} \\
    RMS (mas)               & \multicolumn{2}{c}{5.34} \\
    \hline
  \noalign{\smallskip}
  \multicolumn{2}{c}{Orbital elements EQJ2000} \\
  \noalign{\smallskip}
    $P$ (day)         & 3.40735 & $\pm$ 0.00174 \\
    $a$ (km)          & 583.9 & $\pm$ 17.3 \\
    $e$               & 0.116 & $\pm$ 0.018 \\
    $i$ (\degr)       & 94.1 & $\pm$ 5.5 \\
    $\Omega$ (\degr)  & 347.5 & $\pm$ 3.8 \\
    $\omega$ (\degr)  & 206.8 & $\pm$ 10.6 \\
    $t_{p}$ (JD)      & 2452835.12416 & $\pm$ 0.09474 \\
    $J_2$ & 0.13 & $\pm$ 0.01 \\
    \hline
  \noalign{\smallskip}
  \multicolumn{2}{c}{Derived parameters} \\
  \noalign{\smallskip}
    $M_{\textrm{Emma}}$ ($\times 10^{18}$ kg)      & 1.364 & $\pm$ 0.120 \\
    $\lambda_p,\,\beta_p$ (\degr)  & 257, +18 & $\pm$ 4, 6 \\
    $\alpha_p,\,\delta_p$ (\degr)  & 258, -4 & $\pm$ 4, 6 \\
    $\Lambda$ (\degr)              & 0.8 & $_{-0.8 }^{+8.9}$ \\
    $\rho$ ( g~cm$^{-3}$) & 1.1 &$\pm0.1$ \\
    \hline
  \end{tabular}
\end{center}
\end{table}

\end{document}